\documentclass[prd,preprintnumbers,superscriptaddress,showpacs,showkeys,nofootinbib]{revtex4-1}
\usepackage{amsmath,amsfonts,amssymb,amsthm,amstext,amscd,eucal,srcltx,mathrsfs}
\usepackage{epsfig,graphicx,bm}
\usepackage{epstopdf, epsf}
\usepackage{dcolumn} 
\usepackage{hyperref}
\usepackage{color}
\renewcommand{\leq}{\leqslant}
\renewcommand{\geq}{\geqslant}

\newcommand{\R}{\mathbb{R}}
\newcommand{\abs}[1]{\left|{#1}\right|}
\newcommand{\at}[2]{\left. {#1} \right|_{#2}}
\newcommand\sn[3]{\text{sn}^{#1}\left(#2,\, #3\right)}
\newcommand{\EEx}{\mathcal{E}}
\newcommand{\bigO}[1]{\mathcal{O}\left(#1\right)}

\renewcommand{\epsilon}{\varepsilon}
\newcommand{\x}{x}
\renewcommand{\u}{u}
\newcommand{\normal}[2]{\mathcal{N}\left({#1}, {#2}\right)}
\newcommand{\normalx}[2]{\mathcal{N}^{*}\left({#1}, {#2}\right)}
\newcommand{\Uphi}{{U_{\varphi}}}
\newcommand{\uphi}{{u_{\varphi}}}
\graphicspath{{figures/}}
\newcommand{\be}{\begin{eqnarray}}
\newcommand{\ee}{\end{eqnarray}}

\renewcommand{\d}{\mbox{${\rm d}$}}

\begin{document}
\title{Orbits in a stochastic Schwarzschild geometry}
\author{Roberto Casadio}
\email{casadio@bo.infn.it}
\affiliation{Dipartimento di Fisica e Astronomia,
Alma Mater Universit\`a di Bologna,
via~Irnerio~46, 40126~Bologna, Italy}
\affiliation{I.N.F.N., Sezione di Bologna, IS FLAG
viale~B.~Pichat~6/2, I-40127 Bologna, Italy}
\author{Andrea Giusti}
\email{agiusti@bo.infn.it}
\affiliation{Department of Physics \& Astronomy,
Bishop's University, 
2600 College Street, J1M 1Z7 Sherbrooke, Qu\'{e}bec, Canada}
\author{Andrea Mentrelli}
\email{andrea.mentrelli@unibo.it}
\affiliation{Alma Mater Research Center on Applied Mathematics (AM$^2$),
Dipartimento di Matematica, Universit\`a di Bologna, 
Via Saragozza 8, I-40123 Bologna, Italy}
\begin{abstract}	
We study geodesics in the Schwarzschild space-time affected by an uncertainty in the mass parameter
described by a Gaussian distribution.
This study could serve as a first attempt at investigating possible quantum effects of black hole
space-times on the motion of matter in their surroundings as well as the role of uncertainties in the
measurement of the black hole parameters.
\end{abstract}

\maketitle
\section{Introduction}
\label{S:intro}
Black holes were always one of the characterising predictions of General Relativity (GR)~\cite{Carroll, Wald},
and the recent detection of gravitational waves from the merging of black hole binaries~\cite{ligo}
has further boosted the interest in such astrophysical objects.
The mathematical properties of these vacuum solutions of the Einstein equations are
already problematic at the classical level, where it is well known that no sensible
energy-momentum tensor can be associated with them~\cite{geroch}.
It becomes even more problematic at the quantum level, since our very limited understanding
from semiclassical approaches yield the famous Hawking radiation~\cite{hawking} and a bunch of
paradoxes (see, e.g.~Refs.~\cite{Hawking:1976ra,Page:1993wv,Hawking:2005kf,Hollands:2014eia,stoica,giusti}
and references therein). 
\par
On the other hand, our observational capacities remain relatively weak in determining with precision
what astrophysical black holes are in nature, and a host of alternative, somewhat more exotic, compact
objects are actively being investigated in the present literature
(see e.g.~Refs.~\cite{giusti,Casadio:2013ulk,Casadio:2014vja,Casadio:2015bna,Capozziello:2009jg,Sotiriou:2011dz,Sotiriou:2013qea,Mazur:2001fv,
Chirenti:2007mk,Nicolini:2005vd, Bonanno:2000ep, cardoso}).
In this respect, it is very important to determine the physical consequences of alternative models
of compact objects or alternative descriptions of gravity on observable quantities, regardless of the origin
of such deviations from the simple black hole metrics of GR.
\par
For example, the Horizon Quantum Mechanics~\cite{HQM-1,HQM-2,HQM-3,HQM-4} offers an alternative
perspective to the semiclassical approach to gravity, whose aim is to put under the spotlight the quantum
features of a black hole's geometric structure inherited by a purely quantum mechanical description
of its source.
In this regard, the location of trapping surfaces becomes fuzzy because of the quantum mechanical
nature of the source, thus providing a clear motivation for the study presented here. 
\par
Another motivation can be found in light of the theory of stochastic gravity (see e.g.~\cite{hu}
and references therein), which offers an extension of semiclassical gravity based on the
classical Einstein equations sourced by the stress-energy tensor of quantum matter fields
by including the contribution of quantum fluctuations to the vacuum expectation values
of matter.
Specifically, these fluctuations are accounted for by means of a noise kernel bitensor,
and the semiclassical Einstein equations are replaced by the so called Einstein-Langevin equations.
\par
In this work, after reviewing some generalities on the structure of the orbits in both Newtonian gravity and GR,
we analyze the geodesics of a Schwarzschild space-time~\footnote{We use units with $c=1$ and denote the
Newton constant with $G$.} \cite{Schwarzschild:1916uq, Schwarzschild:1916ae}
\begin{equation}
\label{eq:schw} 
\d s^2
=
-
\left(1-\frac{2\,G\,M}{r}\right) \d t^2
+
\left(1-\frac{2\,G\,M}{r}\right)^{-1} \d r^2
+
r^2\left(\d \theta^2+\sin^2\theta \, \d \varphi^2\right)
\ ,
\end{equation}
affected by an uncertainty in the mass parameter modeled in terms of a Gaussian distribution.
\section{Orbits in Newtonian gravity and in the Schwarzschild geometry}
\label{S:oNS}
The equation that governs the radial motion of a test particle in the Schwarzschild metric~\eqref{eq:schw}
or in Newtonian physics can be written as~\cite{Carroll, Wald}
\begin{equation}
\label{eq:classical} % Eq. (5.65) Carroll
	\frac{1}{2} \left( \frac{\d r}{\d \tau} \right)^2
	=
	\frac{E^2}{2}
	-V(r)
	\equiv
	\EEx - V(r)
	\ ,
\end{equation}
where we denoted with $E$ and $L$, respectively, the conserved energy and angular momentum
per unit mass in the massive case ($\epsilon = 1$), or the conserved energy and angular momentum in the massless case
($\epsilon = 0$).
The potential in the above equation reads
\begin{equation}
	V(r)
	=
	\frac{\epsilon}{2}\left(1 - \frac{2\,G\,M}{r}\right)
	+ \frac{L^2}{2\,r^2} - \gamma\, \frac{G\,M\,L^2}{r^3}
	\ ,
\end{equation}
where the parameter $\gamma = 1$ in GR and $\gamma = 0$ (and $\varepsilon = 1$) in Newtonian gravity.
Eq.~\eqref{eq:classical} formally resembles the equation for a classical particle of unit mass and energy
$\EEx$ moving in a one-dimensional potential~\footnote{The potential $V$ is actually the \textit{potential energy}
(per unit mass).} $V$ (the conserved energy per unit mass is $E$, but the effective potential
corresponds to $\EEx = E^2/2$).
\subsection{Circular orbits}
The stationary points of the potential $V$ represent circular orbits, whose radius $r_{\rm c}$ is thus given by
\begin{equation} % Eq. (5.68) Carroll
	0
	=
	\at{\frac{dV}{dr}}{r=r_{\rm c}}
	\sim
	\epsilon\, G\,M\, r_{\rm c}^2 - L^2\, r_{\rm c} + 3\,\gamma\,G\,M\,L^2
	\ .
	\label{rcEq}
\end{equation}
Of course, these circular orbits are stable (unstable) if they correspond to a minimum (maximum)
of the potential.
\subsubsection{Newtonian gravity}
In Newtonian gravity ($\gamma = 0$), we have the following well-known results:
\begin{itemize}
\item[a)]
For massless particles ($\epsilon = 0$) no circular orbits exist.
In fact, bound orbits do not exist in general (massless particles move on a straight line).
\item[b)]
For massive particles ($\epsilon = 1$) there are stable circular orbits at the radius
\begin{equation}
\label{eq:rc-Newton}
	r_{\rm c}
	=
	\frac{L^2}{G\,M}
	\ ,
\end{equation}
as well as bound orbits that oscillate around the radius $r_{\rm c}$.
In general, if the energy is greater than the asymptotic value~\footnote{This follows from
$V(r \to \infty) = 1/2$ in Eq.~\eqref{eq:classical}.}
$E = 1$, the orbits are unbound (parabolas or hyperbolas), otherwise they are bound
(circles or ellipses).
\end{itemize}
\subsubsection{General Relativity}
\label{sec:GR-circular-orbits}
In GR ($\gamma = 1$), the term $G\,M\,L^2/r^3$ becomes important when $r$ is small and
the potential $V$ vanishes at the Schwarzschild radius $r=2\,G\,M$.
The following general results are also well-known:
\begin{itemize}
\item[a)]
For massless particles ($\epsilon = 0$), we find
\begin{equation}
\label{eq:rc-GR-massless}
	r_{\rm c}
	=
	3\,G\,M
	\equiv
	r_{\rm ph}
	\ ,
\end{equation}
which represents the innermost (unstable) circular orbit of a photon.
Note that the radius of this orbit does not depend on $L$.
\item[b)]
For massive particles ($\epsilon = 1$), the zeros of Eq.~\eqref{rcEq} are given by
\begin{equation}
\label{eq:rc-GR-mass}
	r_{\pm}
	=
	\frac{L^2}{2\,G\,M}\left(1\pm \sqrt{1- \frac{12 \,G^2\, M^2}{L^2}}\right)
	\equiv
	\frac{L^2\left(1\pm\chi\right)}{2\,G\,M}
	\ .
\end{equation}
Hence, when $L > \sqrt{12}\,G\,M \approx 3.46\,GM$ (that is $\chi$ is real),
there is an inner unstable circular orbit ($r_{\rm in}=r_-$)
and a outer stable circular orbit ($r_{\rm out}=r_+$).
For large $L$, we have
\begin{equation}
	\lim_{L\to\infty}
	r_{\rm in}
	=
	3\,G\,M
	\ ,
\end{equation}
so that the unstable orbit approaches the massless orbit~\eqref{eq:rc-GR-massless},
whereas 
\begin{equation}
	r_{\rm out}
	\sim
	\frac{L^2}{G\,M}
\end{equation}
and the stable circular orbit moves farther and farther away, approaching the Newtonian
expression~\eqref{eq:rc-Newton}. 
Conversely, decreasing $L$ the two orbits come closer together and
coincide for $L = \sqrt{12}\,G\,M$.
The common radius of this (stable) circular orbit is~\footnote{This is the Innermost Stable
Circular Orbit (ISCO).}
\begin{equation}
	r_{\rm c}
	=
	2\,r_{\rm ph}
	=
	6\,G\,M
	\ .
\end{equation}
Finally, no circular orbits are possible when $L < \sqrt{12}\,G\,M$ (since $\chi$ becomes imaginary).
\end{itemize}
\subsection{Non-circular orbits}
Non-circular orbits in GR are not perfectly closed ellipses.
Nonetheless, they can be viewed, to a good level of approximation, as ellipses that precess.
Thus, it is rather convenient to describe the evolution of the radial coordinate $r$ as a function
of the angular coordinate $\varphi$, i.e.~$r = r\left(\varphi\right)$.
\par
Recalling that $L = r^2 \, \d \varphi / \d \tau$~\cite{Carroll}, 
from Eq.~\eqref{eq:classical} we obtain
\begin{equation}
\label{eq:dr_phi} % Eq. (5.76) Carroll
\left( \frac{\d r}{\d \varphi} \right)^2
=
\frac{E^2-\epsilon}{L^2}\,r^4
+\frac{2\,\epsilon\,G\,M}{L^2}\, r^3
-r^2
+2\,\gamma\,G\,M\,r
\ .
\end{equation}
Letting $x \equiv {L^2}/{G\,M\,r}$, 
Eq.~\eqref{eq:dr_phi} becomes
\begin{equation}
\label{eq:dx_dphi} % Eq. (5.78) Carroll
\left( \frac{\d x}{\d \varphi}\right)^2
=
\frac{L^2\left(E^2-\epsilon\right)}{G^2\, M^2}
+
2\,\epsilon \,x
-x^2
+\gamma\, \frac{2\,G^2\,M^2}{L^2} \,x^3
\ ,
\end{equation}
which, upon differentiating with respect to $\varphi$, can also be written as
\begin{equation}
\label{eq:d2x_dphi2}
	\frac{\d^2 x}{\d\varphi^2}
	= 
	\epsilon - x + \gamma\, \frac{3\,G^2\,M^2}{L^2} \,x^2
	\ .
\end{equation}
Note that $x = 1$ corresponds to the Newtonian circular orbit~\eqref{eq:rc-Newton}.
%
%
%
%\section{Analytical solutions}
%
%
%\label{S:analytical}
%
%
%
\par
For the purpose of finding analytical solutions to Eq.~\eqref{eq:dx_dphi}, we define
the dimensionless parameters
\begin{equation}
	\alpha
	=
	\frac{G\,M}{L}
	\ ,
	\qquad
	\beta
	=
	\frac{G\,M}{E\,L}
	=
	\frac{\alpha}{E}
	\ ,
	\qquad
	\rho
	=
	\frac{2\,G^2\,M^2}{L^2} 
	=
	2 \, \alpha ^2
	\ ,	
\end{equation}
so that Eq.~\eqref{eq:dx_dphi} will read
\begin{equation}
\label{eq:dx_dphi_2}
	\left(\frac{\d x}{\d\varphi}\right)^2
	=
	\left(\frac{1}{\beta^2} - \frac{\epsilon}{\alpha^2}\right)
	+ 2\,\epsilon\, x
	-x^2
	+\gamma\, \rho \,x^3
	\ .
\end{equation}
Analogusly, Eq.~\eqref{eq:d2x_dphi2} can be rewritten as
\begin{equation}
\label{eq:d2x_dphi2b} % Eq. (5.79) Carroll
	\frac{\d^2 x}{\d\varphi^2}
	=
	\epsilon - x + \frac{3}{2}\,\gamma\, \rho \, x^2
	\ .
\end{equation}
\subsubsection{Newtonian gravity}
For Newtonian gravity ($\gamma = 0$), Eq.~\eqref{eq:dx_dphi_2} 
%and~\eqref{eq:d2x_dphi2b}
reduces to
%, respectively,
%
\begin{eqnarray}
\label{eq:dx_dphi_2_newton}
\left(\frac{\d x}{\d\varphi}\right)^2
&=&
\left(\frac{1}{\beta^2} - \frac{\epsilon}{\alpha^2}\right)
+ 2\,\epsilon \,x 
-x^2
\nonumber
\\
&\equiv&
- \left(\x - \x_1\right) \left(\x - \x_2\right)
\ ,
\end{eqnarray}
%
%and
%
%\begin{equation}
%	\frac{d^2 x}{d\varphi^2}
%	=
%	\epsilon - x
%	\ .
%\end{equation}
%
where the roots $x_1 = \epsilon - \kappa$ and $x_2 = \epsilon + \kappa$,
with
\begin{equation}
\kappa = \sqrt{\frac{1}{\beta^2} - \frac{\epsilon}{\alpha^2} + \epsilon^2}
\ .
%	\qquad \left(\x_1 + \x_2 = 2\epsilon\right),
\end{equation}
%
%Eq.~\eqref{eq:dx_dphi_2_newton} can be written as
%
%\begin{equation}
%	\left(\frac{d\x}{d\varphi}\right)^2 = - \left(\x - \x_1\right) \left(\x - \x_2\right)
%\end{equation}
%
The general solution to Eq.~\eqref{eq:dx_dphi_2_newton} is then given by
\begin{eqnarray}
	x\left(\varphi\right)
	&=&
	x_1 + \left( x_2 -x_1\right) \sin^2\left( \frac{\varphi}{2} + \delta \right)
\nonumber
\\
	&=&
	 \epsilon + \left(x_1 - \epsilon \right) \cos\left( \varphi + 2\,\delta\right),
	\label{eq:sol_Newton}
\end{eqnarray}
where $\delta$ is an integration constant determined by the initial condition $x(\varphi_0) = x_0$.
Eq.~\eqref{eq:sol_Newton} describes a conic of eccentricity 
\begin{equation}
e
=
\frac{x_2 -x_1}{x_2 + x_1}
=
\frac{\kappa}{\epsilon}
\ .
\end{equation}
For massive particles ($\epsilon = 1$), we see that:
\begin{itemize}
\item
	if $x_1 > 0$ and $x_1 \neq x_2$ (i.e. $0 < \kappa < 1$), the orbit is an ellipse where $x_1$ and $x_2$
	represent the distances of furthest and closest approach, respectively;
\item
	if $x_1 = x_2$ (i.e. $\kappa = 0$), the orbit is circular, with radius given by the Newtonian value~\eqref{eq:rc-Newton} (this happens when $E^2 = 1 - G^2\,M^2/L^2$);
\item
	if $x_1 = 0$ (i.e. $\kappa = 1$, namely $\alpha = \beta$), the orbit is a parabola and
	$x_2$ represents the distance of closest approach;
\item
	if $x_1 < 0$ (i.e. $\kappa > 1$), the orbit is a hyperbola, and $x_2$ represents again the distance
	of closest approach.
\end{itemize}
For massless particles ($\epsilon = 0$), $-\kappa\le x\le \kappa$ and the orbit is always an unbound straight line.
\subsubsection{General Relativity}
\label{sec:GR-orbits} 
In the GR case ($\gamma = 1$), the cubic polynomial on the right-hand side of Eq.~\eqref{eq:dx_dphi_2}
admits three roots, which we denote by $x_1$, $x_2$ and $x_3$, so that
\begin{equation}
\label{eq:dx_dphi_sn}
	\left(\frac{\d x}{\d\varphi}\right)^2
	=
	\rho \left(x - x_1\right) \left(x - x_2\right) \left(x - x_3\right)
	\ ,
\end{equation}
with $x_1 + x_2 + x_3 = {1}/{\rho}$.
The general solution of Eq.~\eqref{eq:dx_dphi_sn} is
\begin{equation}
\label{eq:orbit_sol_sn}
	\x\left(\varphi\right)
	=
	x_1 + \left(x_2 - x_1\right) \sn{2}{\frac{\varphi}{2}\, \sqrt{\rho\left(x_3 - x_1\right)} + \delta}{k}
	\ ,
\end{equation}
where $\sn{}{z}{k}$ is the Jacobi elliptic function with argument $z$ and elliptic modulus
\begin{equation}
k = \sqrt{\frac{x_2 - x_1}{x_3 - x_1}}
\ .
\label{k}
\end{equation}
The integration constant $\delta$ is again found from the initial condition $x(\varphi_0) = x_0$.
\par
The roots $x_1$, $x_2$ and $x_3$ can either be all real, or one real and two complex conjugates.
In the first case, we name the roots such that $x_1 \leq x_2 \leq x_3$, while in the second case we
denote the real root as $x_1$.
The following situations are therefore possible:
\begin{itemize}
\item $\bm{x_1 < x_2 < x_3}$:
if all three roots are distinct real numbers, the second derivative
\begin{equation}
	\frac{\d^2 x}{\d\varphi^2}
	=
	\frac{\rho}{2}
	\left[
	\left(x - x_2\right)
	\left(x - x_3\right)
	+ \left(x - x_1\right)
	\left(x - x_3\right)
	+\left(x - x_1\right)
	\left(x - x_2\right)
	\right]
\end{equation}
is positive, negative, and positive at $x=x_1$, $x=x_2$, and $x=x_3$, respectively.
It follows that a graph of $x$ versus $\varphi$ can either oscillate between $x_1$ and $x_2$,
or it can move away from $x_3$ towards infinity (which corresponds to $r \to 0$). 
If $x_1 < 0$, only part of an \textit{oscillation} will actually occur.
This corresponds to the particle coming from infinity, getting near the central mass,
and then moving away again toward infinity, like the hyperbolic trajectory in the Newtonian
case.
\item $\bm{x_1 < x_2 = x_3}$:
if the particle has just the right amount of energy (for a given angular momentum), $x_2$ and $x_3$
will merge.
The radius $r_{\rm in} \sim 1/x_2 = 1/x_3$ is called the \textit{inner radius\/}, and we have
$3/2\,\rho \leq r_{\rm in}\leq 3\,\rho$.
There are three solutions in this case:
\begin{enumerate}
\item
the orbit spirals into $r_{\rm in}$, approaching the inner radius (asymptotically) as 
a decreasing exponential in $\varphi$, $\tau$, or $t$, or
\item
one can have a circular orbit at the inner radius $r_{\rm in}$, or
\item
one can have an orbit that spirals down from the inner radius $r_{\rm in}$ towards the central
singularity.
\end{enumerate}
Since $\sn{}{z}{1} = \tanh\left(z\right)$, in this case Eq.~\eqref{eq:orbit_sol_sn} simplifies to
\begin{equation}
\label{eq:orbit_sol_tanh}
	x\left(\varphi\right)
	=
	x_1
	+\left(x_2 - x_1\right) \tanh^2\left(\frac{\varphi}{2}\,\sqrt{\rho\left(x_2 - x_1\right)} + \delta\right)
\ .
\end{equation}
\item $\bm{x_1 = x_2 < x_3}$:
a circular orbit also results when $x_1 = x_2$.
In this case, the radius $r_{\rm out} \sim 1/x_1 = 1/x_2$, is called the \textit{outer radius\/}. 
\item $\bm{x_2 \, , \, x_3 \in \mathbb{C}}$:
if the particle approaches the center with enough energy and sufficiently low angular momentum
then only $x_1$ will be real.
This corresponds to the particle spiraling and falling into a black hole with a finite change in $\varphi$.
\end{itemize}
Equivalent forms of the above solutions are briefly reviewed in Appendix~\ref{A:equivalent}.
\section{The central mass as a random variable}
\label{S:randomM}
We shall now consider the mass $M$ as a random variable, assumed to be normally distributed with mean value $M_0$ and standard deviation $\sigma_M$ (the variance is $\sigma_M^2$) \cite{Papoulis}.
Equivalently, we may denote the mass $M$ as $M_0 + \delta M$, where $\delta M$ is a random variable
with zero mean value, $\mu_{\delta M} = 0$, and variance $\sigma_{M} ^2$, that is
\begin{equation}
\delta M
\sim
\normal{0}{\sigma_M^2}
\ ,
\end{equation}
where $\normal{\mu_z}{\sigma_z^2}$ is the normalised Gaussian distribution of mean value $\mu_z$.
\par
Since $\delta M$ is a random variable, any trajectory parametrised as $r=r(\varphi)$, will be a functional
of this random variable which, for a given value of the angle $\varphi$, will therefore become a random variable itself.
In order to stress that the angle $\varphi$ is now seen as a parameter, and $\delta M$ is treated as an independent
(random) variable, we shall employ the notation $r=r_\varphi(\delta M)$.
%
%
%
%
%\subsection{General approach \label{sec:pdf} }
%
%
\par
Since Eq.~\eqref{eq:dx_dphi} has the known analytical solution~\eqref{eq:orbit_sol_sn},
it is conceivable to carry out the analysis of the probability distribution of $r_\varphi$, representing
the position of the particle at a given angle $\varphi$, without relying on any perturbation methods.
\par
In order to show explicitly the dependence on the mass $M$, we first rewrite Eq.~\eqref{eq:dx_dphi_sn}
[equivalent to Eq.~\eqref{eq:dx_dphi}] as
\begin{equation}
\label{eq:dx_dphi_sn-bis}
	\left(\frac{\d u}{\d \varphi}\right)^2
	=
	\frac{\rho}{M} \left(u - u_1\right) \left(u - u_2\right) \left(u - u_3\right)
	\ ,
\end{equation}
where we introduced the (dimensionful) variable
\begin{equation}
u 
=
M\, x
=
 \frac{L^2}{G\, r}
 \ , 
\label{eq:u}
\end{equation}
and $u_i = M\, x_i$, for $i=1,2,3$.
The solution of Eq.~\eqref{eq:dx_dphi_sn-bis} [equivalent to the solution~\eqref{eq:orbit_sol_sn}]
is now given by
\begin{equation}
\label{eq:orbit_sol_sn-bis}
	u(\varphi)
	=
	u_1
	+
	\left(u_2 - u_1\right) \sn{2}{\frac{\varphi }{2}\,\sqrt{\frac{\rho}{M}\left(u_3 - u_1\right)} + \delta}{k}
	\ ,	
\end{equation}
where 
\begin{equation}
k
=
\sqrt{\frac{u_2 - u_1}{u_3 - u_1}},
\end{equation}
and $u(\varphi_0) = u_0$.
It is important to remark that $u_1$, $u_2$, $u_3$, $\delta$, $\rho$ and $k$ are now all functions
of $M$.
Since we are assuming $M$ to be a random variable with a given probability density function $f_M$, the variable
$\uphi \equiv u(\varphi)$ will turn into a random variable with probability distribution $f_{\Uphi}$. 
\par
As customary in probability theory, from now on we shall indicate the random variables with capital letters
($M$ and $\Uphi$, respectively for the mass and the dependent variable $\uphi$) and with lowercase letters
($m$ and $\uphi$, respectively), the values taken on by the random variables.
We can therefore write Eq.~\eqref{eq:dx_dphi_sn-bis} as 
\begin{equation} \label{eq:orbit_sol_sn-tris}
	\Uphi
	=
	u_1 + \left(u_2 - u_1\right)
	\sn{2}{\frac{\varphi}{2}\, \sqrt{\frac{2\,G^2\,M}{L^2}\left(u_3 - u_1\right)} + \delta}{k}
	\ .
\end{equation}
In order to determine the probability density function $f_{\Uphi}$ of the random variable $\Uphi$,
it is useful to work with its cumulative distribution function $\mathcal{F}_{\Uphi}\left(\uphi\right)$,
defined as the probability $\mathcal{P}$ of $\Uphi$ taking on values smaller then or equal to $\uphi$.
Since $\Uphi$ is a monotonically increasing function of $M$ (which we shall denote with the symbol $g$,
$\Uphi = g(M)$), we have
\begin{eqnarray}
	\mathcal{F}_{\Uphi}(\uphi)
	&=&
	\mathcal{P}(\Uphi \leq \uphi)
	=
	\mathcal{P}(0 \leq M \leq m)
	\nonumber
	\\
	&=&
	\int_{0}^{m} f_M(\xi)\, \d\xi
	\nonumber
	\\
	&=&
	\int_{0}^{g^{-1}(\uphi)} f_M(\xi)\,\d\xi
	\ ,
\end{eqnarray}
where $\uphi = g(m)$.
Making use of the fundamental theorem of calculus, we obtain
\begin{eqnarray} \label{proof-prob}
	f_{\Uphi}(\uphi)
	&=& 
	\frac{\d\mathcal{F}_{\Uphi}(\uphi)}{\d\uphi} 
	=
	\frac{\d\mathcal{F}_{\Uphi}(\uphi)}{\d g^{-1}(\uphi)}\,\frac{\d g^{-1}(\uphi)}{\d\uphi}
	\nonumber
	\\
	&=& 
	f_M(m)\, \frac{\d g^{-1}(\uphi)}{\d \uphi}
	\nonumber
	\\
	&=&
	\frac{f_M(m)}{g^{\prime}(m)}
	=
	\frac{f_M(m)}{\abs{g^{\prime}(m)}}
	\ .
\end{eqnarray}
Were the function $g$ monotonically decreasing, the previous derivation could be repeated just changing
the sign of the inequality, that is
\begin{eqnarray}
	\mathcal{F}_{\Uphi}(\uphi)
	&=&
	\mathcal{P}(\Uphi \leq \uphi)
	=
	\mathcal{P}(M > m)
	\nonumber
	\\
	&=&
	\int_{m}^\infty f_M(\xi)\, \d\xi
	\nonumber
	\\
	&=&
	-\int_{\infty}^{g^{-1}(\uphi)} f_M(\xi)\, \d\xi
	\ ,
\end{eqnarray}
which would lead to
\begin{equation}
	f_\Uphi(\uphi)
	= 
%	\frac{d\mathcal{F}_\Uphi\left(\uphi\right)}{d\uphi} = 
%	\frac{d\mathcal{F}_\Uphi\left(\uphi\right)}{dg^{-1}\left(\uphi\right)} \frac{dg^{-1}\left(\uphi\right)}{d\uphi} = 
%	- f_M\left(m\right) \frac{dg^{-1}\left(\uphi\right)}{d\uphi} =
	- \frac{f_M(m)}{g^{\prime}(m)}
	=
	\frac{f_M(m)}{\abs{g^{\prime}(m)}}
	\ .
\end{equation}
\par 
The case of a non-monotonic function $g$ can be treated similarly, just splitting the integration range
so as to have a monotonic function in each interval.
It is then easy to see that, if the function $f_\Uphi$ has $N$ roots, which we denote as $u_\varphi = g(m_k)$, for $k=1, 2, \ldots, N$, we obtain the general formula
\begin{equation}
\label{eq:pdfu}
f_\Uphi(\uphi)
=
\sum_{k=1}^{N} \frac{f_M(m_k)}{\abs{g^{\prime}(m_k)}}
\ ,
%\qquad
%\uphi = g(m_1) = g(m_2) = \ldots = g(m_N)
%\ . 
\end{equation}
The validity of the previous relation is not limited to $\Uphi$ and $M$.
In order to stress this, and for future use, the previous relation can be written as
\begin{equation}
\label{eq:pdf}
	f_Y(y)
	=
	\sum_{k=1}^{N} \frac{f_X(x_k)}{\abs{g^{\prime}(x_k)}}
	\ ,
%	\qquad
%	y
%	=
%	g(x_1) = g(x_2) = \ldots = g(x_N)
%	\ . 
\end{equation}
for a generic random variable $Y$ function of $X$ (with whatever probability density function $f_X$),
such that $Y = g(X)$.
%
%
%
%
%\subsection{Further generalisation}
%
%
\par
It is possible to extend the idea described above and to assume other parameters as random variables as well,
not just the mass $M$.
For instance, the initial position $u_0$ of the particle can be assumed as a random variable.
In such a case, Eq.~\eqref{eq:pdf} can be applied, where the dependent and independent random variables
are $\Uphi$ and $U_0$, respectively, and $\Uphi = g(U_0)$.
\par
The next step, is to assume more parameters simultaneously as random variables.
For example, the energy $E$ and the angular momentum $L$ might actually be considered as
random variables with mean values equal to their nominal values (say $\mu_E$ and $\mu_L$, respectively)
and variances accounting for their uncertainties (say $\sigma_E^2$ and $\sigma_L^2$, respectively).
The random variables $E$ and $L$ would be in this case independent and follow normal distributions,
$E \sim \normal{\mu_E}{\sigma_E^2}$, $L \sim \normal{\mu_L}{\sigma_L^2}$.
In this case it is also conceivable that the variables $E$ and $L$ are not independent.
In this scenario, it could be more accurate to introduce a joint probability density function $f_{EL}(e,l)$,
instead of two independent probability density functions $f_E(e)$ and $f_L(l)$ for the energy
and angular momentum separately.
This approach is of course more general, and the case of independent variables could be
recovered when $f_{EL}(e,l) = f_{E}(e)\, f_{L}(l)$.
\par
This idea can be generalised even further.
If the energy $E$, the angular momentum $L$, and the initial position $u_0$ are all affected
by uncertainties and should be treated as random variables, we could have a joint probability
density function $f_{ELU_0}(e,l,u_0)$.
In all these cases, the variable $\Uphi$ representing the quantity $u$ at some angle $\varphi$
would be a multivariate random variable, i.e.~a function of more than one random variable.
We shall continue to use the symbol $g$ for this function.
However, a straightforward generalisation of Eq.~\eqref{eq:pdf} to this case does not appear
possible. 
\par
In the case of two random variables, say $E$ and $L$, with joint probability density function $f_{EL}$,
a procedure similar to the one presented for the case where $M$ is treated as a random variable would yield
\begin{eqnarray}
	\mathcal{F}_\Uphi(\uphi)
	&=&
	\mathcal{P}(\Uphi \leq \uphi)
	=
	 \mathcal{P}\left( g(E,L) \leq \uphi\right)
	\nonumber 
	\\
	&=&
	\int_{\mathcal{D}(\uphi)} f_{EL}(\xi, \eta)\, \d\xi\, \d\eta
	\ ,
\end{eqnarray}
where $\mathcal{D}(\uphi) \subset \R^2$ is the region of the $EL$ plane over which
$g(E,L) \leq \uphi$.
Once the cumulative distribution function $\mathcal{F}_\Uphi(\uphi)$ is known,
the probability density function $f_\Uphi(\uphi)$ can be finally obtained by means of the definitions of $\mathcal{F}_\Uphi$ and $f_\Uphi$, i.e.
\begin{equation}
	f_\Uphi(\uphi)
	=
	\frac{\d \mathcal{F}_\Uphi(\uphi)}{\d\uphi}
	\ .
\end{equation}
In contrast with the case of a single random variable, this procedure for obtaining $f_\Uphi\left(\uphi\right)$
is unlikely to be feasible analytically, due to the complexity of both the function $g$ and the domain of integration $\mathcal{D}$.
It can probably be done numerically, though. 
\par
Following this approach, it should be possible to investigate the uncertainty on the orbit,
described by a random variable $\Uphi$, due to the uncertainties on the parameters
$E$, $L$, $U_0$ (and possibly others), and to compare this uncertainty on the orbit
to the one produced by an uncertainty of the central mass $M$.
The basis of this approach is laid on two key ingredients:
\begin{enumerate}
\item
Finding the (analytical or numerical) solution of the equation of motion~\eqref{eq:d2x_dphi2};
\item
Computing the probability density function of the orbit, once the probability density
functions of the parameters are given, following the method described above.
\end{enumerate}
\section{Mass uncertainty and non-circular orbits}
\label{S:circular}
In Section~\ref{sec:GR-orbits}, solutions of the equation of motion~\eqref{eq:classical} were reviewed,
and qualitatively different kinds of orbit have been identified.
For these orbits, the effect of assuming the mass $M$ as a normally distributed random variable
is now investigated for orbits with set initial conditions for the position and velocity. 
Let us first recall that we shall be using the (dimensionful) variable $u$ in Eq.~\eqref{eq:u}, like in the previous Section, 
and that the orbits are divided into the following four groups ($u_0$ represents the initial position of the particle):
\begin{itemize}
\item
If the third degree polynomial on the right-hand side of Eq.~\eqref{eq:dx_dphi_sn-bis} has three distinct real roots ($u_1 < u_2 < u_3$), we can have the following behaviours:
	\begin{itemize}
	\item
	if $0 < u_1 < u_0 < u_2 < u_3$, the graph of $u$ versus $\varphi$ oscillates between $u_1$ and $u_2$
	(\textbf{first case});
	\item
	if $0 < u_1 < u_2 < u_3 < u_0$ , the graph of $u$ versus $\varphi$ moves away from $u_3$ towards infinity
	(\textbf{second case});
	\item
	if $u_1 < 0 < u_0 < u_2 < u_3$ only part of an oscillation occurs, with the particle first approaching the central
	mass and then moving away towards infinity (\textbf{third case});
	\end{itemize}
\item 
If the polynomial has one real root and two complex conjugates roots, the particle falls into the black hole
and the orbit is a spiral with a finite change in $\varphi$ (\textbf{fourth case}).
\end{itemize}
It is important to recall that, if the energy and the angular momentum of the particle are such that
$u_1 < u_2 = u_3$, three behaviours are possible:
the orbit may be circular at the radius $r_{\rm in} \sim 1/u_2 = 1/u_3$
(the inner radius), or it may spiral in asymptotically approaching the inner radius
or spiral down from the inner radius to the central point.
If instead $u_1 = u_2 < u_3$, a circular orbit at the radius $r_{\rm out} \sim 1/u_1 = 1/u_2$ (the outer radius)
is also possible.
In all of these cases, after the slightest perturbation of the central mass $M$, we fall back in one of the four
previous cases.
For example, when the stable orbit is perturbed, we fall in the first case.
\par
In the following, we shall analyse the above four cases.
For the purpose of making the results more readable, we will first show plots of the orbits corresponding
to fixed numerical values of the parameters of relevance for the each distribution of the mass $M$,
and specifically chosen for the purpose of displaying clearly the qualitative features of the orbits in the case under
consideration.
We then display the chosen distribution for the mass $M$ and the final distribution of the quantity of relevance
again for each specific case ({\em e.g.}~amplitude of the oscillation, minimum distance from the centre, etc).
In particular, we shall use units with $G=1$.
\subsection{First case}
\begin{figure}[h]
\centering
	\includegraphics[scale=0.44]{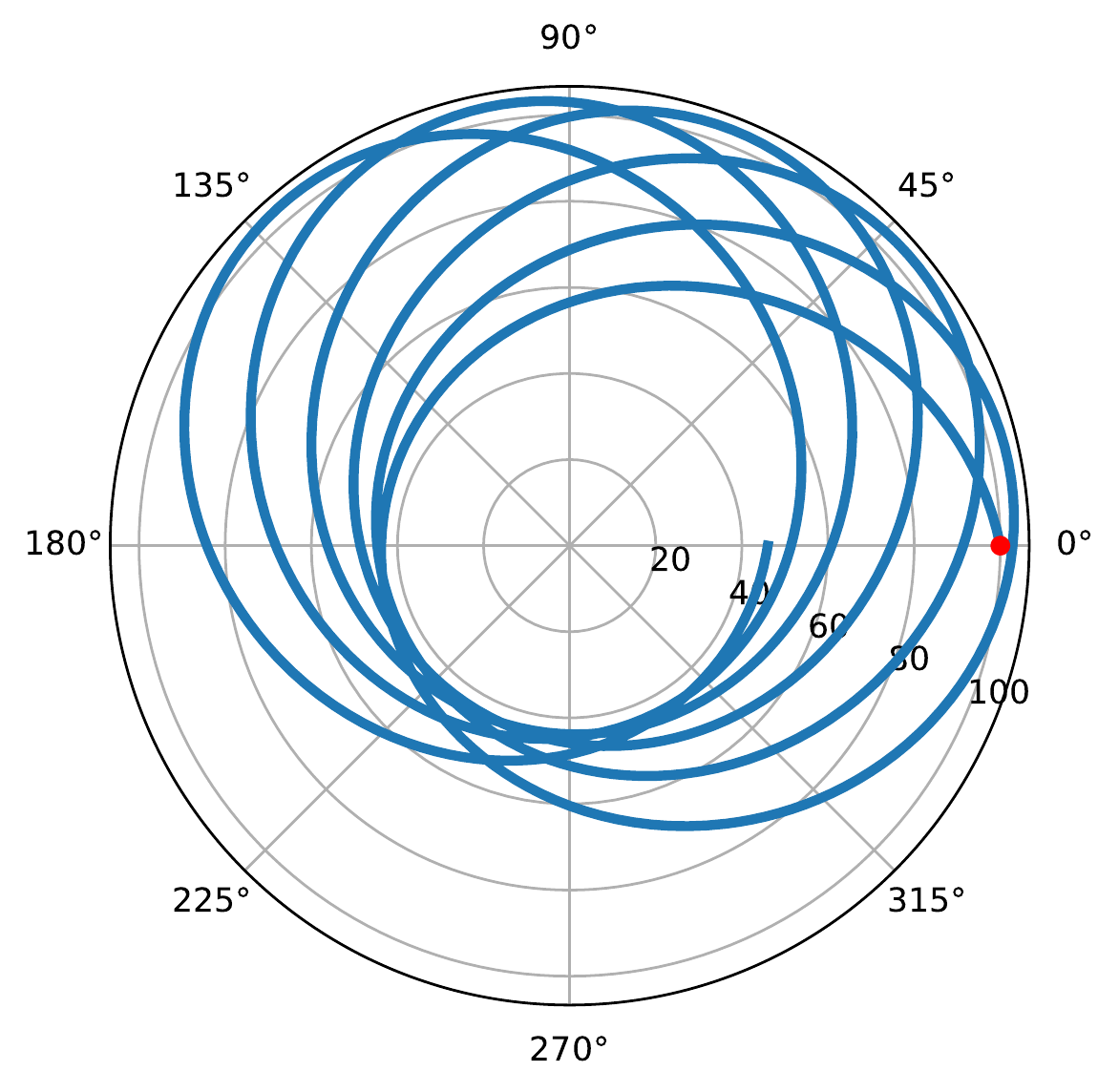} %
	\includegraphics[scale=0.44]{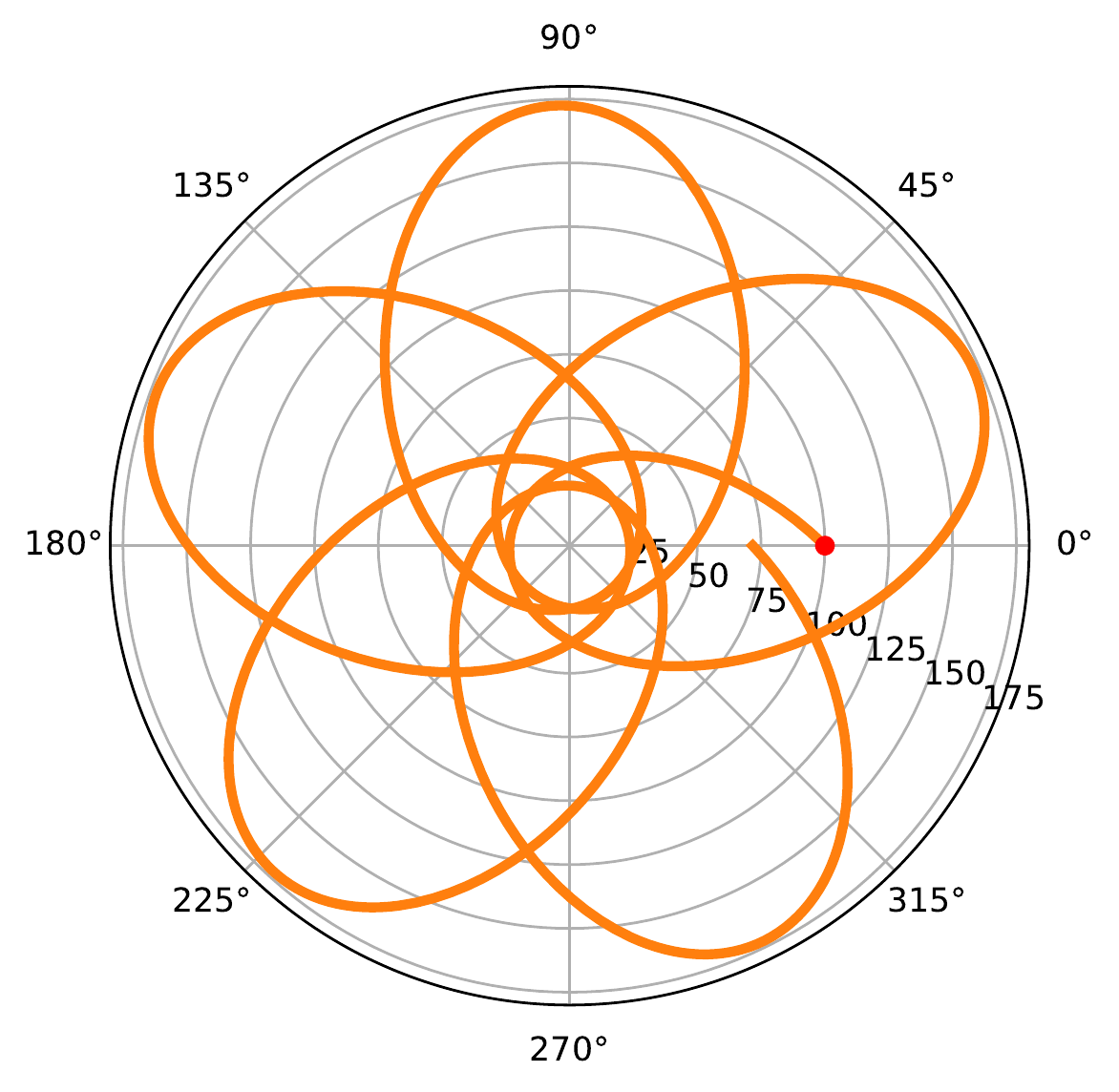} %
	\includegraphics[scale=0.44]{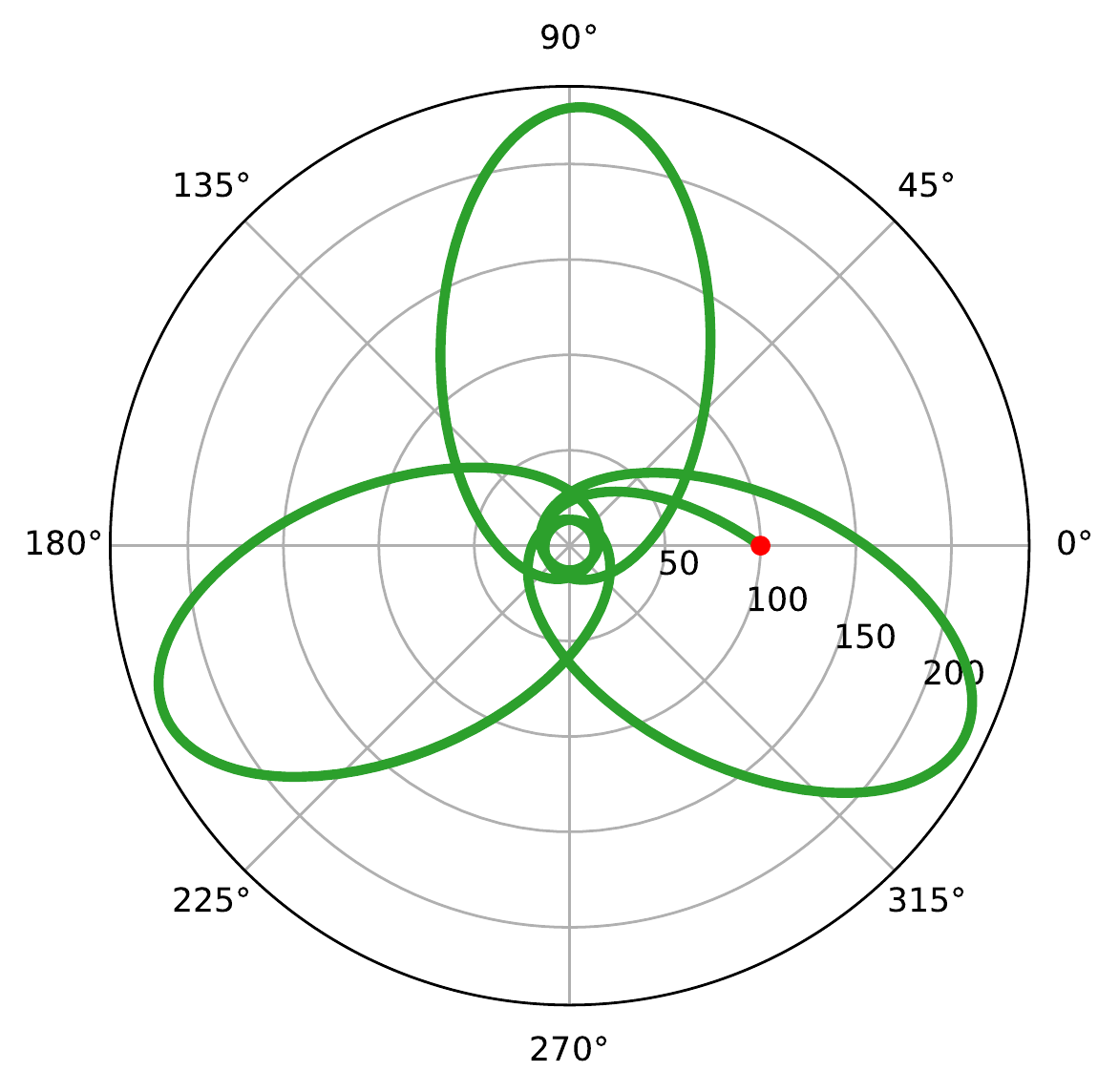}
	\caption{(First case) Orbits for different values of $M=1.5$, $2$, $2.5$
	(left to right, with $G=1$, $E=0.9$, $L=10$, $u_0=1$.)
	The red dot is the starting point and colours are the same as in Fig.~\ref{fig:case1a_1}.
	\label{fig:case1a_0} }
\end{figure}
\begin{figure}
\centering
	\includegraphics[scale=0.53]{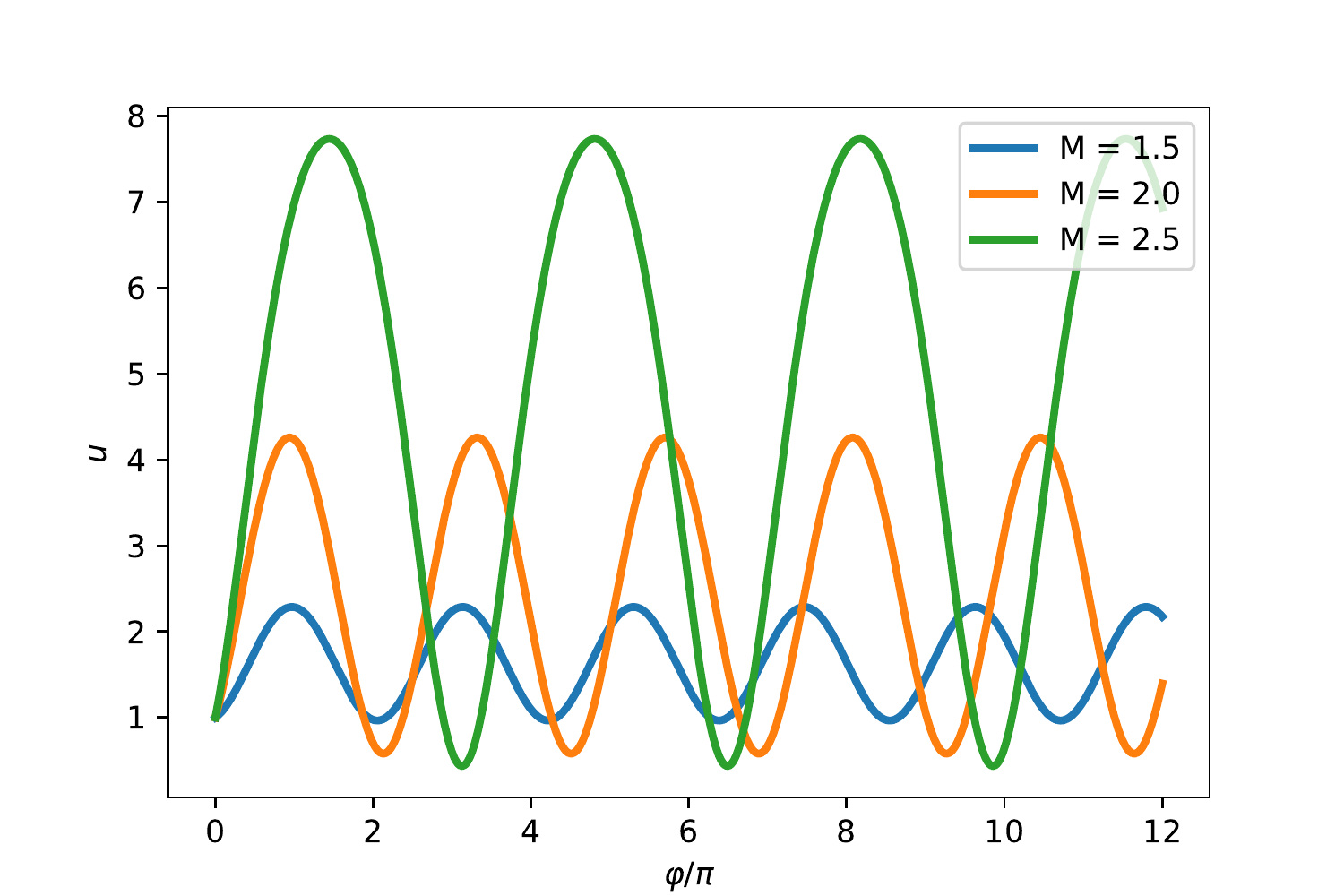}
	\includegraphics[scale=0.53]{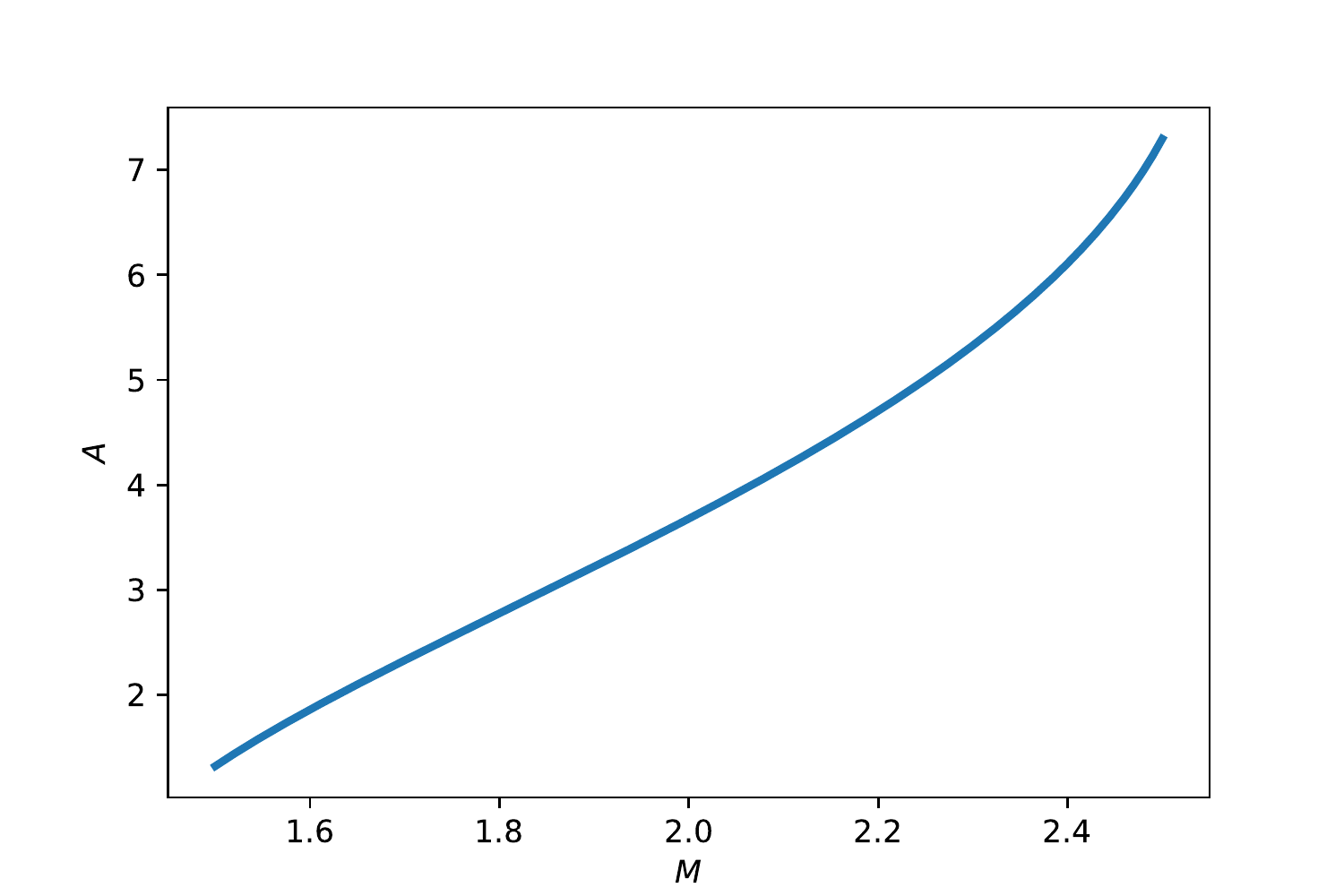}
	\caption{(First case) Graphs of $u(\varphi)$ for different values of $M=1.5$, $2$, $2.5$
	(with $G=1$, $E=0.9$, $L=10$, $u_0=1$) (left panel),
	and maximum amplitude of the oscillations as a function of the mass $M$ (right panel).
	\label{fig:case1a_1} }
\end{figure}
When the orbit oscillates between $u_1$ and $u_2$, it is interesting to analyse the amplitude $A$
(and the period) of the oscillations as a function of the random variable $M$.
For example, the trajectories $r=r(\varphi)$ corresponding to three different values of $M$ are shown in
Fig.~\ref{fig:case1a_0} in polar form.
Cartesian plots of $u(\varphi)$ are shown in Fig.~\ref{fig:case1a_1}, together with the amplitude
of the oscillations $A$, as a function of the mass $M$.
(When comparing Fig.~\ref{fig:case1a_0} with Fig.~\ref{fig:case1a_1}, it should be noted
that $r = L^2/G\,u = 100/u$.)
\begin{figure}
\centering
	\includegraphics[scale=0.53]{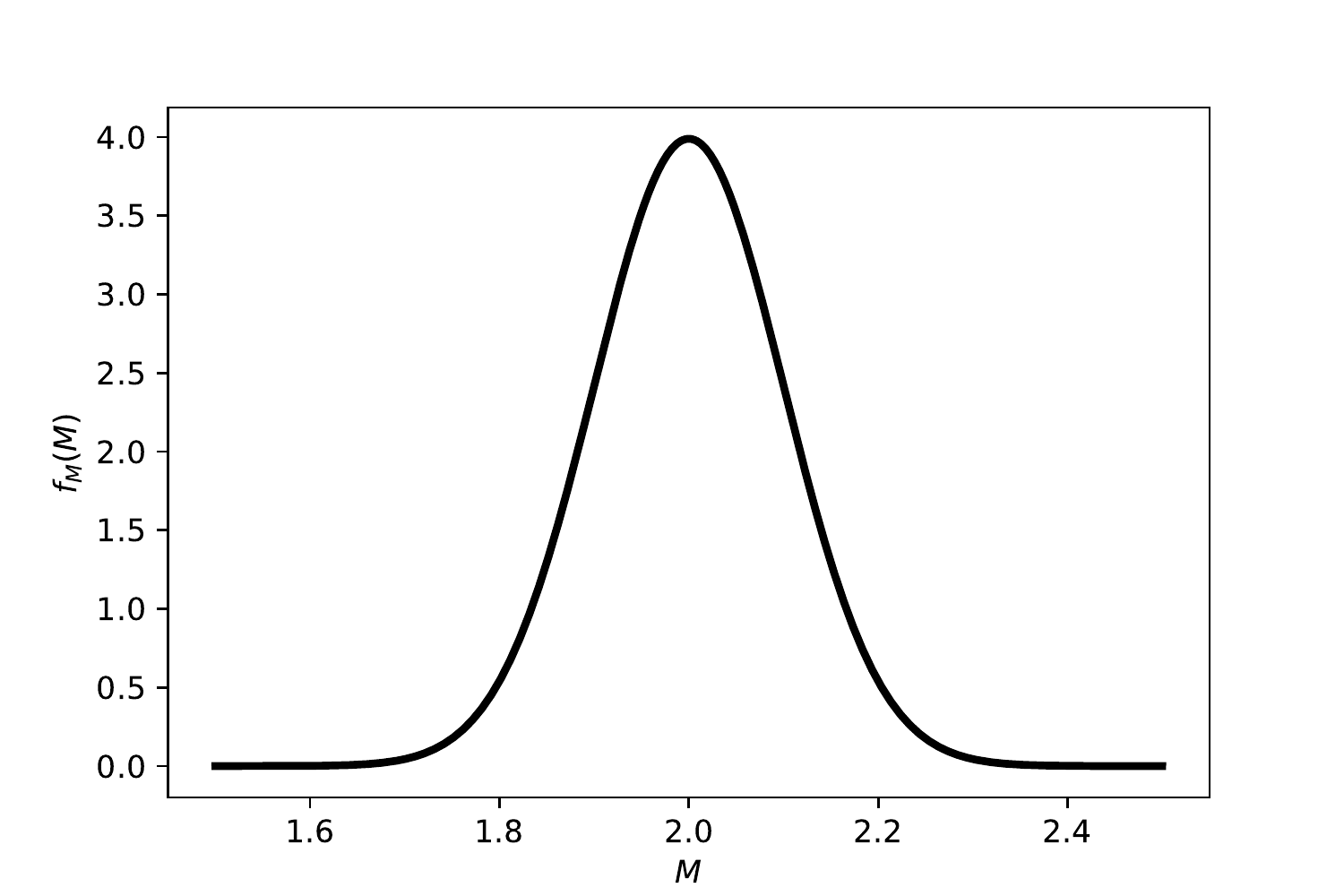}
	\includegraphics[scale=0.53]{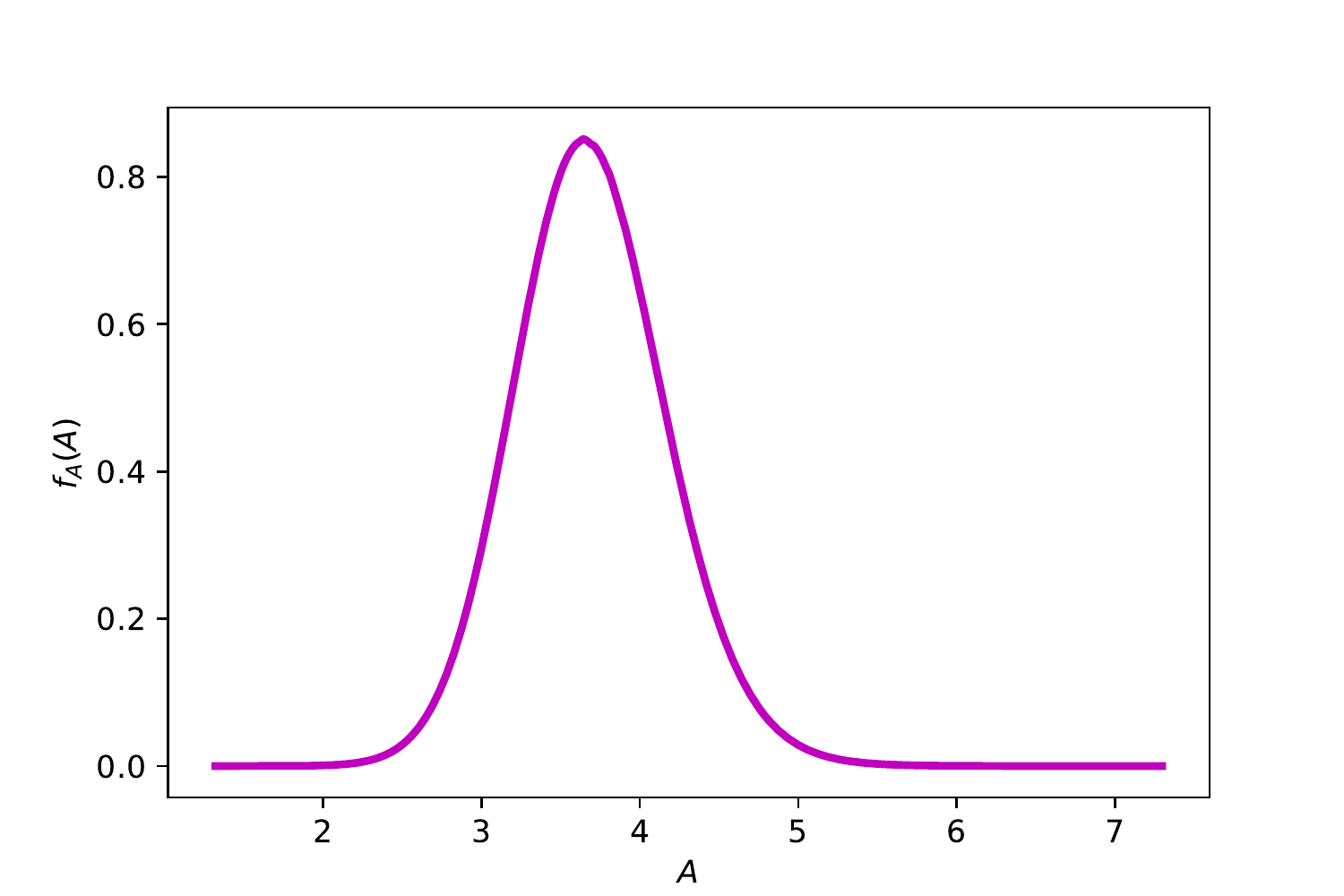}
	\caption{(First case) Probability density function for the mass $M$ (left panel) and for the amplitude $A$
	(right panel).
	The amplitude corresponding to the mean value of $M$ is $\sim 3.68$; the value of the amplitude
	corresponding to the maximum of the distribution of $A$ if $\sim 3.64$.
	(The random variable $A$ is not normally distributed, and its distribution is not symmetric).
	\label{fig:case1a_2} }
\end{figure}
\par
If the random variable $M$ is normally distributed with mean value $\mu_M = 1.5$ and standard
deviation $\sigma_M = 0.1$, the probability density function $f_A$ of the random variable $A$
can be obtained by means of Eq.~\eqref{eq:pdf} once the probability density function $f_M$ and
the function $A = g(M)$ are known.
The probability density functions $f_M$ and $f_A$ are shown in Fig.~\ref{fig:case1a_2}.
\subsection{Second case}
\begin{figure}[h]
\centering
	\includegraphics[scale=0.33]{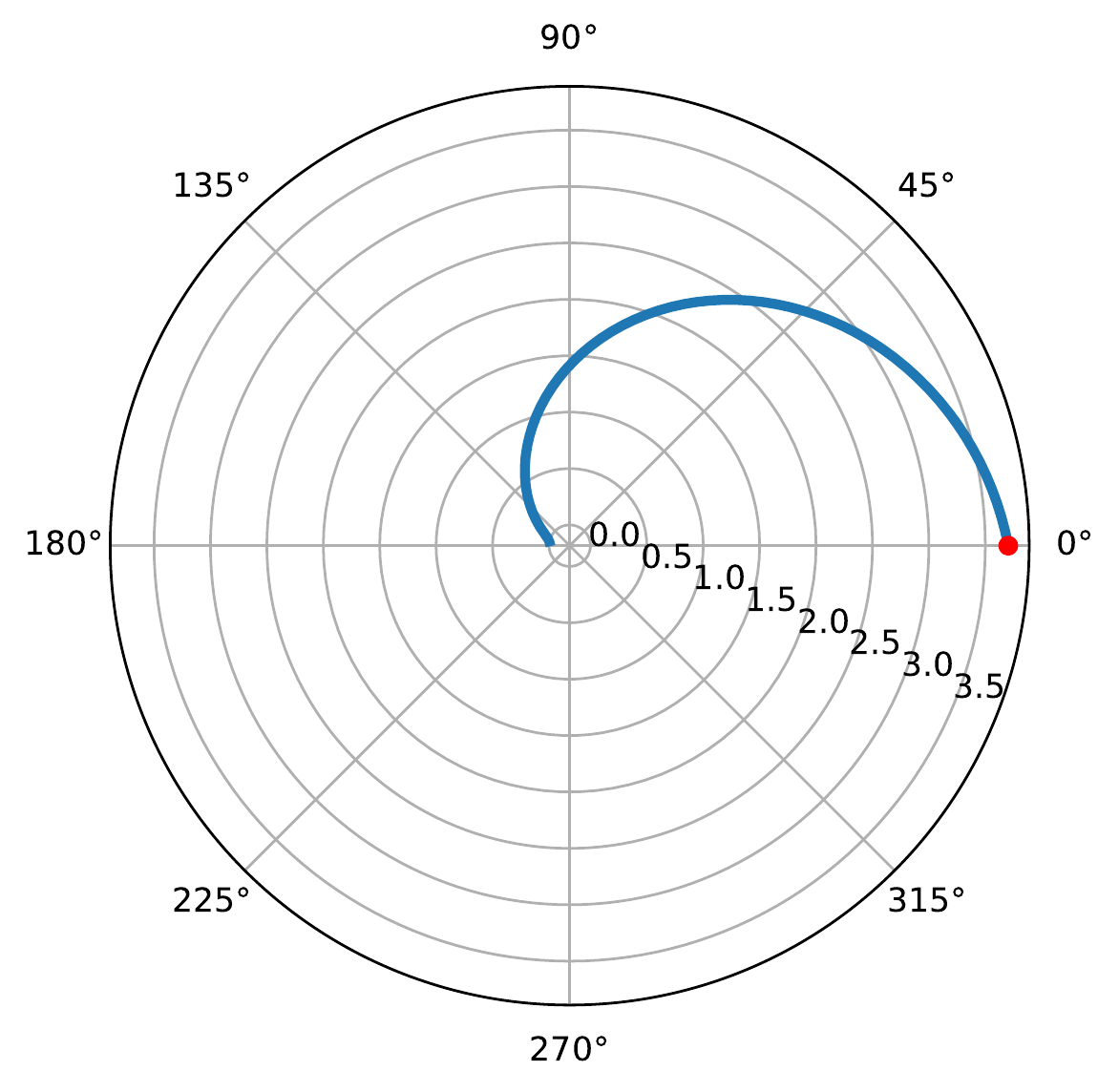} %
	\includegraphics[scale=0.33]{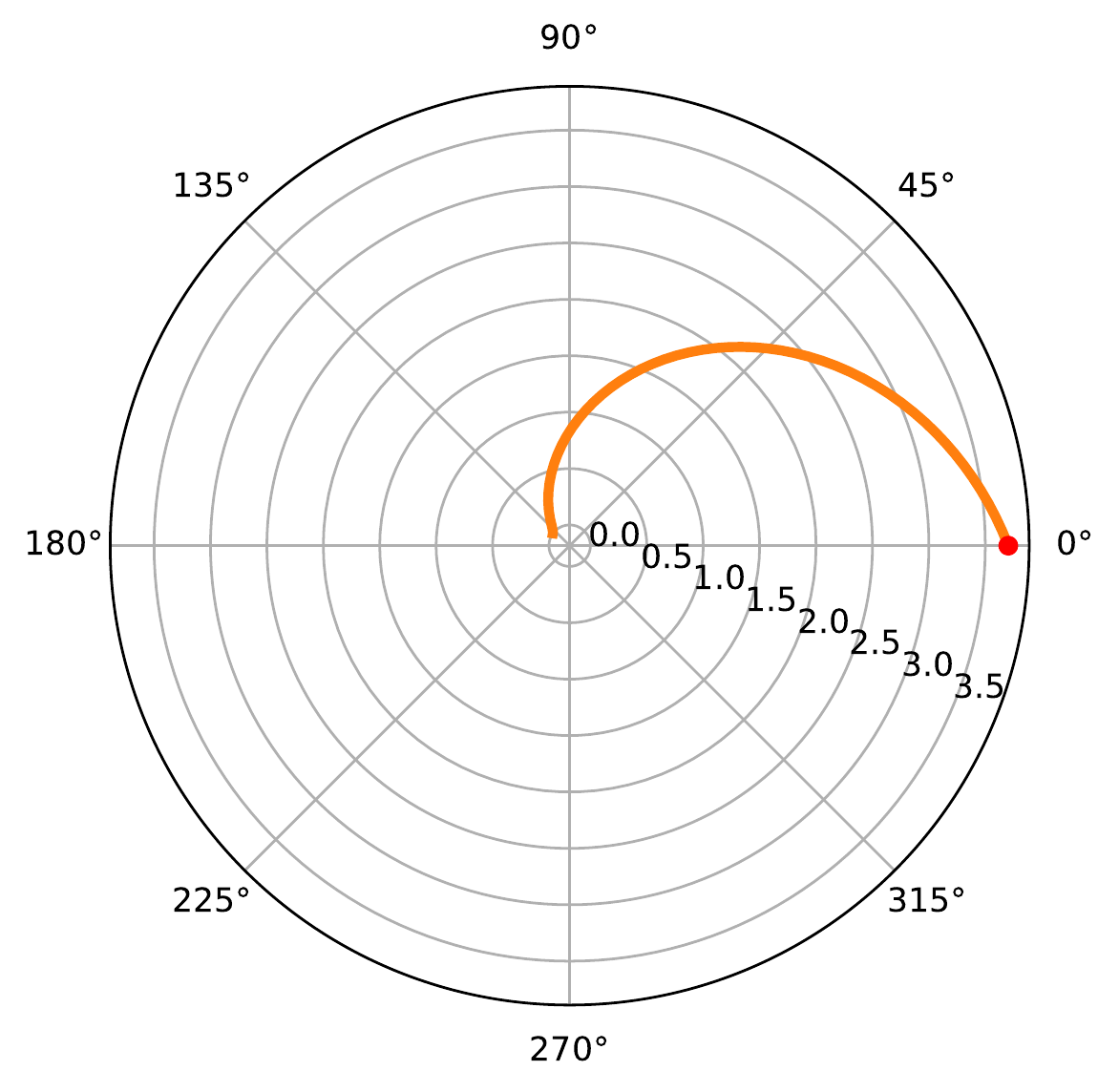} %
	\includegraphics[scale=0.33]{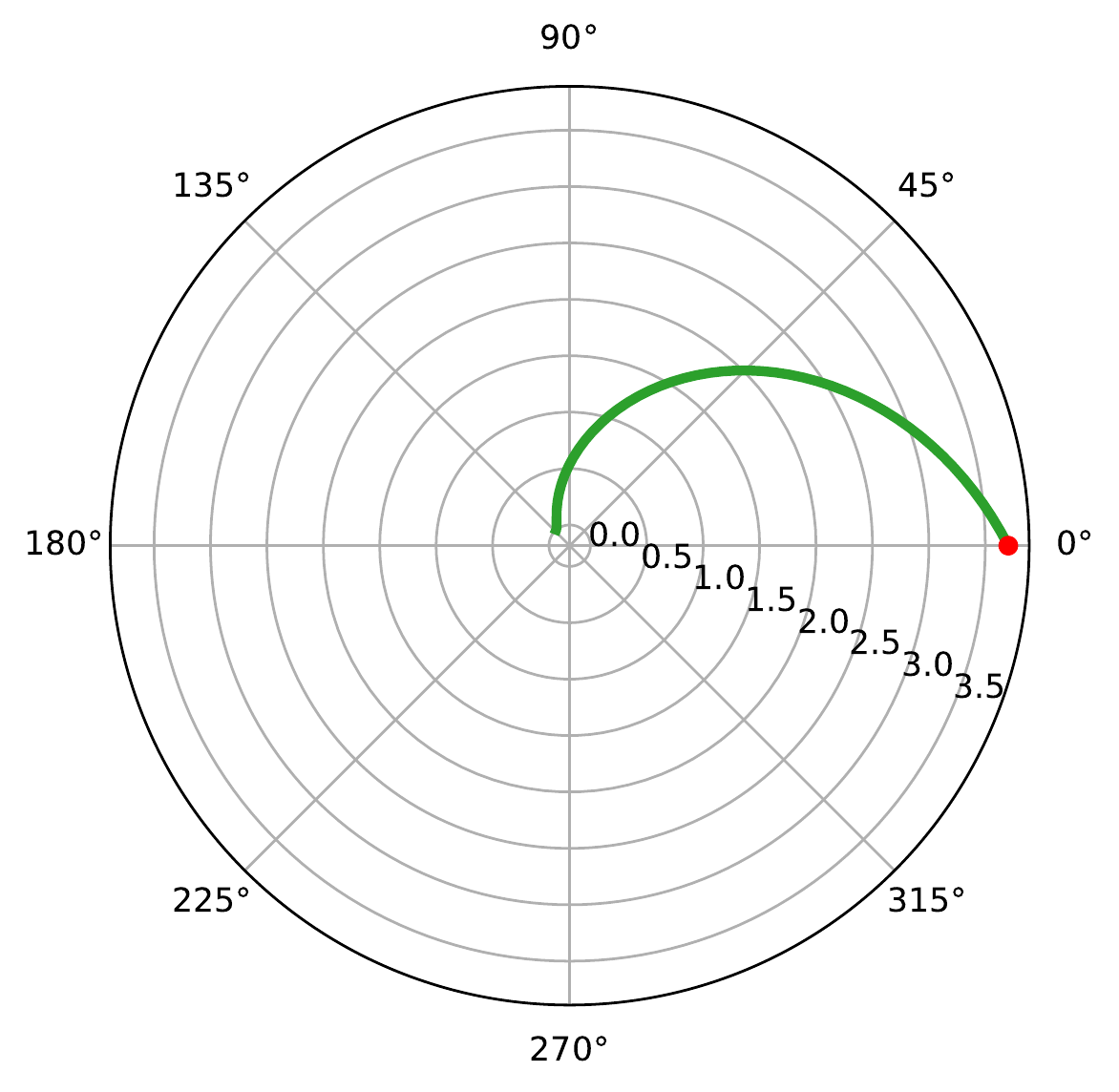} %
	\includegraphics[scale=0.33]{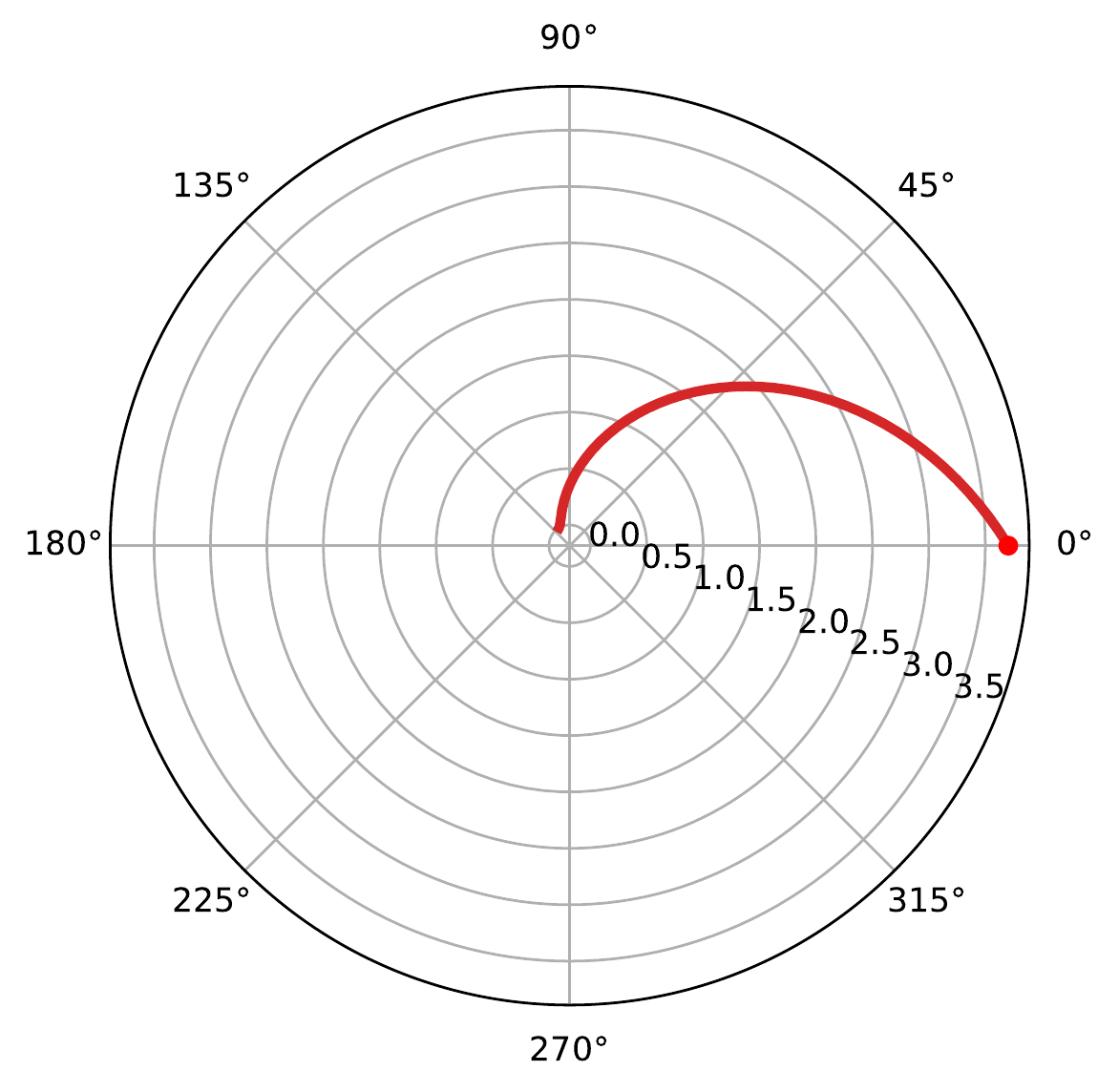}
	\caption{(Second case) Orbits for values of $M=1.7$, $1.9$, $2.1$, $2.3$ (left to right, with
	$G=1$, $E=0.99$, $L=10$, $u_0=27$.)
	The red dot is the starting point and colours are the same as in Fig.~\ref{fig:case1b_1}(a).
	\label{fig:case1b_0} }
\end{figure}
\begin{figure}[t]
\centering
	\includegraphics[scale=0.53]{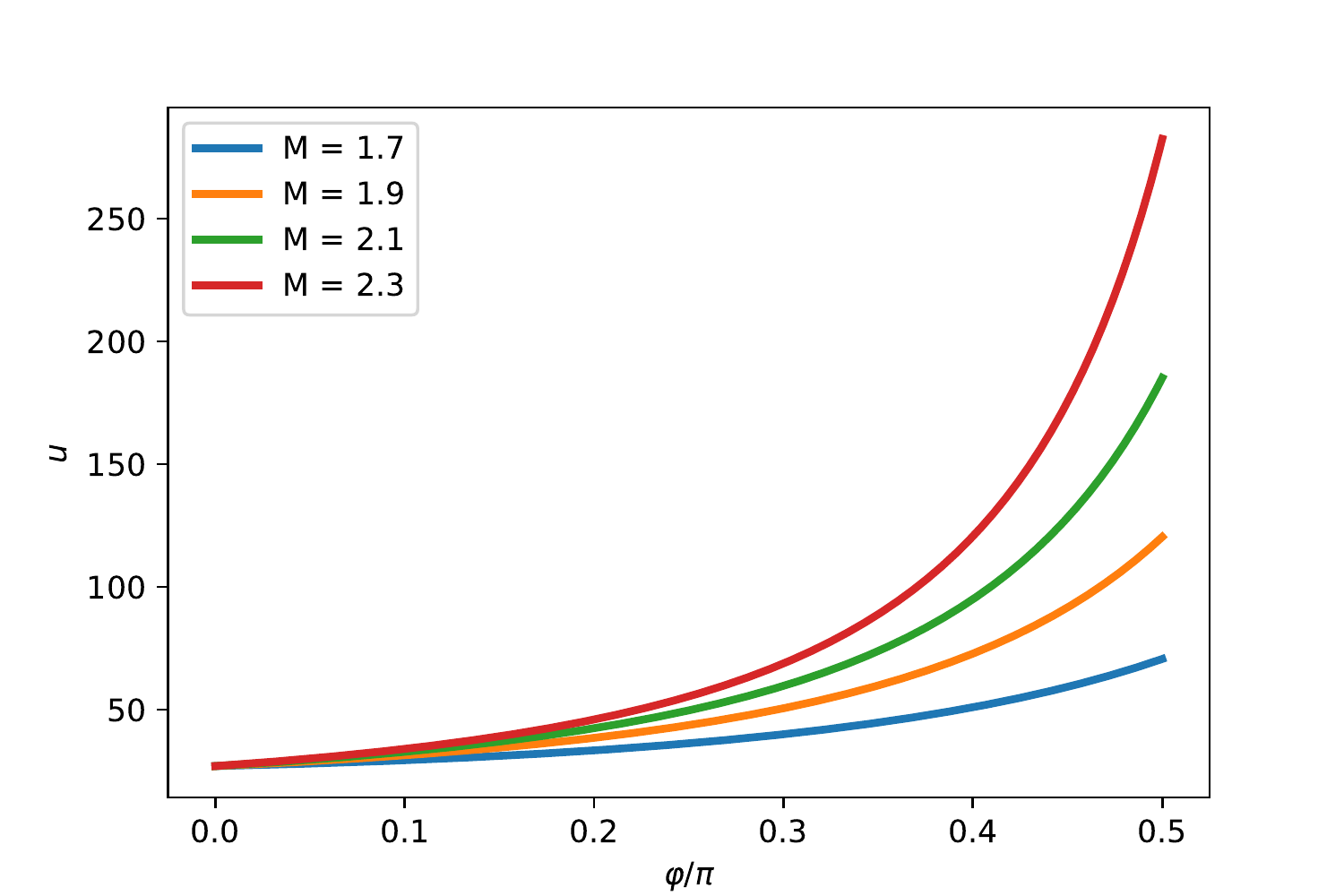}
	\includegraphics[scale=0.53]{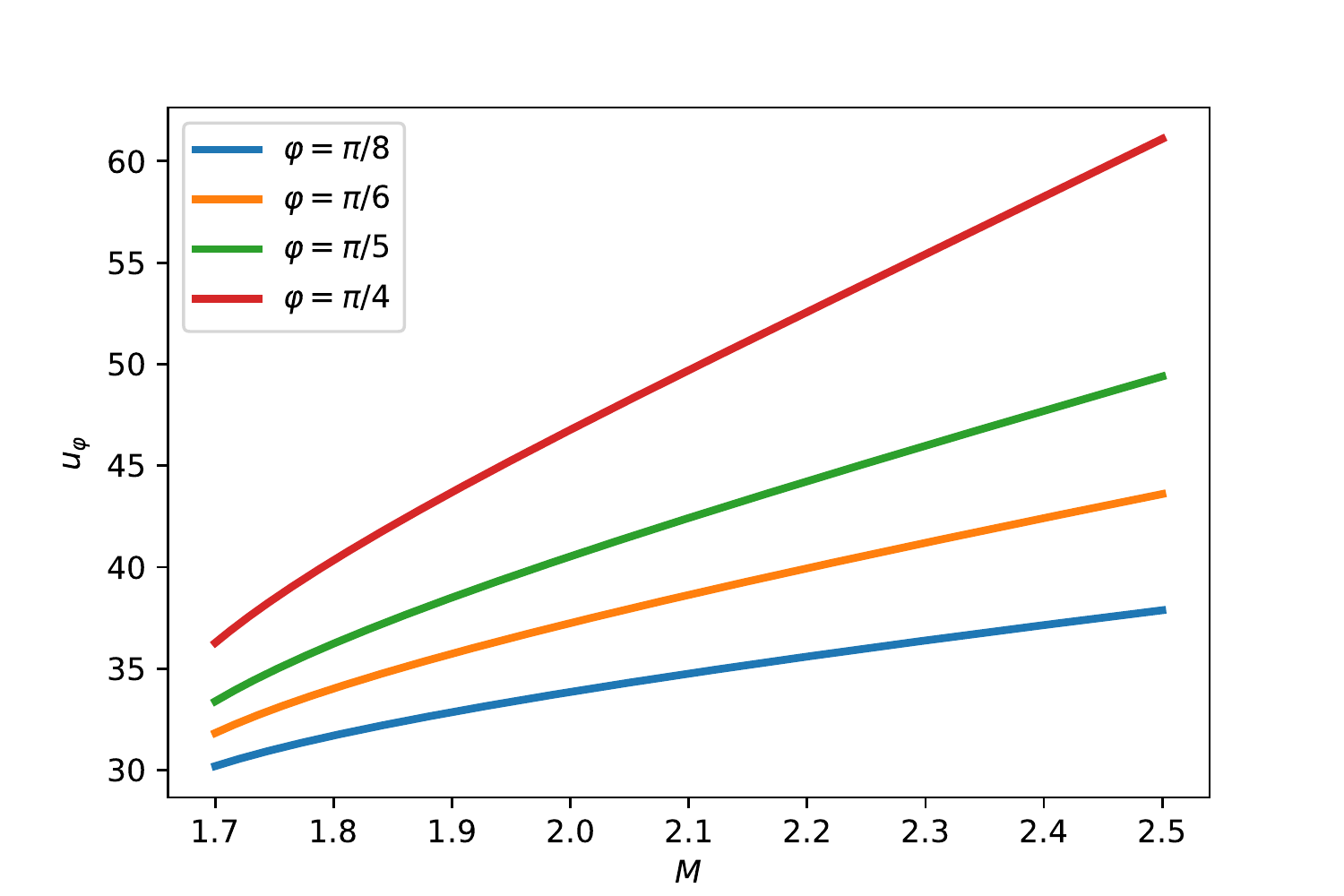}
	\caption{(Second case) Graphs of $u(\varphi)$ for values of $M=1.7$, $1.9$, $2.1$, $2.3$
	(left panel, with $G=1$, $E=0.99$, $L=10$, $u_0=27$), and $\uphi$ as a function of the mass $M$
	for fixed angles $\varphi$ (right panel).
	\label{fig:case1b_1} }
\end{figure}
The second possibility listed above is that the particle will monotonically move towards the central mass ($u\to \infty$),
as is shown by the polar trajectories $r(\varphi)$ represented in Fig.~\ref{fig:case1b_0}.
For such a type of orbits, the function $u\left(\varphi\right)$ is plotted for several values of $M$
in Fig.~\ref{fig:case1b_1}.
For this case, it is therefore interesting to analyse the behaviour of $\uphi$ as a function of $M$ for different values
of the angle $\varphi$ that parameterizes the orbit.
\begin{figure}
\centering
	\includegraphics[scale=0.53]{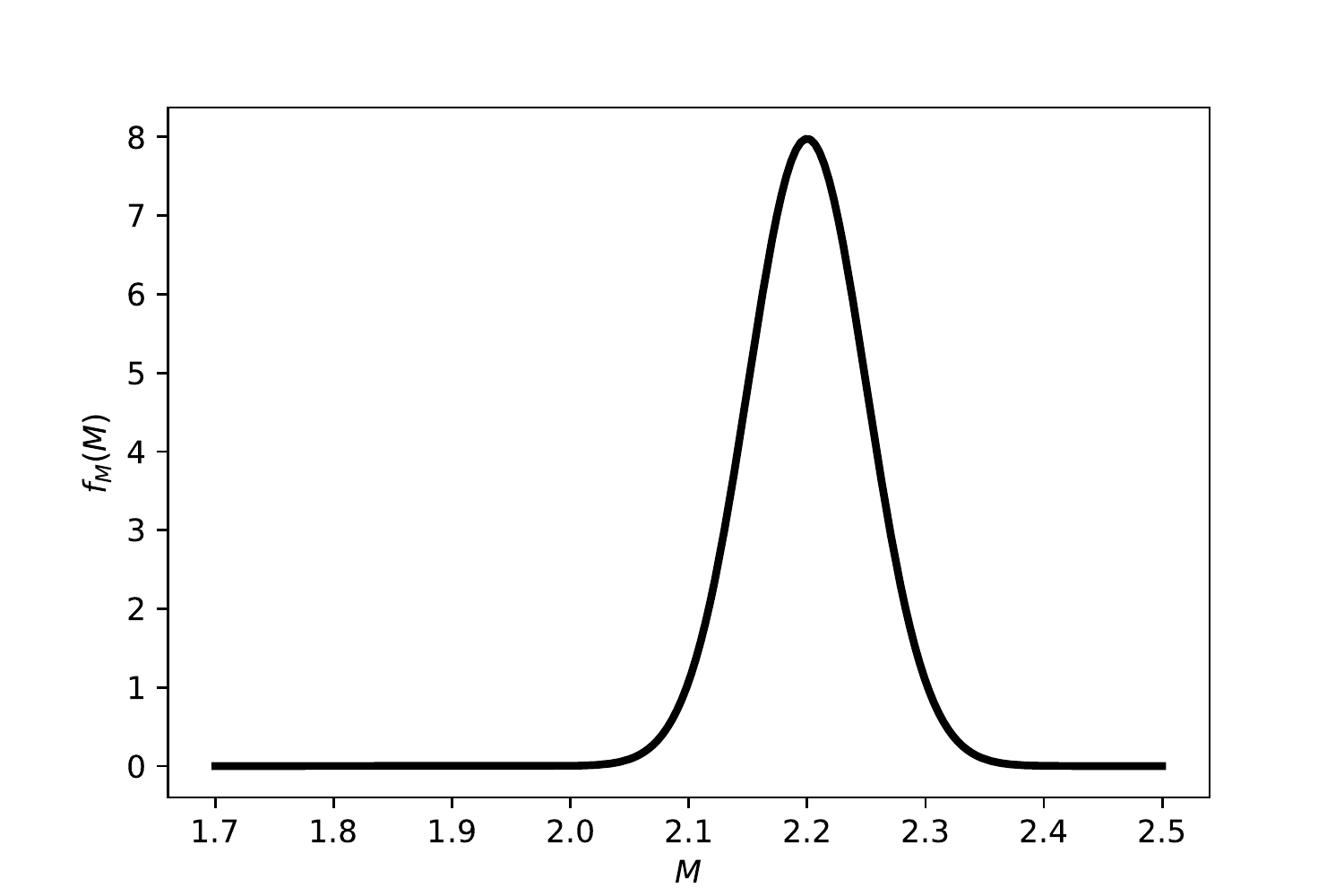}
	\includegraphics[scale=0.53]{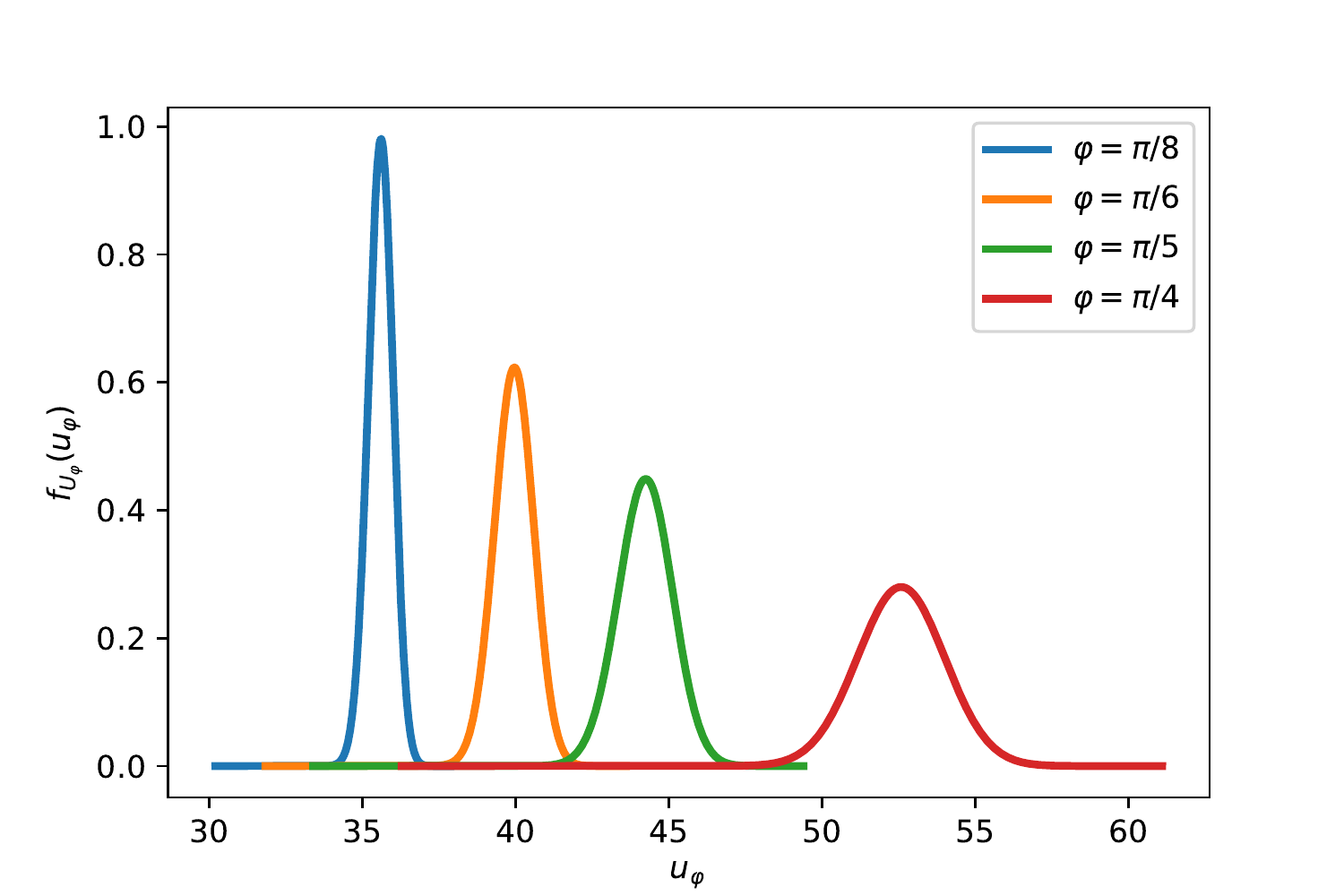}
	\caption{(Second case) Probability density function for the mass $M$ (left panel) and for the variable
	$\uphi$ for different angles $\varphi$ (right panel).
	\label{fig:case1b_2} }
\end{figure}
\par
If the random variable $M$ is normally distributed with mean value $\mu_M = 1$ and standard deviation
$\sigma_M = 0.05$, the resulting probability density function $f_{U}$ of the random variable $\uphi$,
for some exemplary values of $\varphi$, is shown in Fig.~\ref{fig:case1b_2} along with
the corresponding probability density function $f_M$.
As expected the variance of the probability density function of $\uphi$ increases with $\varphi$.
\subsection{Third case}
\begin{figure}
\centering
	\includegraphics[scale=0.44]{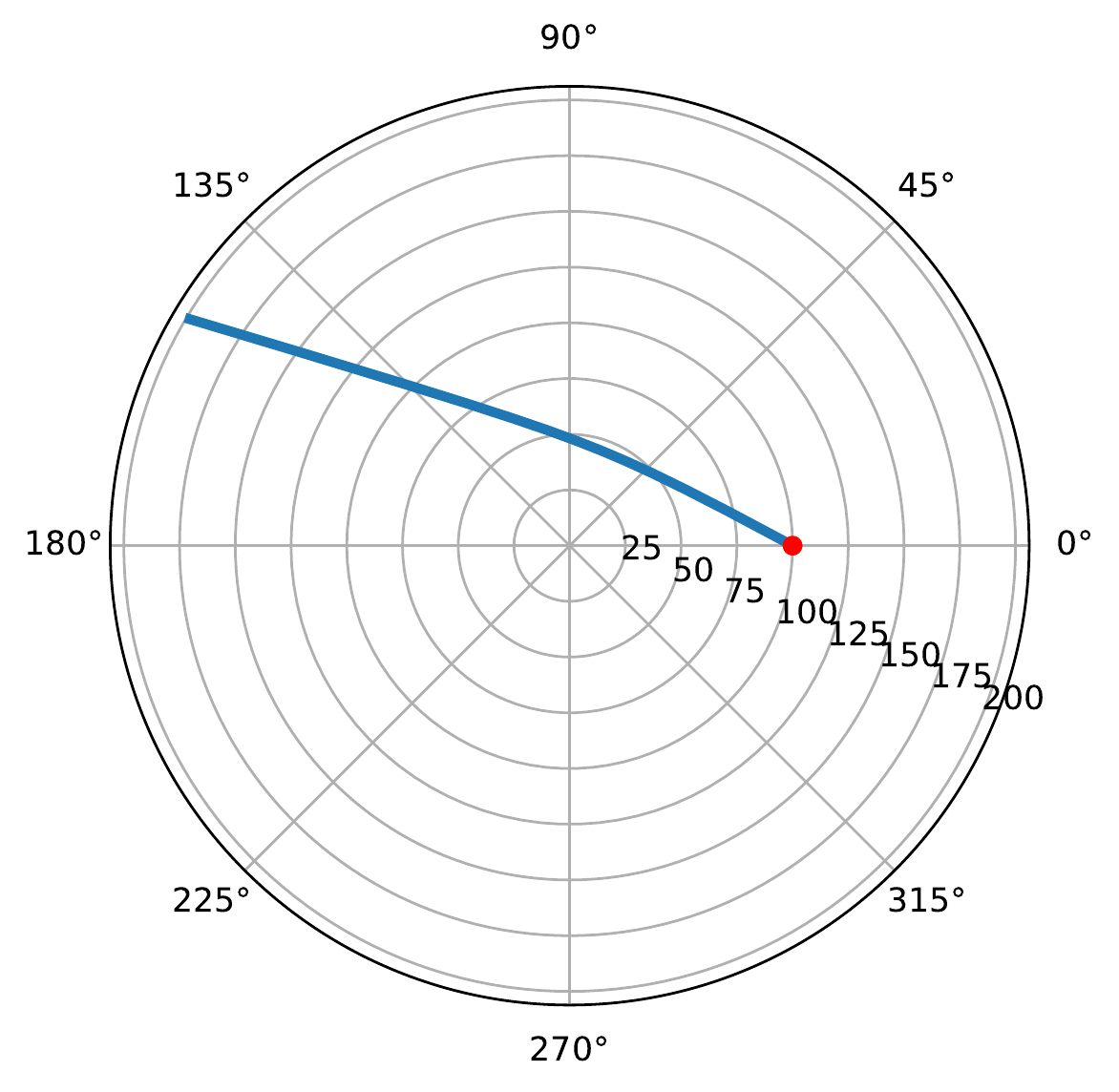}%
	\includegraphics[scale=0.44]{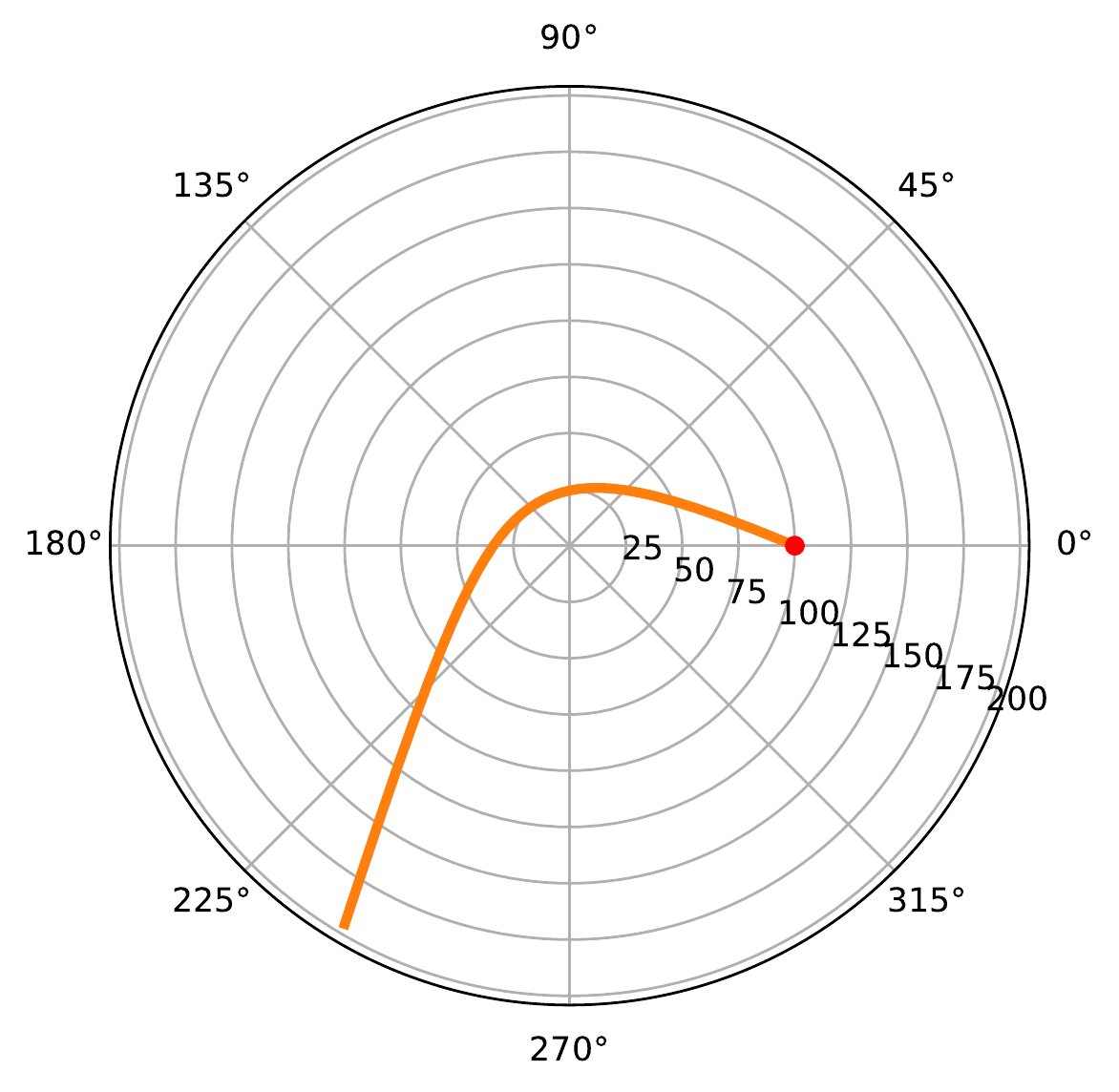}%
	\includegraphics[scale=0.44]{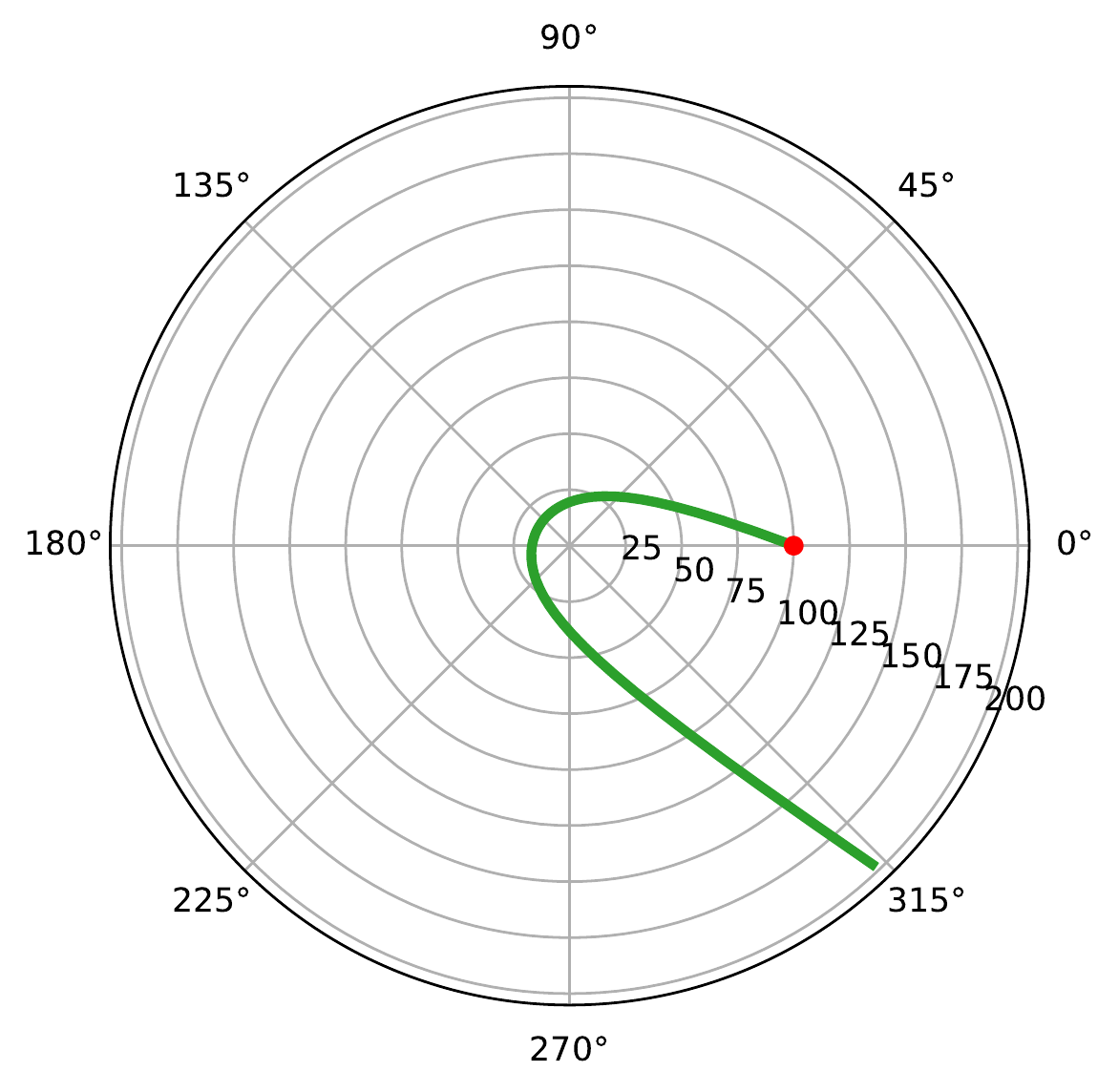}
	\includegraphics[scale=0.44]{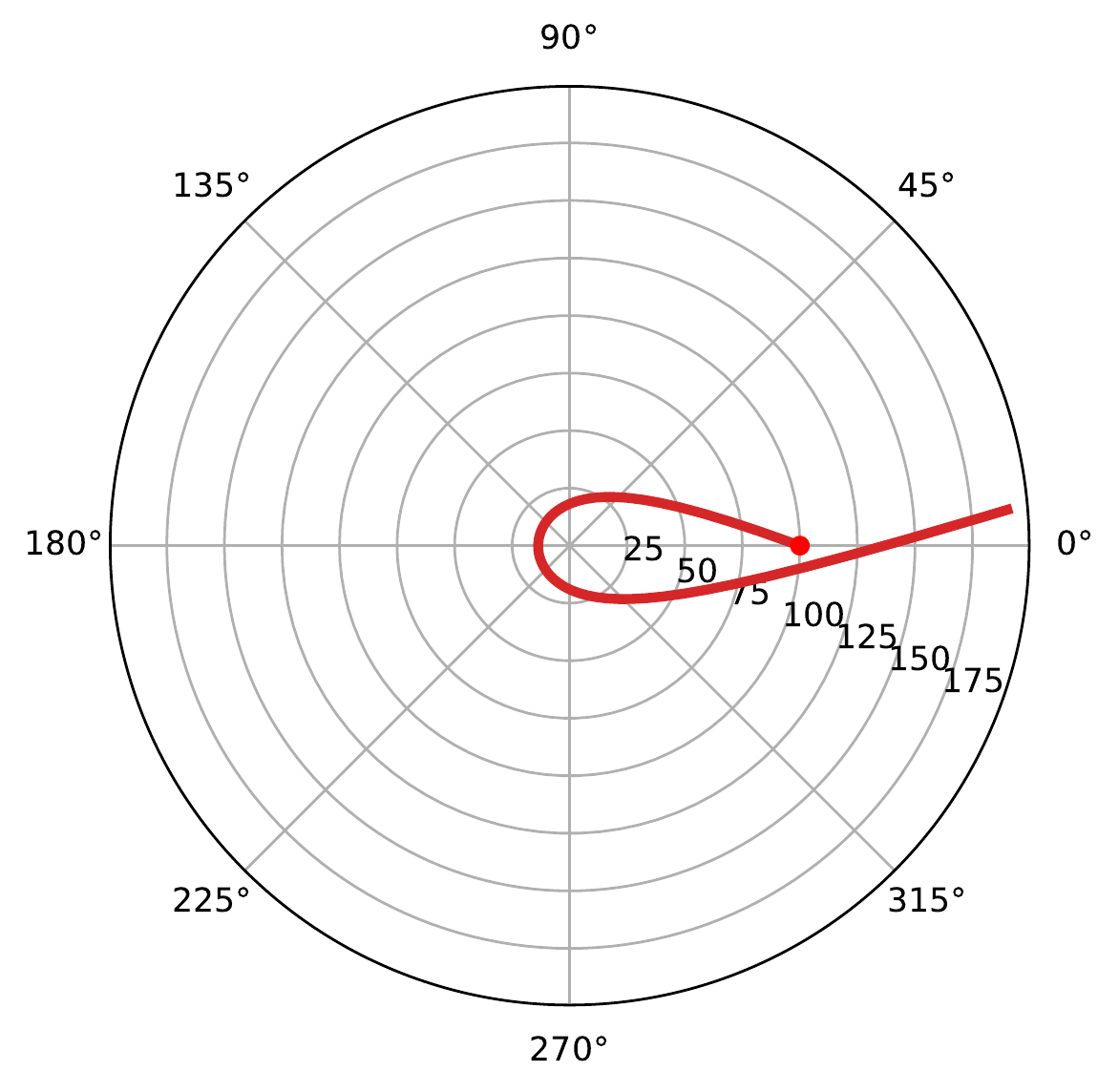}%
	\includegraphics[scale=0.44]{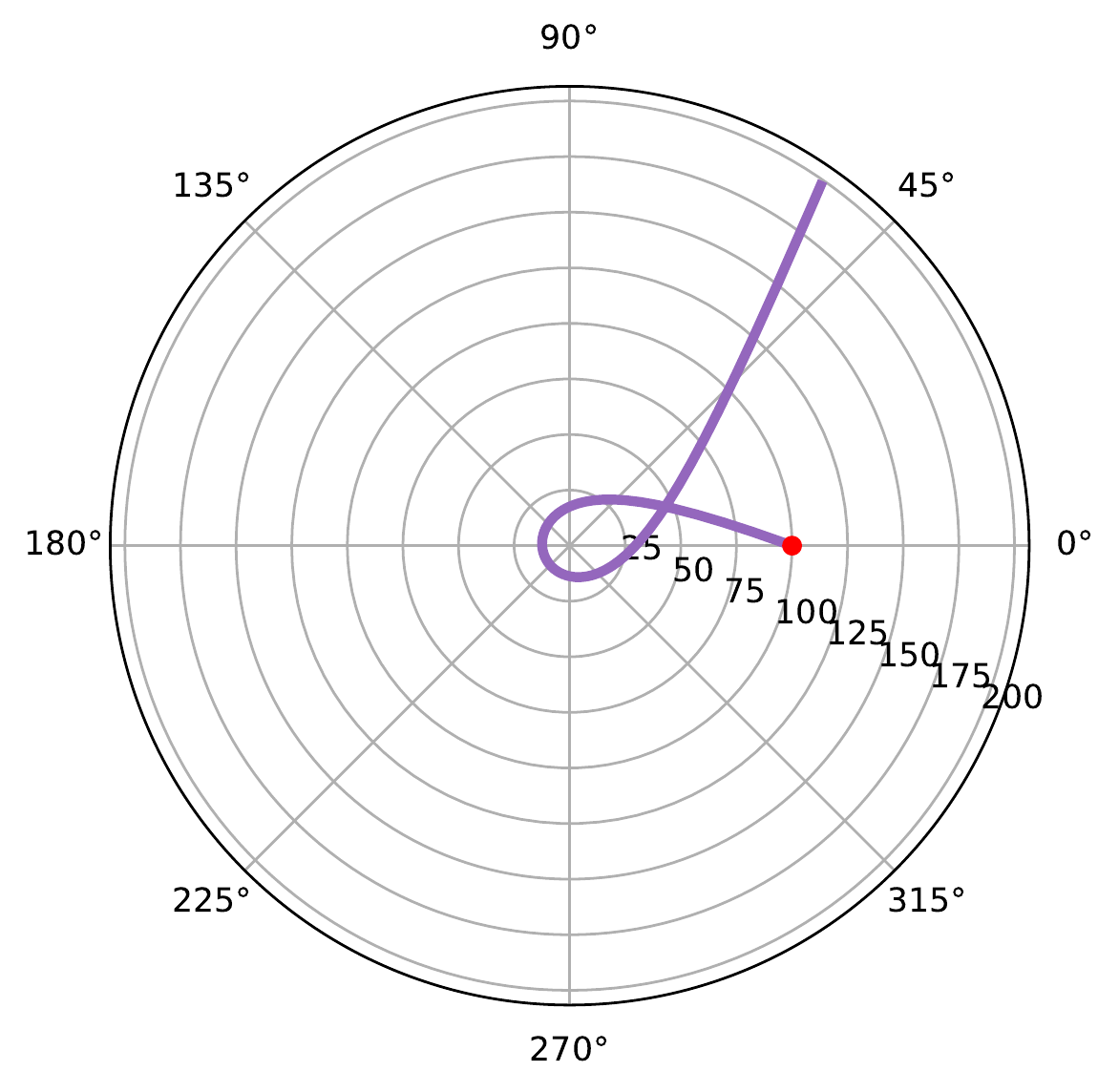}%
	\includegraphics[scale=0.44]{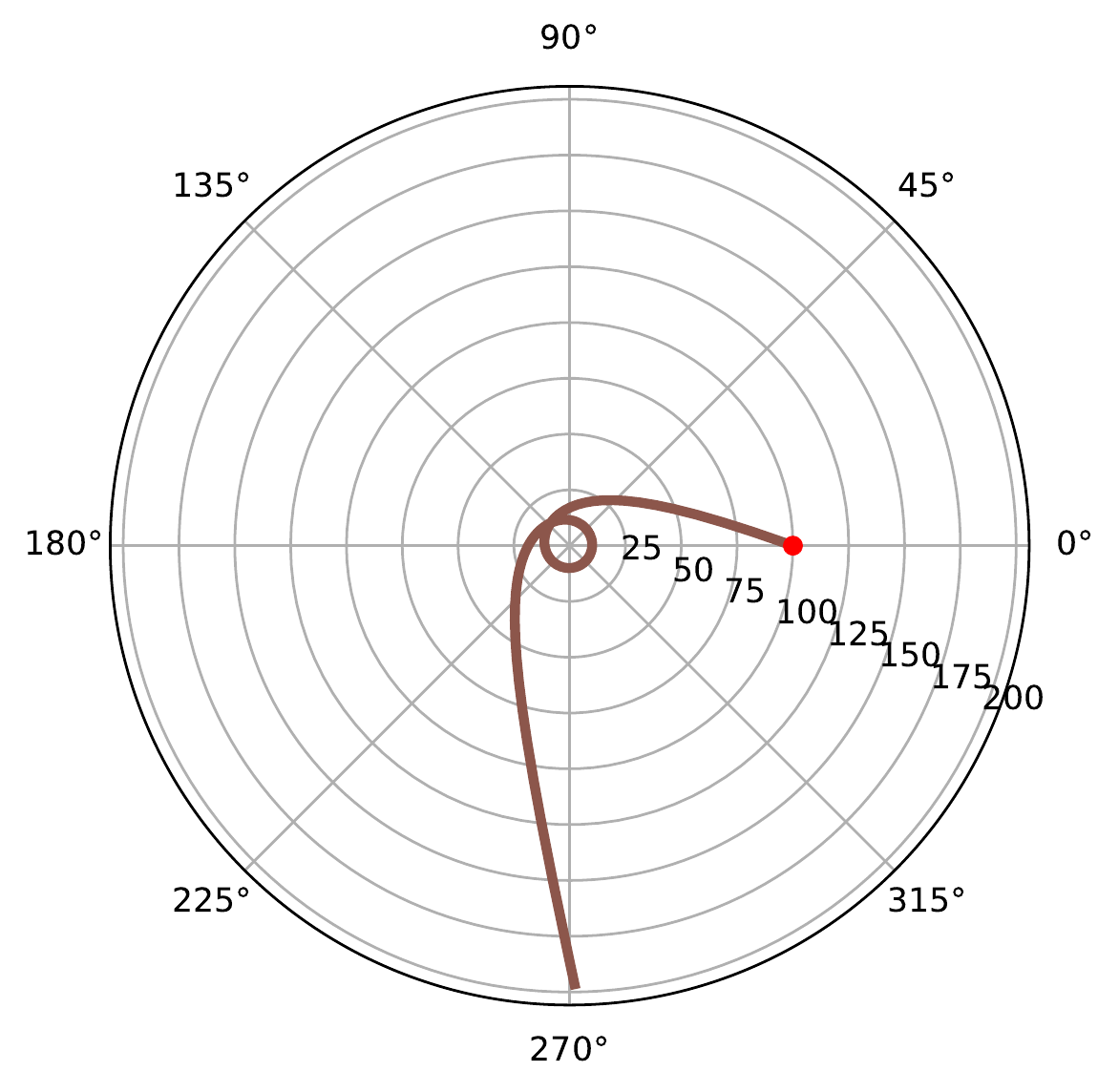}
	\caption{(Third case) Orbits for values of $M$ ranging from $0.2$ to $2.4$
	(tope left to bottom right, with $G=1$, $E=1.02$, $L=10$, $u_0=1$.)
	The red dot is the starting point and colours are the same as in Fig.~\ref{fig:case1c_1}.
	\label{fig:case1c_0} }
\end{figure}
\begin{figure}
\centering
	\includegraphics[scale=0.53]{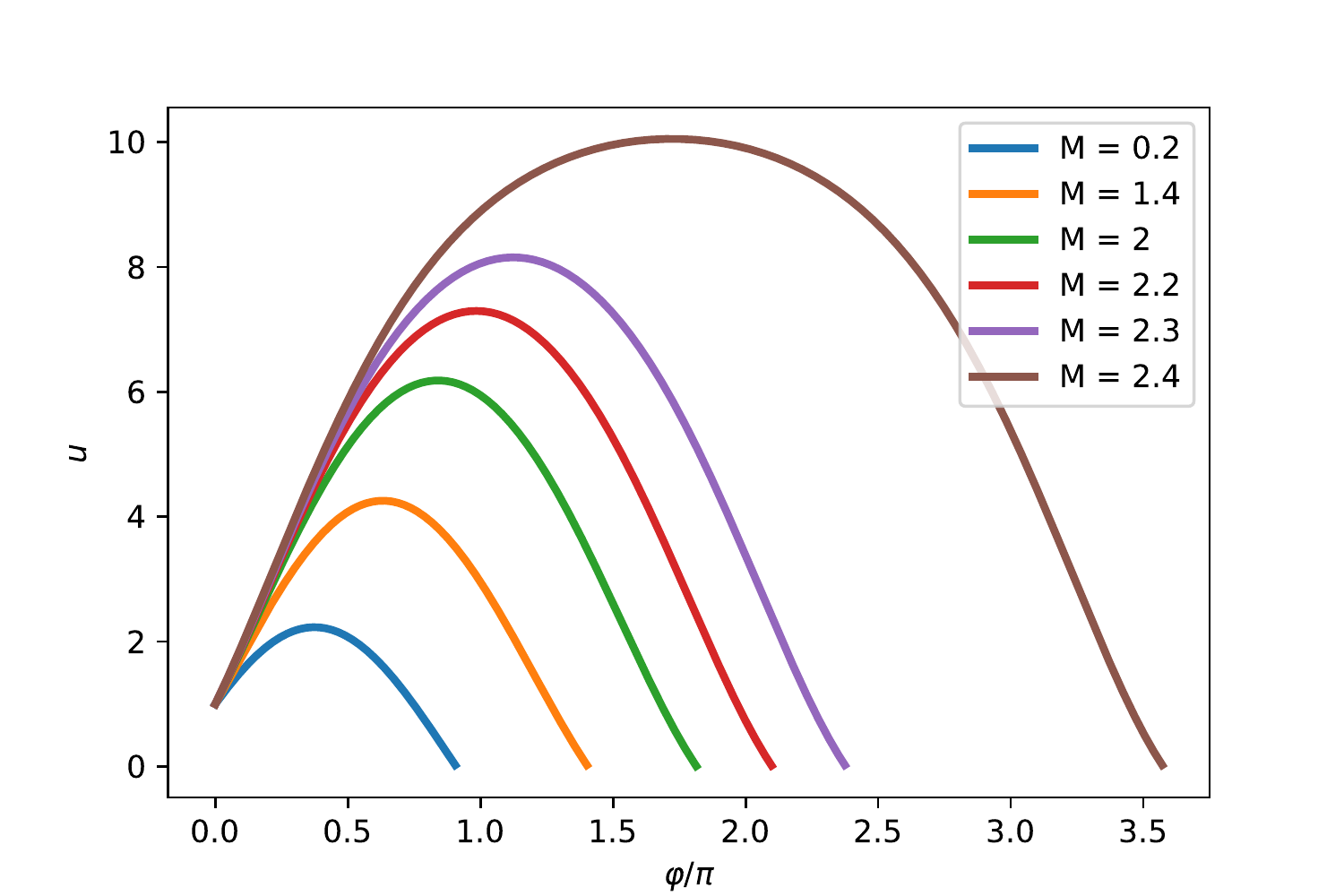}
	\includegraphics[scale=0.53]{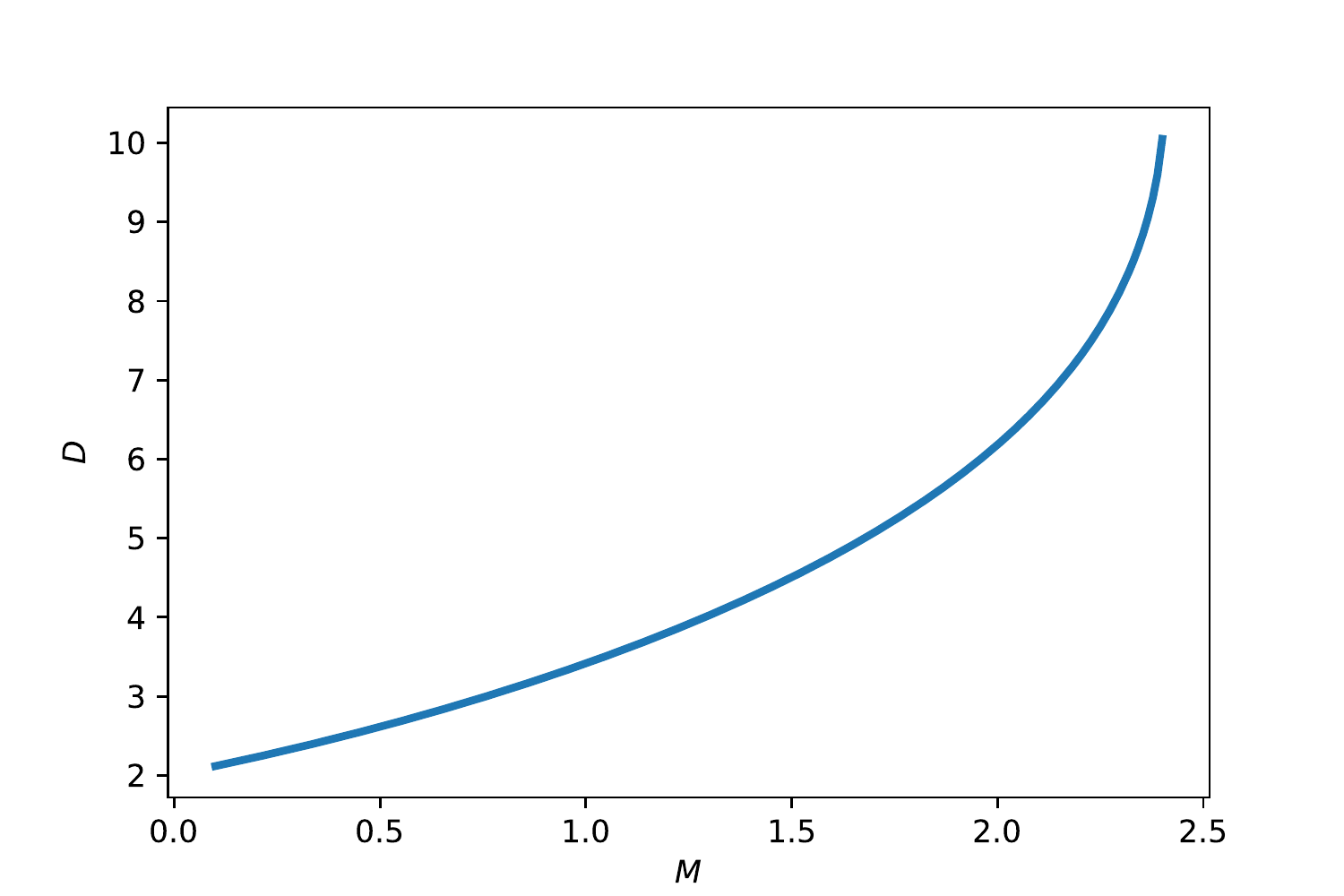}
	\caption{(Third case) Graphs of $u(\varphi)$ for values of $M$ ranging from $0.2$ to $2.4$
	(left panel, with $G=1$, $E=1.02$, $L=10$, $u_0=1$) and maximum value $D$ of $u$ 
	(inverse minimum distance of the particle from the central mass) as a function of the mass $M$
	(right panel).
	\label{fig:case1c_1} }
\end{figure}
Another type of solution is the one representing a particle coming from infinity towards the central mass
and then moving back towards infinity ($r \to \infty$, i.e.~$u \to 0$).
Some typical polar trajectories $r(\varphi)$ of this kind are shown in Fig.~\ref{fig:case1c_0} 
and the corresponding function $u(\varphi)$ are displayed in Fig.~\ref{fig:case1c_1}.
The maximum value $D$ of $u$ reached along the trajectory (proportional to the inverse of the minimum
distance from the central mass) is also plotted as a function of $M$.
\begin{figure}
\centering
	\includegraphics[scale=0.53]{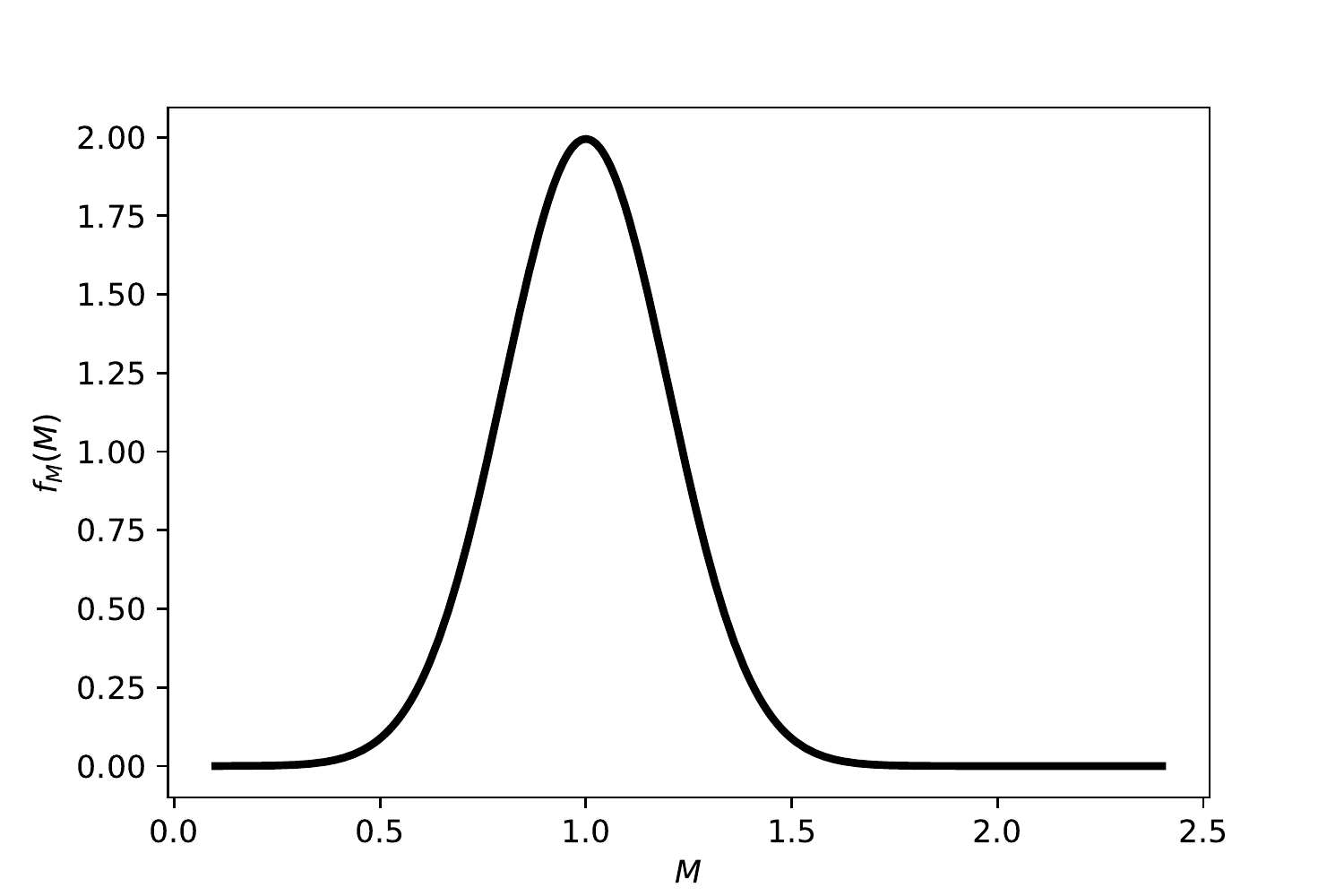}
	\includegraphics[scale=0.53]{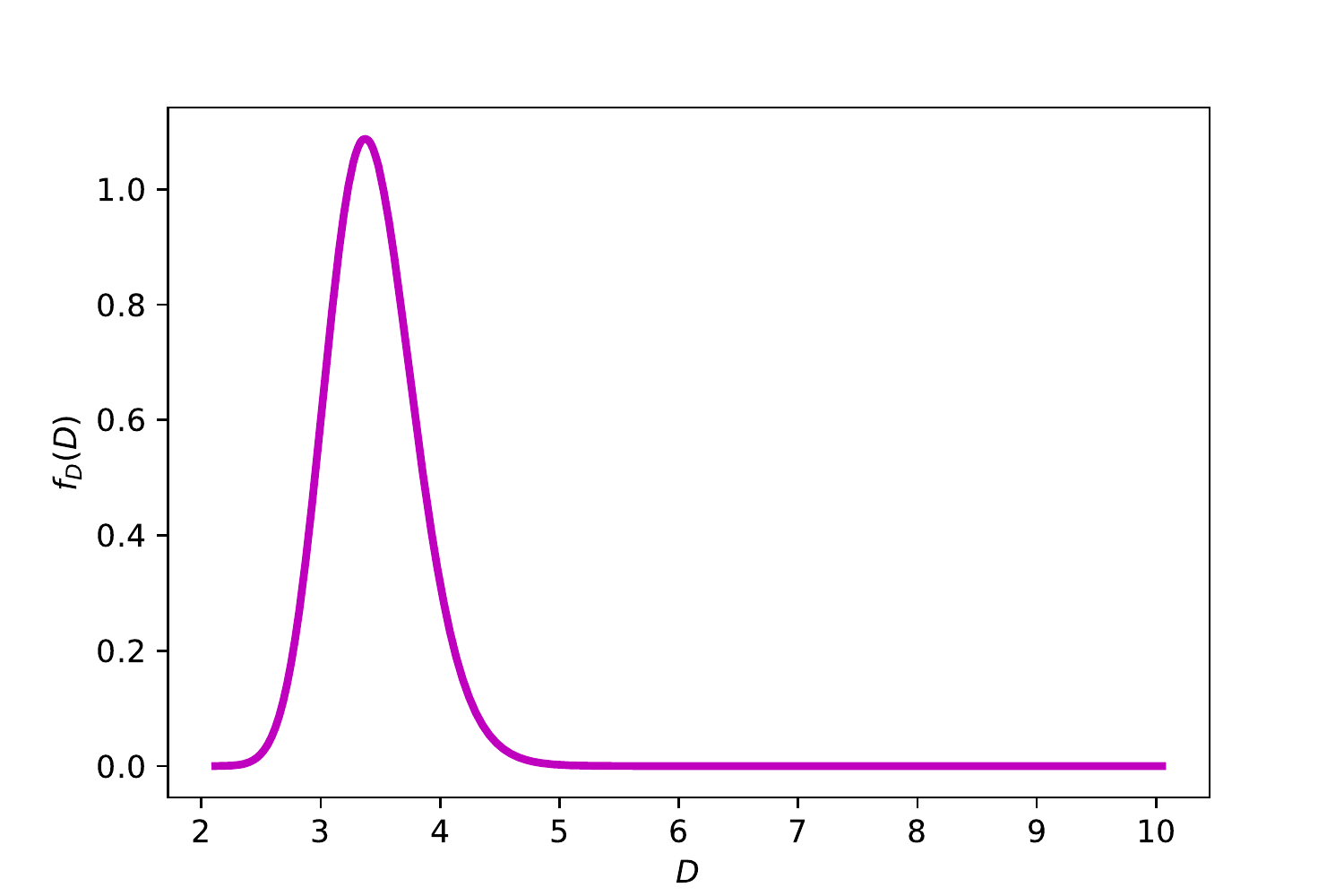}
	\caption{(Third case) Probability density function for the mass $M$ (left panel) and for the amplitude $D$
	inversely proportional to the minimum distance from the central mass (right panel).
	\label{fig:case1c_2} }
\end{figure}
\par
If the random variable $M$ is normally distributed with mean value $\mu_M = 1$ and standard deviation
$\sigma_M = 0.2$, the probability density function $f_D$ of the random variable $D$ is again obtained
by means of Eq.~\eqref{eq:pdf}.
The probability density functions $f_M$ and $f_D$ are shown in Fig.~\ref{fig:case1c_2}.
\subsection{Fourth case}
\begin{figure}
\centering
	\includegraphics[scale=0.33]{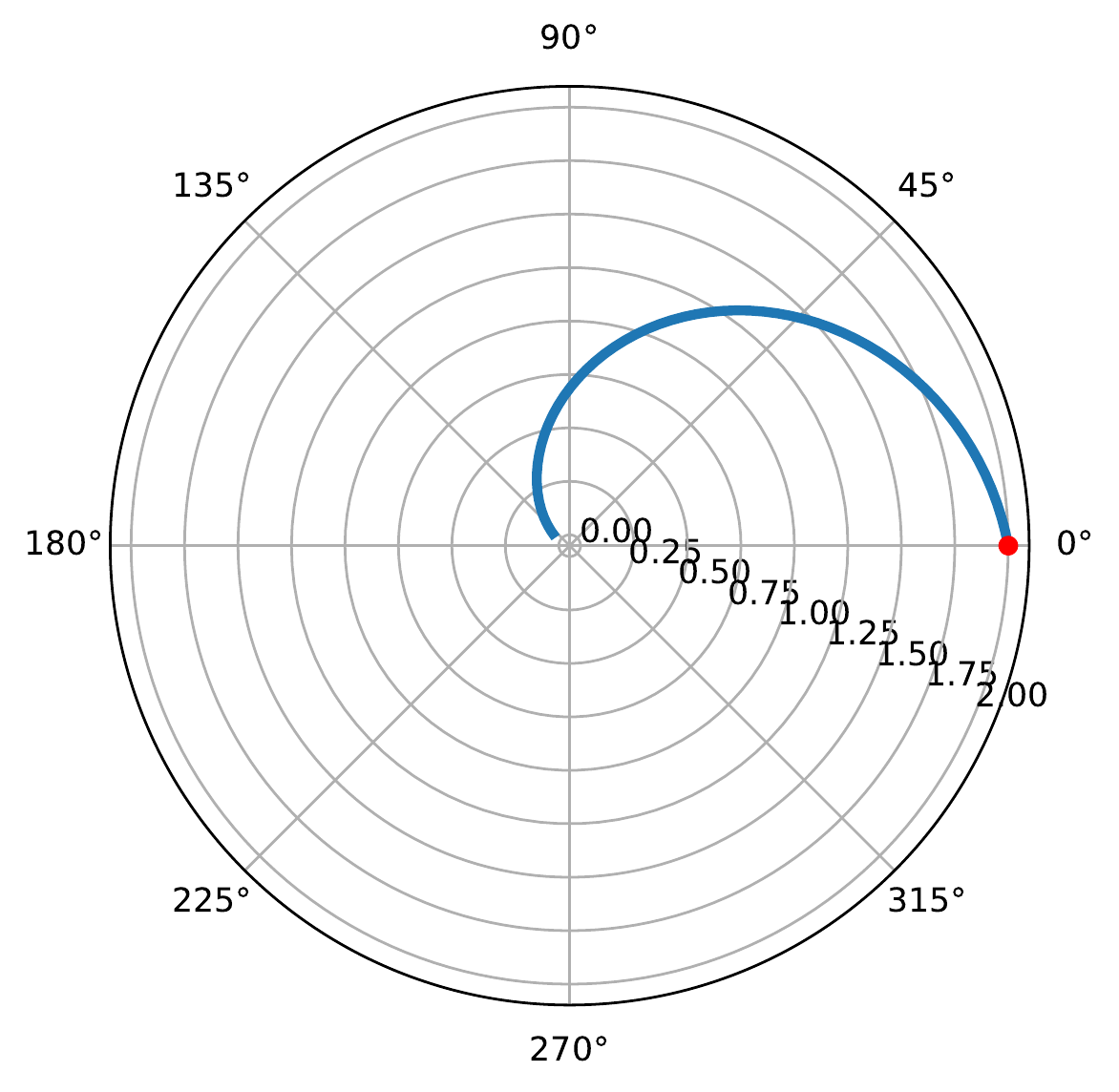}%
	\includegraphics[scale=0.33]{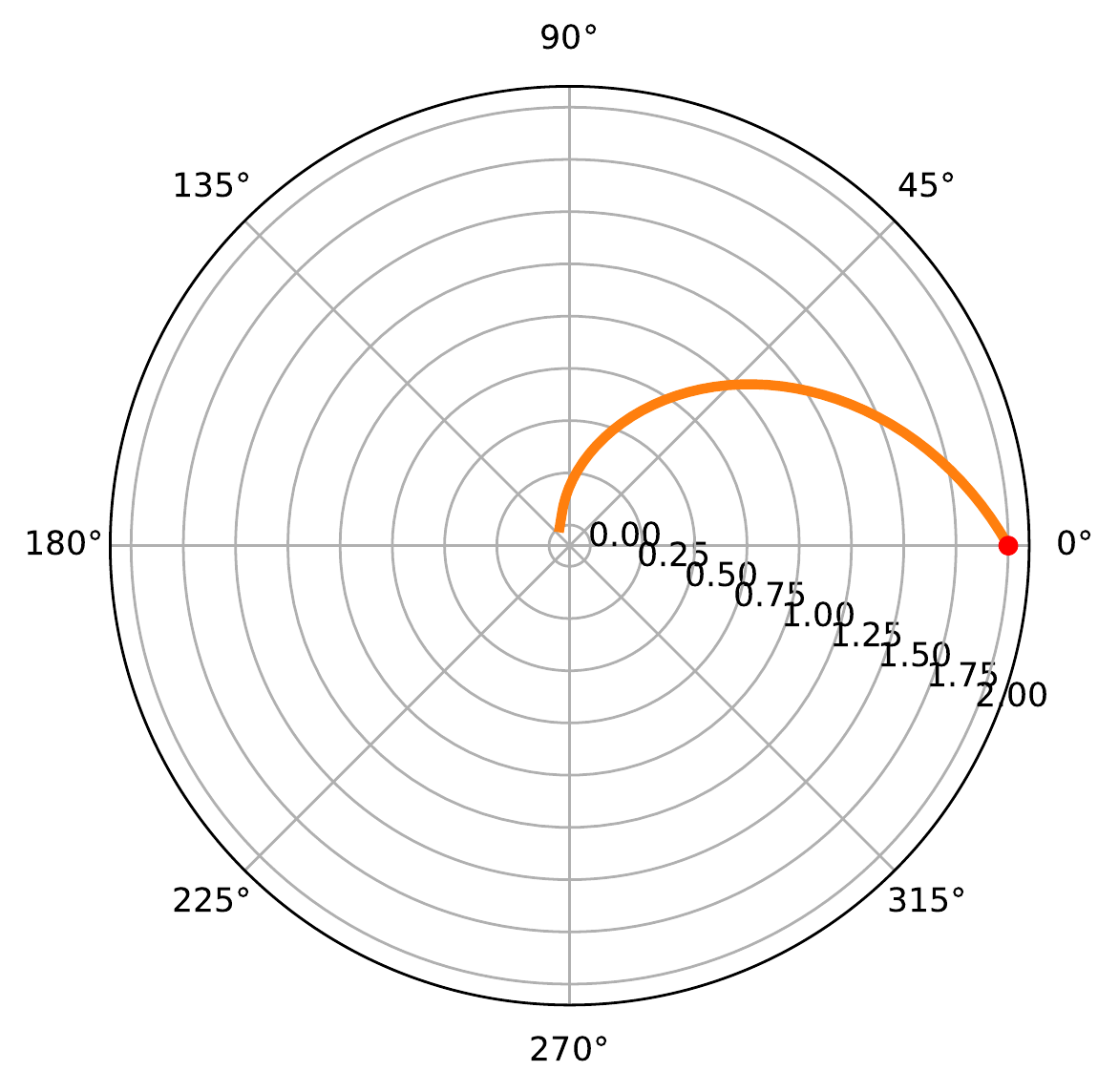}%
	\includegraphics[scale=0.33]{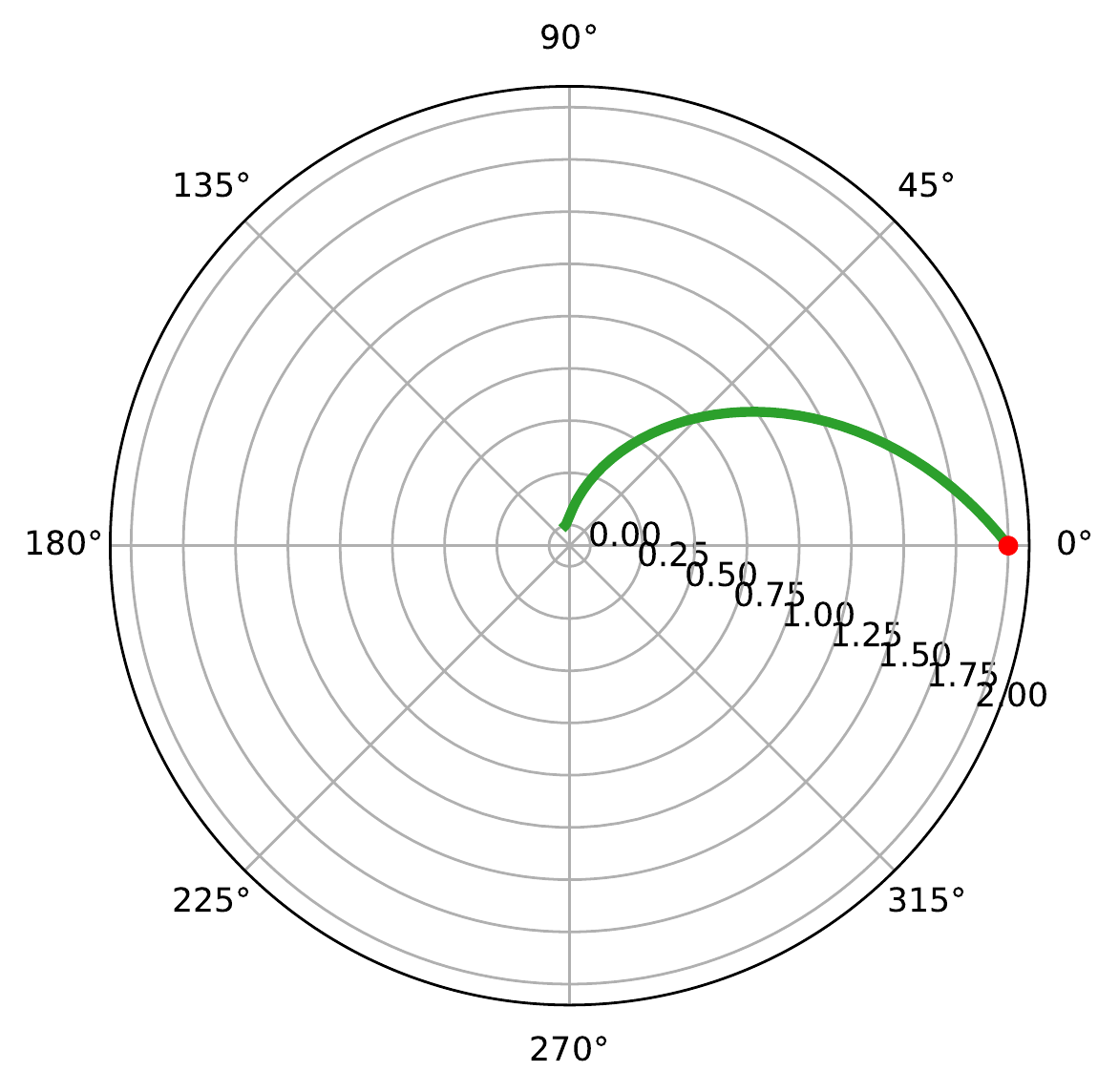}%
	\includegraphics[scale=0.33]{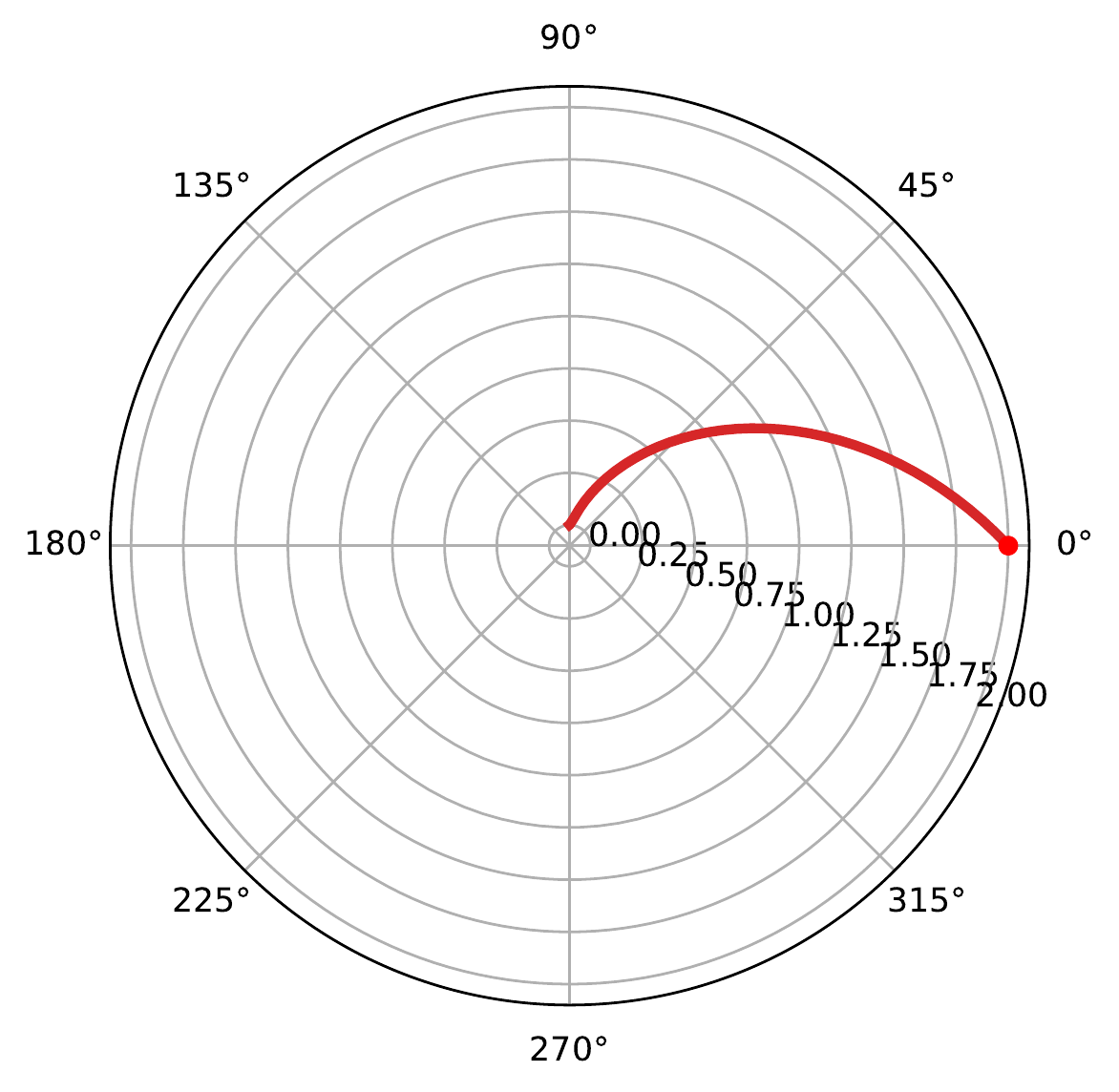}%
	\caption{(Fourth case) Orbits for values of $M=1$, $1.3$, $1.6$, $1.9$
	(left to right, with $G=1$, $E=0.98$, $L=10$, $u_0=50$.
	The red dot is the starting point and colours are the same as in Fig.~\ref{fig:case4_1}.
	\label{fig:case4_0} }
\end{figure}

\begin{figure}
\centering
	\includegraphics[scale=0.53]{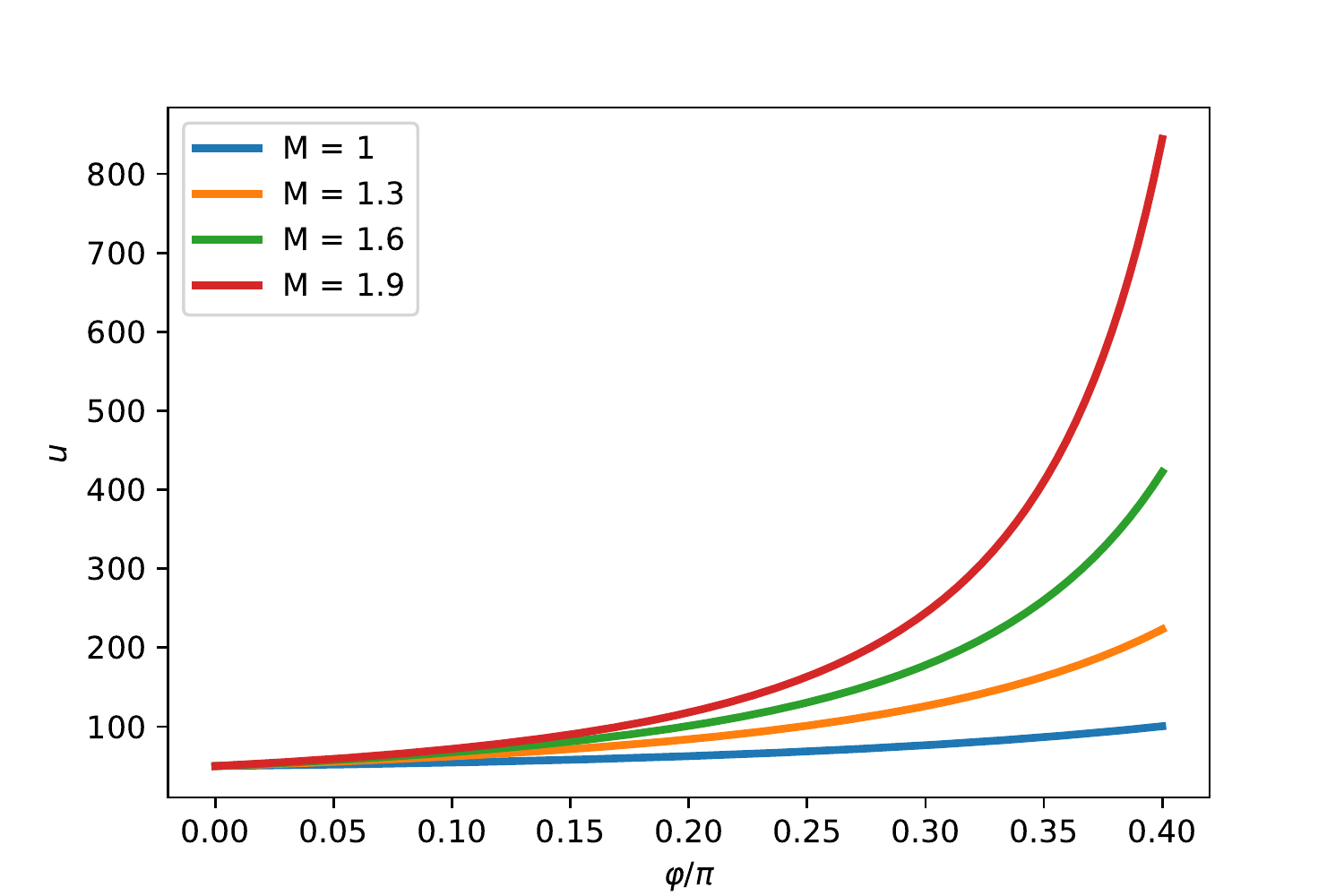}
	\includegraphics[scale=0.53]{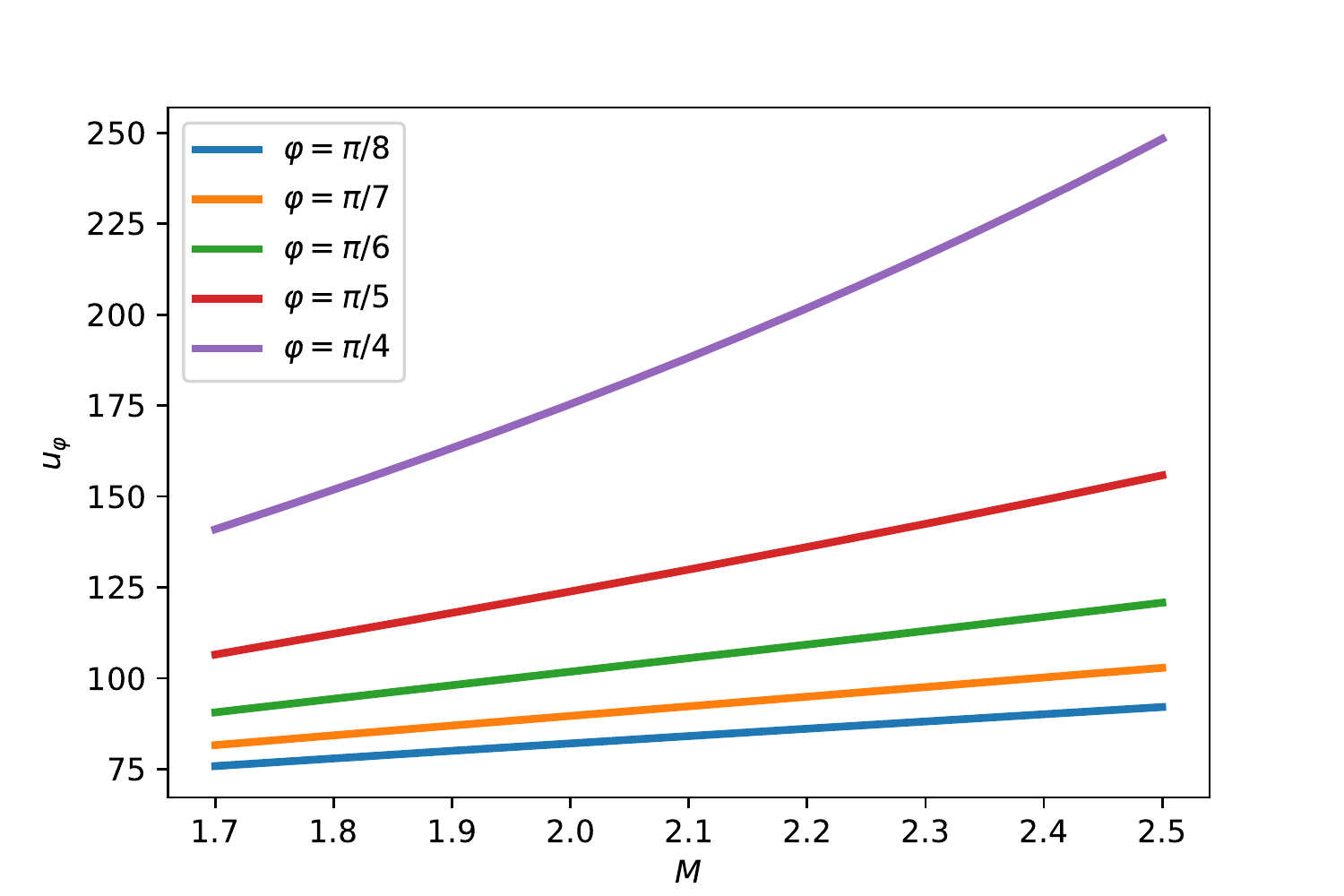}
	\caption{(Fourth case) Graphs of $u(\varphi)$ for values of $M=1$, $1.3$, $1.6$, $1.9$
	(left panel, with $G=1$, $E=0.98$, $L=10$, $u_0=50$) and maximum value $D$ of $u$
	(inverse minimum distance of the particle from the central mass)
	as a function of the mass $M$ (right panel)
	.\label{fig:case4_1} }
\end{figure}
The last type of solution is the one representing a particle with enough energy and sufficiently low angular
momentum which spirals into a black hole with a finite change in $\varphi$.
Some examples of these orbits are shown in Fig.~\ref{fig:case4_0} and the corresponding function
$u(\varphi)$ is plotted in Fig.~\ref{fig:case4_1}, along with the behaviour of $\uphi$ as a function of $M$
for different values of the angle $\varphi$.
\begin{figure}
\centering
	\includegraphics[scale=0.53]{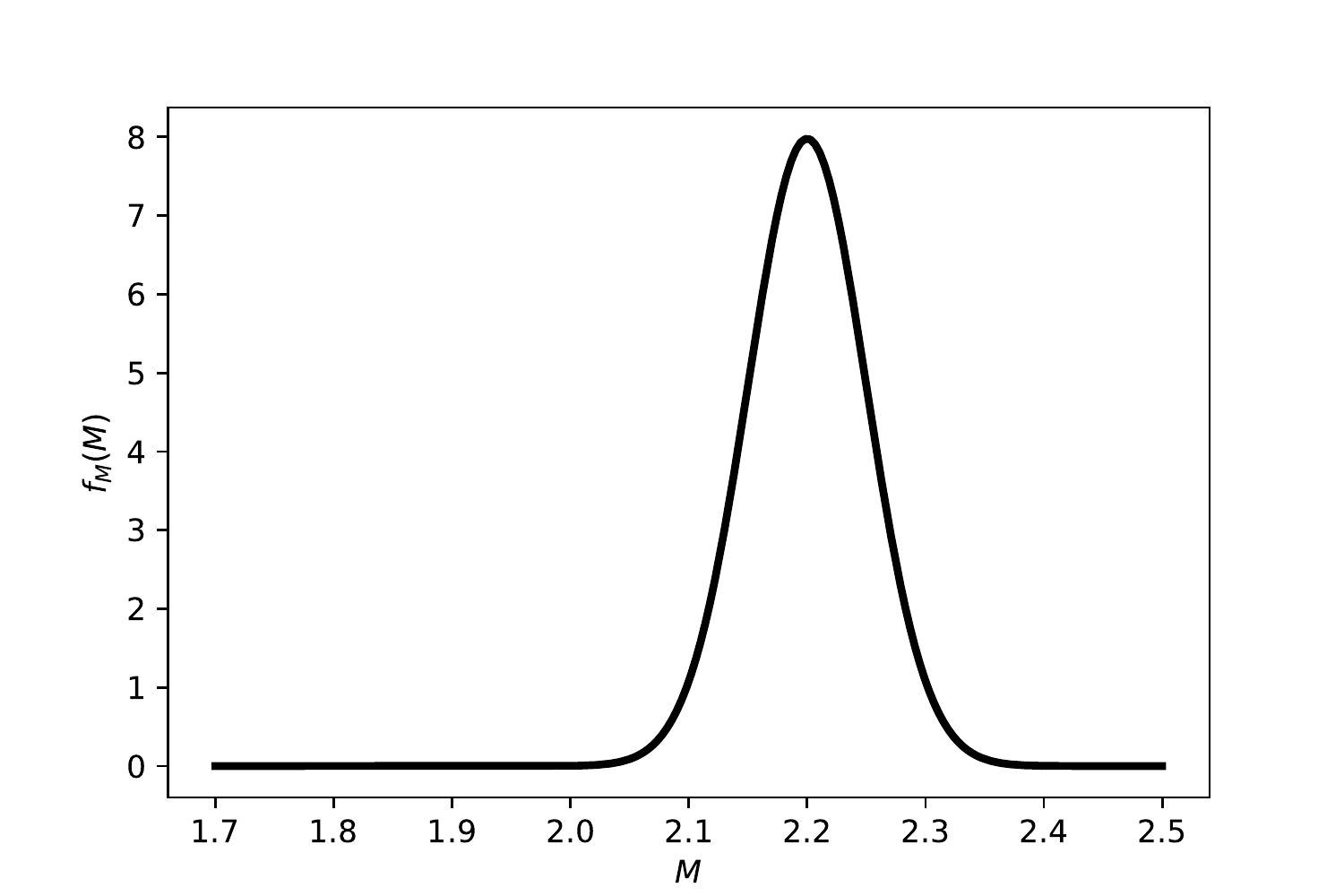}
	\includegraphics[scale=0.53]{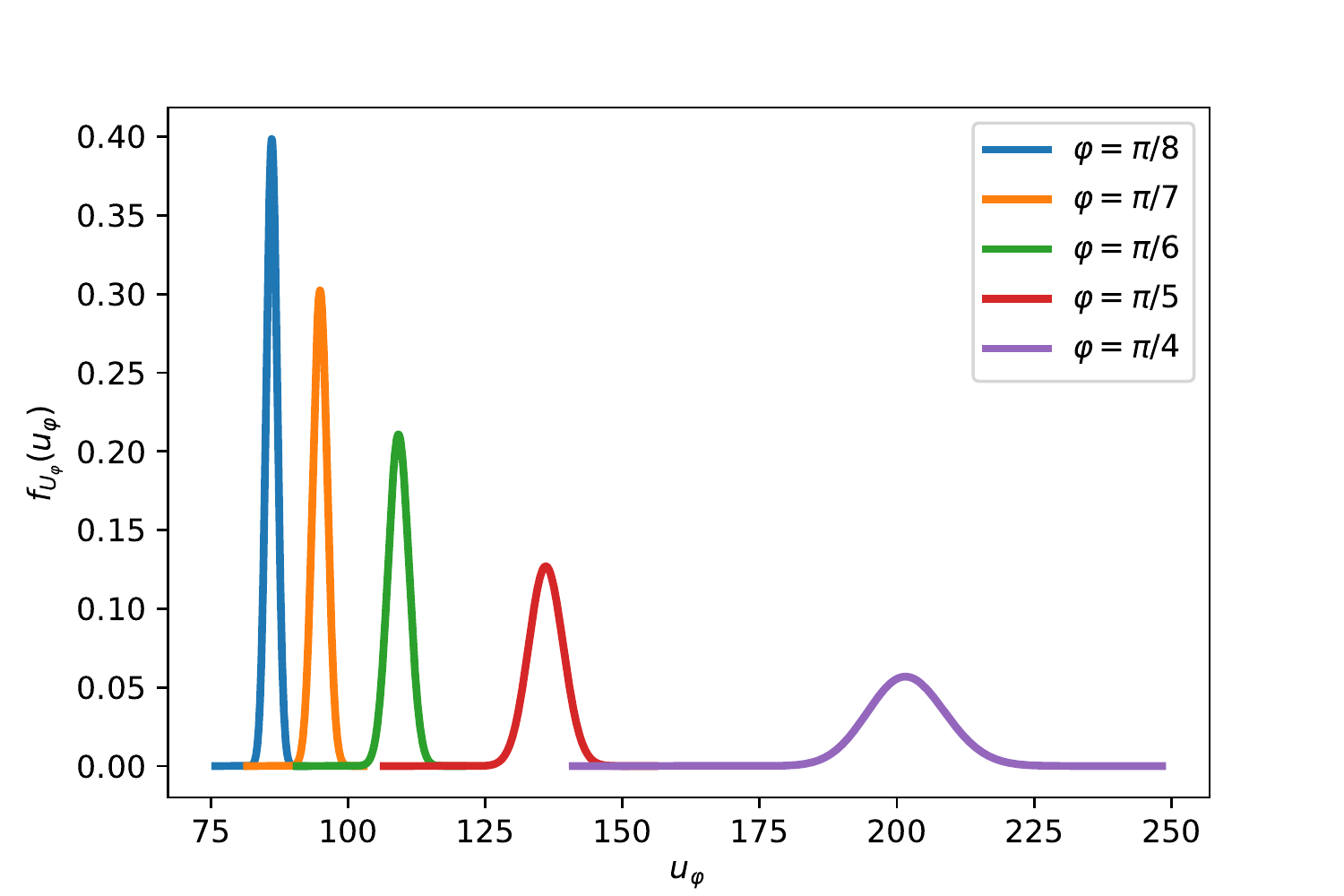}
	\caption{(Fourth case) Probability density function for the mass $M$ (left panel) and for the amplitude
	$U$ (right panel).
	\label{fig:case4_2} }
\end{figure}
\par
For $M$ normally distributed with mean value $\mu_M = 2.2$ and standard deviation $\sigma_M = 0.05$, 
the resulting probability density functions $f_{U}$ for some angles $\varphi$ is shown in Fig.~\ref{fig:case4_2}.
\section{Perturbation of circular orbits}
\label{S:perturbedCircular}
We can now study circular orbits again using the variable $u$ defined in Eq.~\eqref{eq:u},
so that Eq.~\eqref{eq:dx_dphi} reads 
\begin{equation}
\label{du_dphi} % Eq. (5.78) Carroll
	\left( \frac{\d u}{\d\varphi}\right)^2
	=
	\frac{E^2\, L^2}{G^2} 
	- \epsilon\, \frac{L^2}{G^2}
	+2\,\epsilon\,M\, u
	- u^2
	+
	\gamma\, \frac{2\,G^2\,M}{L^2}\, u^3
\end{equation}
and Eq.~\eqref{eq:d2x_dphi2} reads
\begin{equation}
\label{eq:d2u_dphi2}
	\frac{\d^2 u}{\d \varphi^2}
	= 
	\epsilon\, M 
	- \u
	+\gamma\, \frac{3\, G^2\, M}{L^2} \,u^2
	\ .
\end{equation}
We wish to study the effect on circular orbits of small perturbations affecting the central mass $M$
and the initial position of the test particle.
We shall first use an analytical perturbative approach for Eq.~\eqref{eq:d2u_dphi2}
with the initial conditions
\begin{equation} \label{eq:perturbed-ic}
	u(0)
	=
	\u_{\rm c}
	\equiv
	\frac{L^2}{G\,r_{c}}
	\ ,
	\qquad
	\frac{\d u}{\d \varphi}(0)
	= 0
	\ ,
\end{equation}
where $r_{\rm c}$ is the radius of a circular orbit (either internal or external).
Note that the initial condition on the derivative implies that an instantaneous perturbation
of the mass does not instantaneously alter the tangential component of the particle velocity.
\subsection{Mass perturbation}
If we denote with $\delta M$ the perturbation on the mass $M$, the perturbed solution can
be written as 
\begin{equation}
\label{eq:perturbation-M}
 u 
 = u^{(0)}
 + \frac{\delta M}{M} \,u^{(1)}
 + \bigO{\frac{\delta M^2}{M^2}}
 \ .
\end{equation}
Upon replacing into Eq.~\eqref{eq:d2u_dphi2} and keeping terms up to order $\delta M/M$, we find
\begin{equation}
	\frac{\d ^2 u^{(0)}}{\d\varphi^2}
	+
	\frac{\delta M}{M}\, \frac{\d^2 u^{(1)}}{\d\varphi^2}
	= 
	\epsilon\, \left(M +\delta M\right)
	-u^{(0)}
	- \frac{\delta M}{M} \,u^{(1)}
	+ 
	\gamma\, \frac{3 \,G^2}{L^2}
	\left(M + \delta M\right)
	\left(u^{(0)} + \frac{\delta M}{M} \,u^{(1)}\right)^2
	\ .
\end{equation}
Assuming $u^{(0)}$ satisfies the unperturbed equation~\eqref{eq:classical} with fixed $M$,
we obtain that the first order correction must satisfy
\begin{equation}
\label{eq:perturbed-M-circular}
	\frac{\d^2 u^{(1)}}{\d\varphi^2}
	=
	\alpha_{\rm c} \,u^{(1)}
	+\beta_{\rm c}
	\ ,
\end{equation}
where
\begin{equation}
	\alpha_{\rm c}
	\equiv
	\gamma\, \frac{6\, G^2\, M}{L^2} \,u^{(0)} - 1
\label{eq:ac}
\end{equation}
and
\begin{equation}
	\beta_{\rm c}
	\equiv
	\gamma\, \frac{3\, G^2\, M}{L^2} \left(u^{(0)}\right)^2
	+
	\epsilon\, M
	\ .
\end{equation}
Moreover, the initial conditions for $u^{(1)}$ are given by
\begin{equation}
	u^{(1)}(0)
	= 
	\frac{\d u^{(1)}}{\d\varphi}(0)
	 =
	 0
	 \ .
\end{equation}
Focusing on the case $\gamma=1$, and restricting ourselves, from now on, to the perturbation
of circular orbits ($u^{(0)} = u_{\rm c}$), we treat the massive and massless cases separately.
\par
In the massive case ($\epsilon = 1$), two circular orbits exist [see Eq.~\eqref{eq:rc-GR-mass}]:
\begin{enumerate}
\item
For the internal orbit we have $r_{c}= r_{\rm in}=r_-$ in Eq.~\eqref{eq:rc-GR-mass}, that is
\begin{equation} \label{uin}
u_{c}
=
\u_{\rm in}
=
\frac{L^2 + \sqrt{L^4 - 12\, G^2\, M^2\, L^2}}{6\, G^2\, M}
\ ,
\end{equation}
for which $\alpha_{\rm c}=\alpha>0$ [see Eq.~\eqref{eq:rc-GR-mass}] and
$\beta_{\rm c}=u_{\rm in}$.
The analytical solution of Eq.~\eqref{eq:perturbed-M-circular} is then
\begin{equation}
u^{(1)}
=
\frac{\beta_{c}}{\alpha_{c}}
\left[ \cosh\left(\sqrt{\alpha_{c}}\, \varphi\right) - 1 \right]
\ ,
\end{equation}
and it is easy to see that, as expected, this orbit is unstable.
In fact, the perturbed solution~\eqref{eq:perturbation-M} to first order reads
\begin{equation}
u(\varphi)
\simeq
u_{\rm in}
\left[ 1 + \frac{\delta M}{M}\, \frac{ \cosh\left(\sqrt{\alpha}\, \varphi\right) - 1 }{\alpha} \right]
\ ,
\end{equation}
from which we can see that increasing the mass ($\delta M > 0$) leads to $r \to 0$ ($\u \to \infty$).
On the other hand, reducing the mass ($\delta M < 0$), leads to $r \to \infty$ ($\u=0$)
for some finite value of $\varphi$.
\item
For the external orbit we have $r_{c}= r_{\rm out}=r_+$ in Eq.~\eqref{eq:rc-GR-mass}, or
\begin{equation} \label{uout}
u_{c}
=
u_{\rm out}
=
\frac{L^2 - \sqrt{L^4 - 12\, G^2\, M^2\, L^2}}{6\, G^2\, M}
\ ,
\end{equation}
so that $\alpha_{\rm c}=-\alpha<0$ and $\beta_{\rm c}=u_{\rm out}$.
The analytical solution of Eq.~\eqref{eq:perturbed-M-circular} is now
\begin{equation}
u^{(1)}
=
\frac{\beta_{c}}{\alpha_{c}} \left[\cos\left(\sqrt{\abs{\alpha_{c}}}\, \varphi\right) - 1\right]
\ ,
\end{equation}
which implies that the outer orbit is stable, with $u^{(1)}$ oscillating between $0$ and
$2\,\beta_{\rm c} / \abs{\alpha_{\rm c}}$.
The perturbed solution to first order is given by
\begin{equation}
u(\varphi)
\simeq
u_{\rm out}
\left[ 1 + \frac{\delta M}{M}\, \frac{ 1-\cos\left(\sqrt{\alpha}\, \varphi\right)}{\alpha} \right]
\ ,
\end{equation}
so that $\delta M>0$ leads to periodic oscillations internal ($u\ge u_{\rm out}$) to the stable orbit $r=r_{\rm out}$,
whereas $\delta M<0$ leads to periodic oscillations external ($u\le u_{\rm out}$) to the stable orbit $r=r_{\rm out}$.
This is potentially quite interesting in practical terms because the oscillations are proportional to the perturbation
$\delta M$ (assuming that the oscillations are large enough to be detectable).
\end{enumerate}
\par	
For the massless case ($\epsilon = 0$), there is one possible circular orbit $r_{\rm c} = r_{\rm ph}$
in Eq.~\eqref{eq:rc-GR-massless}, or
\begin{equation} \label{UPH}
u_{\rm c}
=
u_{\rm ph}
=
\frac{L^2}{3\, G^2\, M}
\ .
\end{equation}
In this case Eq.~\eqref{eq:perturbed-M-circular} becomes
\begin{equation}
	\frac{\d^2 u^{(1)}}{\d\varphi^2}
	=
	\u^{(1)}
	+u_{\rm ph}
\end{equation}
and its analytical solution is
\begin{equation}
	u^{(1)}
	=
	u_{\rm c}
	\left[\cosh(\varphi) - 1\right]
	\ .
\end{equation}
It is clear that, since $\u^{(1)} \to \infty$ as $\varphi \to \infty$, the orbit 
\begin{equation}
	u(\varphi)
	\simeq
	u_{\rm ph}
	\left\{ 1 + \frac{\delta M}{M}  \left[\cosh(\varphi) - 1\right]
	\right\}
\end{equation}
is unstable, and it will fall into the singularity ($u\to \infty$) if $\delta M>0$ or escape to infinity ($u\to 0$)
if $\delta M<0$.
\subsection{Initial position perturbation}
Let us now consider a perturbation $\delta u_0$ on the initial position $u_0$, so that
\begin{equation}
	u
	=
	u^{(0)}
	+
	\frac{\delta u_0}{u_0} \,u^{(1)}
	+
	\bigO{\frac{\delta u_0^2}{u_0^2}}
	\ .
\end{equation}
From Eq.~\eqref{eq:d2u_dphi2}, we have
\begin{eqnarray}  
	\frac{\d^2 u^{(0)}}{\d\varphi^2}
	+
	\frac{\delta u_0}{u_0}\, \frac{\d^2 \u^{(1)}}{\d\varphi^2}
	&=& 
	\gamma\, \frac{3\, G^2\, M }{L^2}
	\left[
	\left(u^{(0)}\right)^2 + \frac{\delta u_0}{u_0} \left(u^{(1)}\right)^2+ 2\, \frac{\delta u_0}{u_0} \,u^{(0)} \,u^{(1)}
	\right]
	\nonumber
	\\
	&&
	+	\epsilon\, M
	-u^{(0)}
	-\frac{\delta u_0}{u_0} \,u^{(1)}
\end{eqnarray}
so that the first order perturbation must satisfy
\begin{equation}
\label{eq:perturbed-u0}
	\frac{\d^2 u^{(1)}}{\d\varphi^2} 
	=
	\alpha_{\rm c} \,u^{(1)}
	\ ,
\end{equation}
with $\alpha_{\rm c}$ still given by Eq.~\eqref{eq:ac}.
Moreover, the proper initial conditions are given by
\begin{equation}
	\u^{(1)}(0)
	=
	\delta u_0
	\ , 
	\qquad 
	\frac{\d \u^{(1)}}{\d \varphi}(0)
	=
	0
	\ .
\end{equation}
Similarly to the case of mass perturbations, we shall only consider circular orbits
($u^{(0)} = u_{\rm c}=u_0$ and $\delta u_0 \equiv \delta u_{\rm c}$)
for $\gamma=1$ and distinguish the massive and massless cases.
\par
In the massive case ($\epsilon = 1$), we distinguish the internal and the external orbits:
\begin{enumerate}
\item
For the internal orbit $\u_{\rm c} = u_{\rm in}$, we can write the perturbation as
\begin{equation}
u^{(1)}
=
\delta \u_{\rm c} \cosh\left(\sqrt{\alpha_{\rm c}}\, \varphi\right)
\ ,
\end{equation}
where $\alpha_{\rm c} = \alpha$.
The perturbed solution
\begin{equation}
u(\varphi)
\simeq
u_{\rm in}
\left[1
+
\frac{\delta \u_{\rm c}}{u_{\rm in}}\,
\cosh\left(\sqrt{\alpha}\, \varphi\right)
\right]
\end{equation}
is therefore unstable.
\item
For the external orbit $\u_{\rm c} = u_{\rm out}$, the perturbation is given by 
\begin{equation}
	u^{(1)}
	=
	\delta u_{\rm c}\, \cos\left(\sqrt{\abs{\alpha_{\rm c}}}\, \varphi\right)
	\ ,
	\end{equation}
with $\alpha_{\rm c}=-\alpha$.
The perturbed solution reads
\begin{equation}
	u(\varphi)
	=
	u_{\rm out}
	\left[1
	+ \frac{\delta u_{\rm c}}{u_{\rm out}}\,\cos\left(\sqrt{\alpha}\, \varphi\right)
	\right]
	\ ,
\end{equation}
which is stable, with oscillations of amplitude $2\,\delta \u_{\rm c}$ around the circular orbit.
\end{enumerate}
\par	
For the massless case ($\epsilon = 0$), there is one possible circular orbit~\eqref{eq:rc-GR-massless}
with $\u_{\rm c} = u_{\rm ph}=L^2/3\, G^2\,M$ and Eq.~\eqref{eq:perturbed-u0} reads
\begin{equation}
\frac{\d^2 u^{(1)}}{\d\varphi^2}
=
u^{(1)}
\ ,
\end{equation}
with solution
\begin{equation}
u^{(1)}
=
\delta \u_{\rm c}
\cosh\left(\varphi\right)
\ .
\end{equation}
As expected, the orbit is unstable,
\begin{equation}
u(\varphi)
\simeq
\u_{\rm ph}
\left[1 + \frac{\delta \u_{\rm c}}{u_{\rm ph}}\, \cosh\left(\varphi\right)
\right]
\ .
\end{equation}
\par
We can now easily compare the effect of a perturbation in the initial position
with the one produced by a perturbation in the central mass by looking at
Table~\ref{T1}.
It appears that the two effects are very similar, in fact, and in the following we
shall focus on the first case in the table. 
\begin{table}[t]
\centering
\begin{tabular}{|c | c | c | c |}
\hline
$\epsilon$ & $u_{\rm c}$ & $u(\varphi;\delta M)$ & $u(\varphi;\delta u_0)$
\\
\hline
\hline
1
& $\phantom{\strut\displaystyle\frac{\frac{A}{B}}{\frac{C}{D}}}
u_{\rm in}
\phantom{\strut\displaystyle\frac{\frac{A}{B}}{\frac{C}{D}}}$
in Eq.~\eqref{uin}
&
$\phantom{\strut\displaystyle\frac{\frac{A}{B}}{\frac{C}{D}}}
u_{\rm in}
\left[ 1 + \frac{\delta M}{M}\, \frac{ \cosh\left(\sqrt{\alpha}\, \varphi\right) - 1 }{\alpha} \right]
\phantom{\strut\displaystyle\frac{\frac{A}{B}}{\frac{C}{D}}}$
&
$u_{\rm in}\left[1+\frac{\delta \u_{\rm c}}{u_{\rm in}}\,\cosh\left(\sqrt{\alpha}\, \varphi\right)\right]$
\\
\hline
1
&
$\phantom{\strut\displaystyle\frac{\frac{A}{B}}{\frac{C}{D}}}
u_{\rm out}
\phantom{\strut\displaystyle\frac{\frac{A}{B}}{\frac{C}{D}}}$
in Eq.~\eqref{uout}
&
$u_{\rm out}\left[ 1 + \frac{\delta M}{M}\, \frac{ 1-\cos\left(\sqrt{\alpha}\, \varphi\right)}{\alpha} \right]$ 
&
$u_{\rm out}\left[1+\frac{\delta u_{\rm c}}{u_{\rm out}}\,\cos\left(\sqrt{\alpha}\, \varphi\right)\right]$
\\
\hline
0
&
$\phantom{\strut\displaystyle\frac{\frac{A}{B}}{\frac{C}{D}}}
u_{\rm ph}
\phantom{\strut\displaystyle\frac{\frac{A}{B}}{\frac{C}{D}}}$
in Eq.~\eqref{UPH}
&
$u_{\rm ph}\left\{ 1 + \frac{\delta M}{M}  \left[\cosh(\varphi) - 1\right]\right\}$
&
$\u_{\rm ph}\left[1 + \frac{\delta \u_{\rm c}}{u_{\rm ph}}\, \cosh\left(\varphi\right)\right]$
\\
\hline
\end{tabular}
\caption{Overview of perturbed solutions.}
\label{T1}
\end{table}
\section{Unstable circular orbits}
\label{S:unstableCircular}
In this section we want to analyse more in details the unstable inner circular orbit $r_{\rm in}$ for
massive particles.
Since the analysis will be performed numerically, it is convenient to introduce the dimensionless
variables
\begin{equation}
\tilde L
\equiv
\frac{L}{G\,M}
\ ,
\qquad
\tilde r_{\rm in}
\equiv
\frac{r_{\rm in}}{G\,M}
\ ,
\end{equation}
and recall that (two) circular orbits exist under the condition $\tilde L \ge \sqrt{12}$.
In particular, as discussed in Section~\ref{sec:GR-circular-orbits}, the inner radius
\begin{equation}
	3
	=\tilde r_{\rm in}(\tilde L\to\infty)
	\le
	\tilde r_{\rm in}
	\le
	\tilde r_{\rm in}(\tilde L=\sqrt{12})
	= 6
	 \ .
\end{equation}
\par
Since the energy and the angular momentum of test particles moving on the internal
circular orbits are connected by the relation
\begin{equation}
E
=
108\left[36 - \tilde L^{-1} \left(\tilde L^2 - 12\right)^{3/2} + \tilde L\right]
\ ,
\end{equation}
the effects of a perturbation of the mass $M$ can be determined 
by considering $L$ as the only free parameter.
\par
\begin{figure}
\centering
	\includegraphics[scale=0.60]{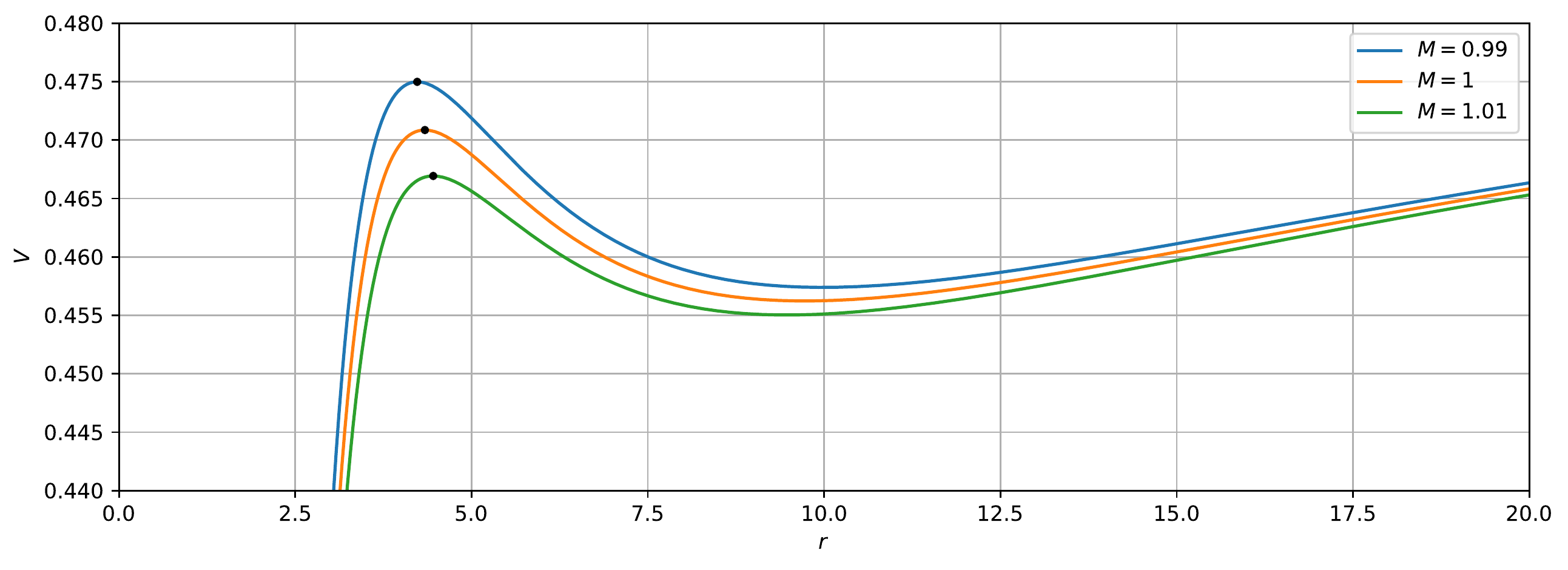}
	\caption{Potential $V$ as a function of the radial coordinate $r$ for values of the central
	mass $M=0.99, 1, 1.01$ (with $G=1$, $L=3.75$).
	Black dots represent the local maximum of the potential, corresponding to the unstable
	circular orbits.
	\label{fig:V} }
\end{figure}

It is worth noticing that, in our analysis, an instantaneous reduction of the mass $M$,
i.e.~a perturbation $\delta M<0$, leads to a smaller radius for the circular orbit and to
an increase of the corresponding potential $V$ (see Fig.~\ref{fig:V}).
After the perturbation, the particle orbiting on the circular orbit would happen to have
an energy lower than the potential energy $V$.
We shall therefore consider only perturbations $\delta M>0$, i.e.~instantaneous increases
of the mass $M$.
(Note that in the analytical perturbation method, solutions are found under the condition
$d u/d \varphi(0) = 0$, see Eq.~\eqref{eq:perturbed-ic} and following comment.)
\par
\begin{figure}
\centering
	\includegraphics[scale=0.60]{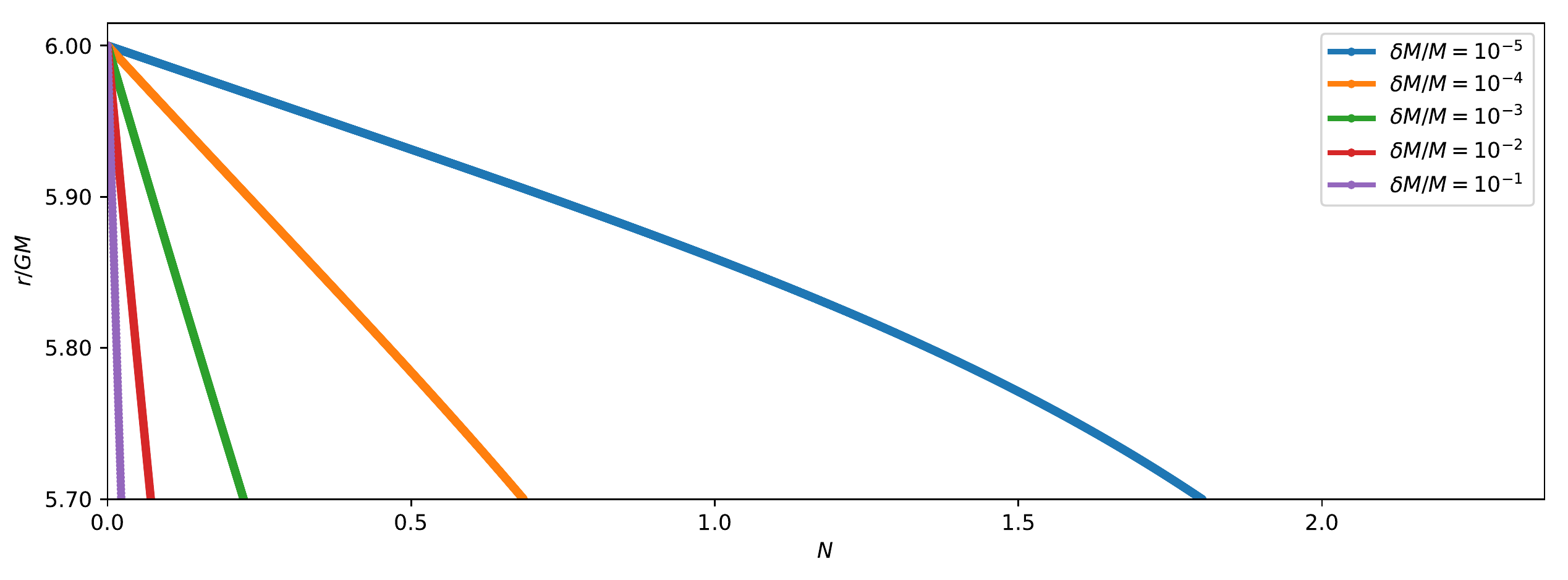}
	\\
	\includegraphics[scale=0.44]{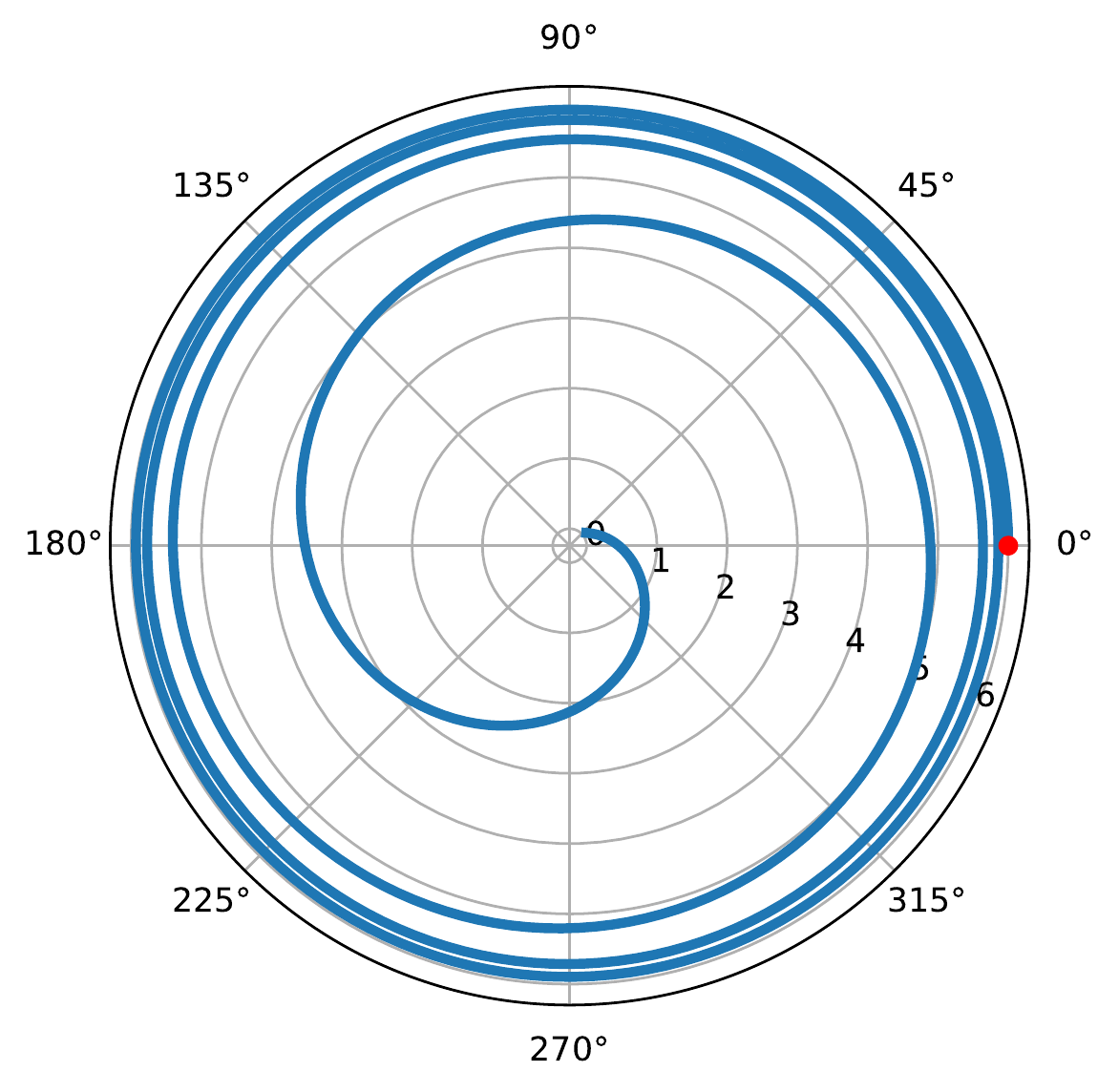}%
	\includegraphics[scale=0.44]{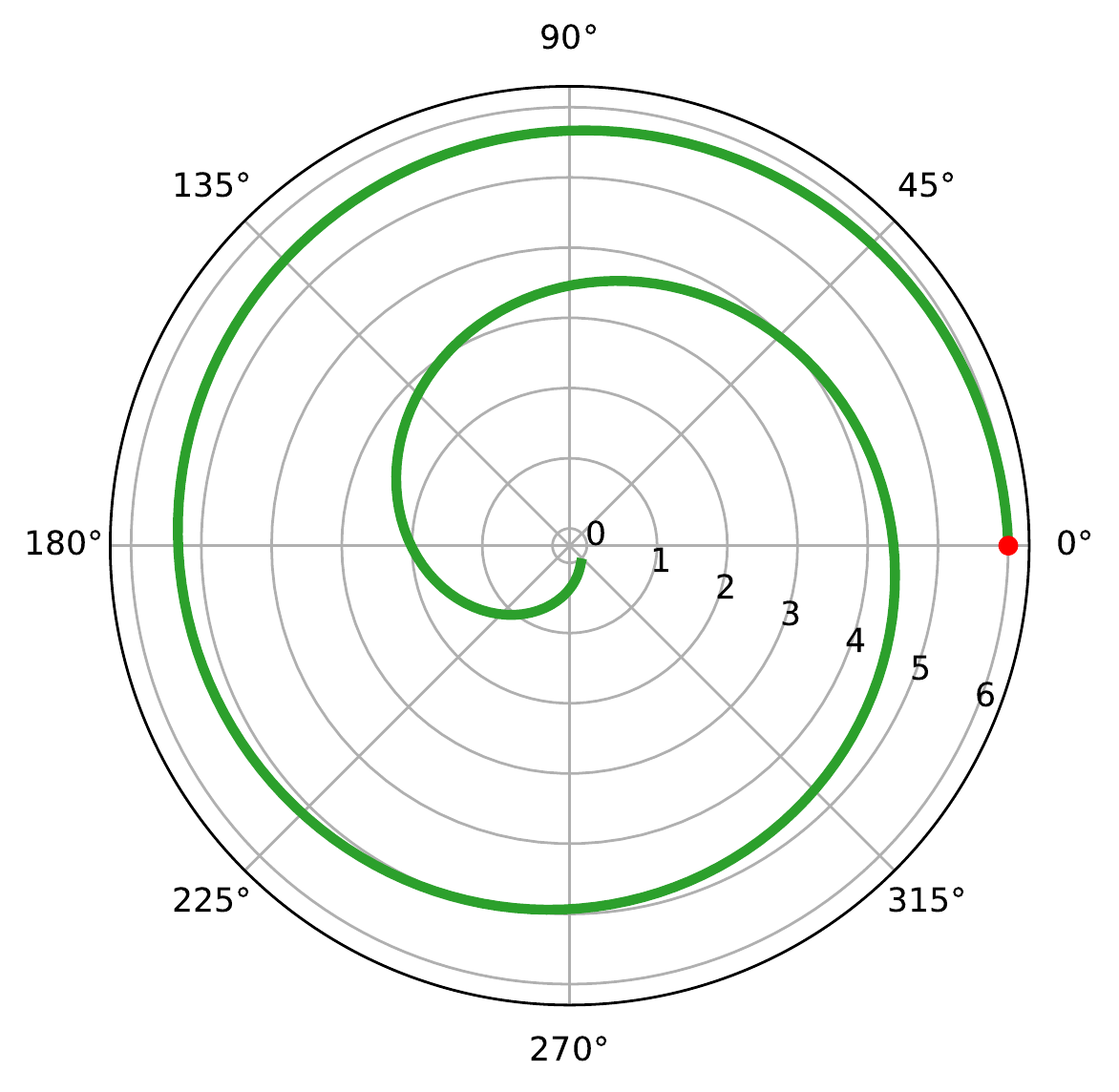}%
	\includegraphics[scale=0.44]{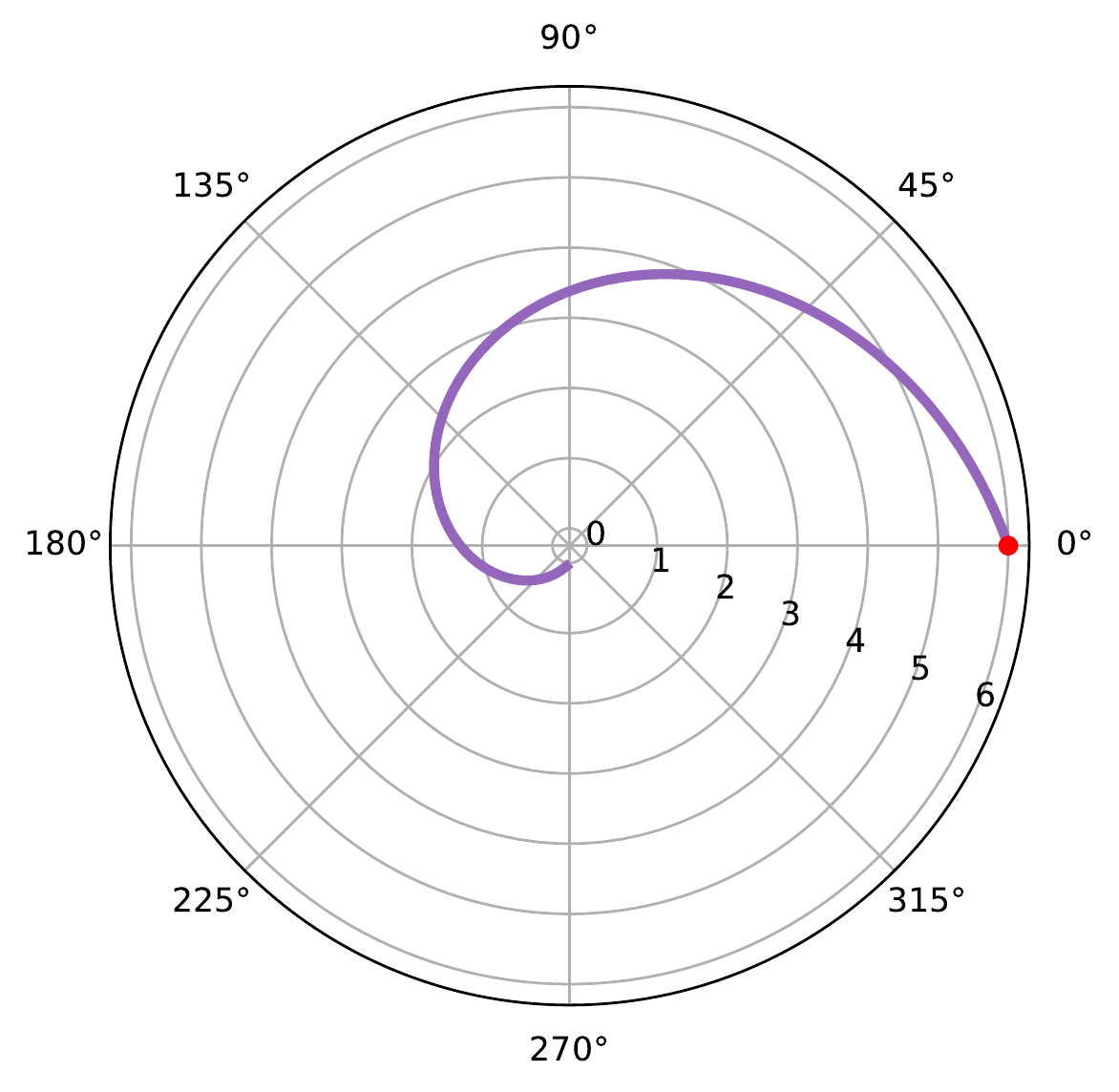}
	\caption{Radial coordinate $\tilde r = r/G\,M$ of trajectories subject to perturbations
	$\delta M$ starting from the unstable circular orbit for $\tilde L = L / G\,M = \sqrt{12}$
	(top panel); trajectories subject to the perturbation $\delta M / M = 10^{-5}$ (bottom left panel),
	$\delta M / M = 10^{-3}$ (bottom center panel) and $\delta M / M = 10^{-1}$ (bottom right panel).
	\label{fig:caseN_1a} }
\end{figure}
\begin{figure}
\centering
	\includegraphics[scale=0.60]{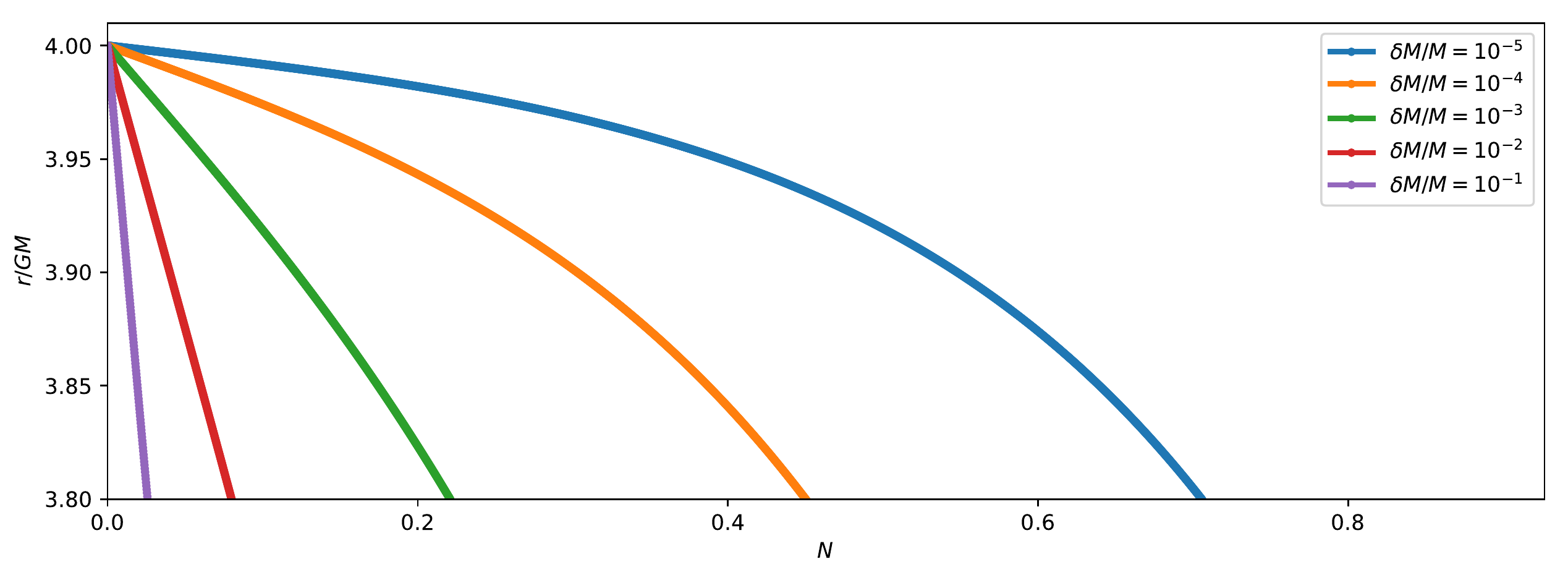}
	\\
	\includegraphics[scale=0.44]{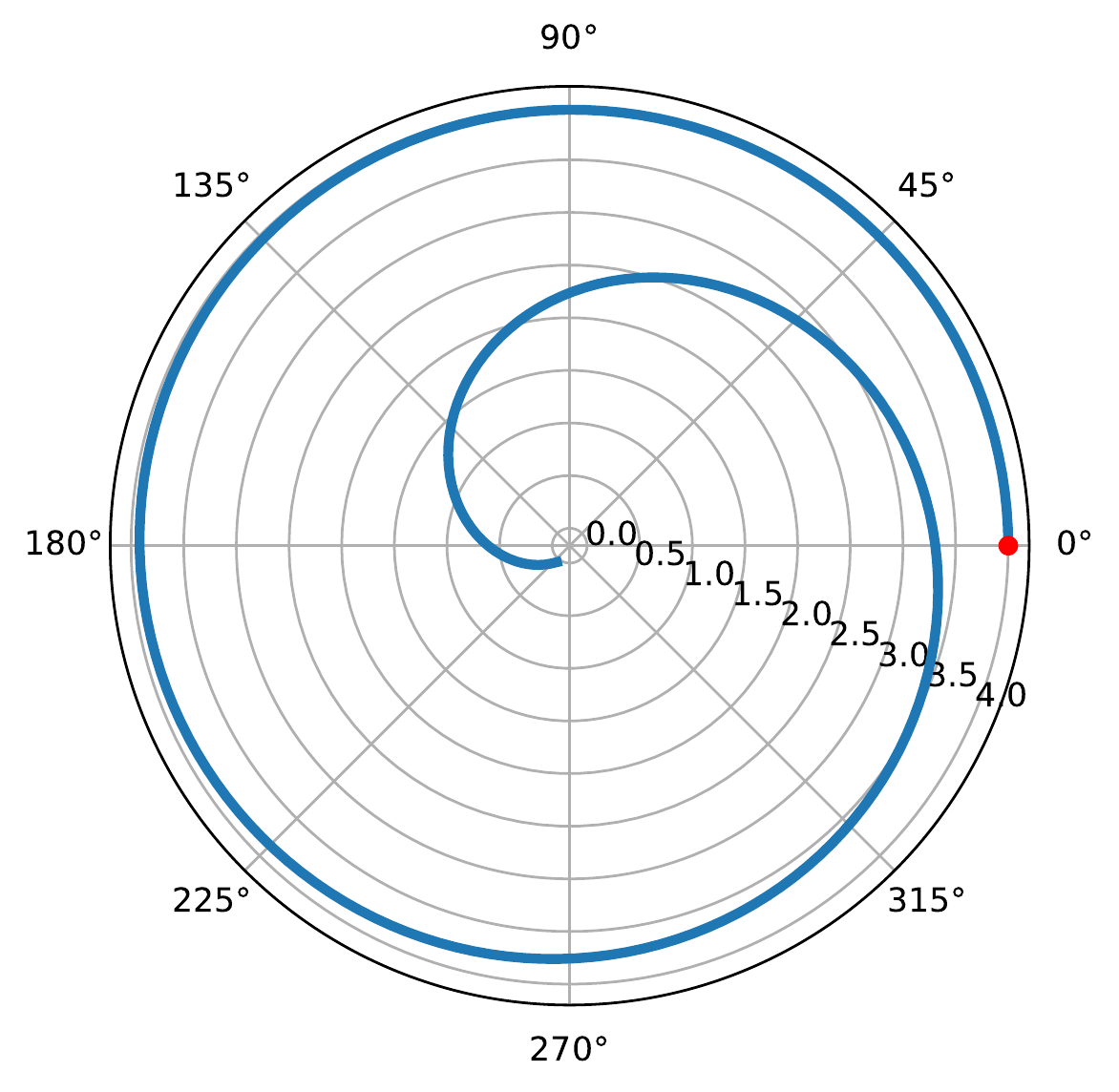}%
	\includegraphics[scale=0.44]{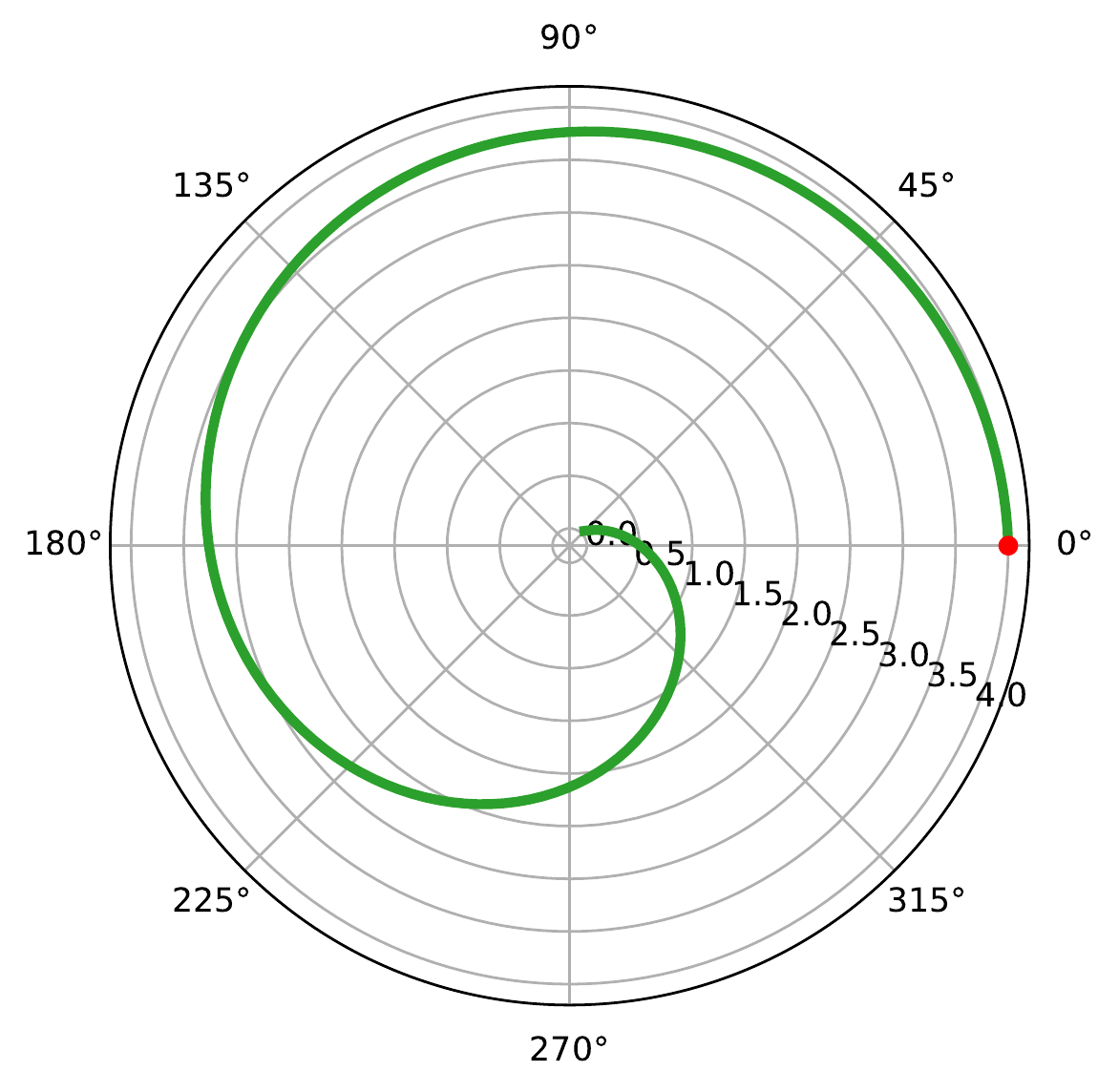}%
	\includegraphics[scale=0.44]{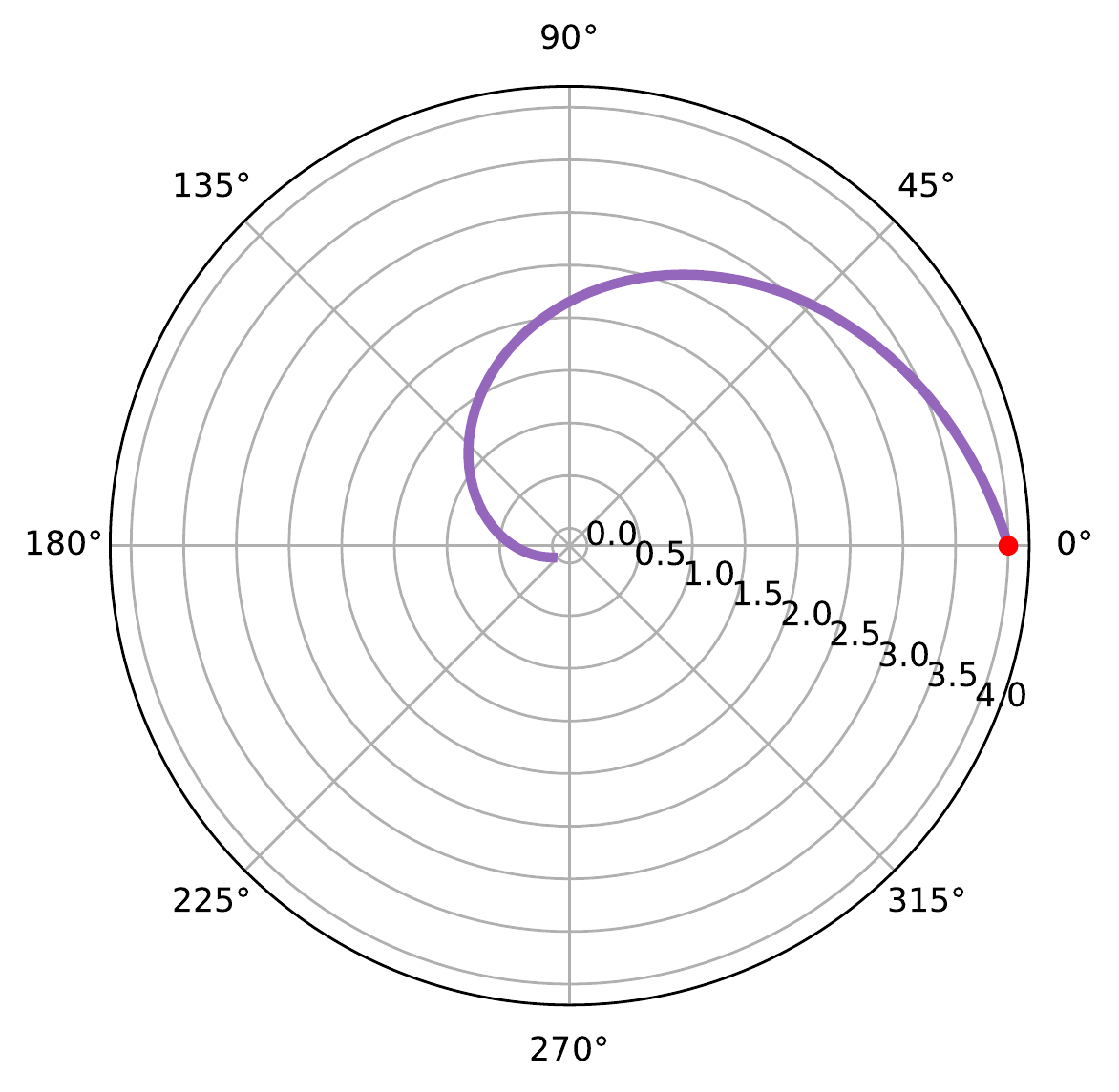}
	\caption{Radial coordinate $\tilde r = r/G\,M$ of trajectories subject to perturbations
	$\delta M $ starting from the unstable circular orbit for $\tilde L = L / G\,M = 4$ (top panel);
	trajectories subject to the perturbation $\delta M / M = 10^{-5}$ (bottom left panel),
	$\delta M / M = 10^{-3}$ (bottom center panel) and $\delta M / M = 10^{-1}$ (bottom right panel).
	\label{fig:caseN_1b} }
\end{figure}
\begin{figure}
\centering
	\includegraphics[scale=0.60]{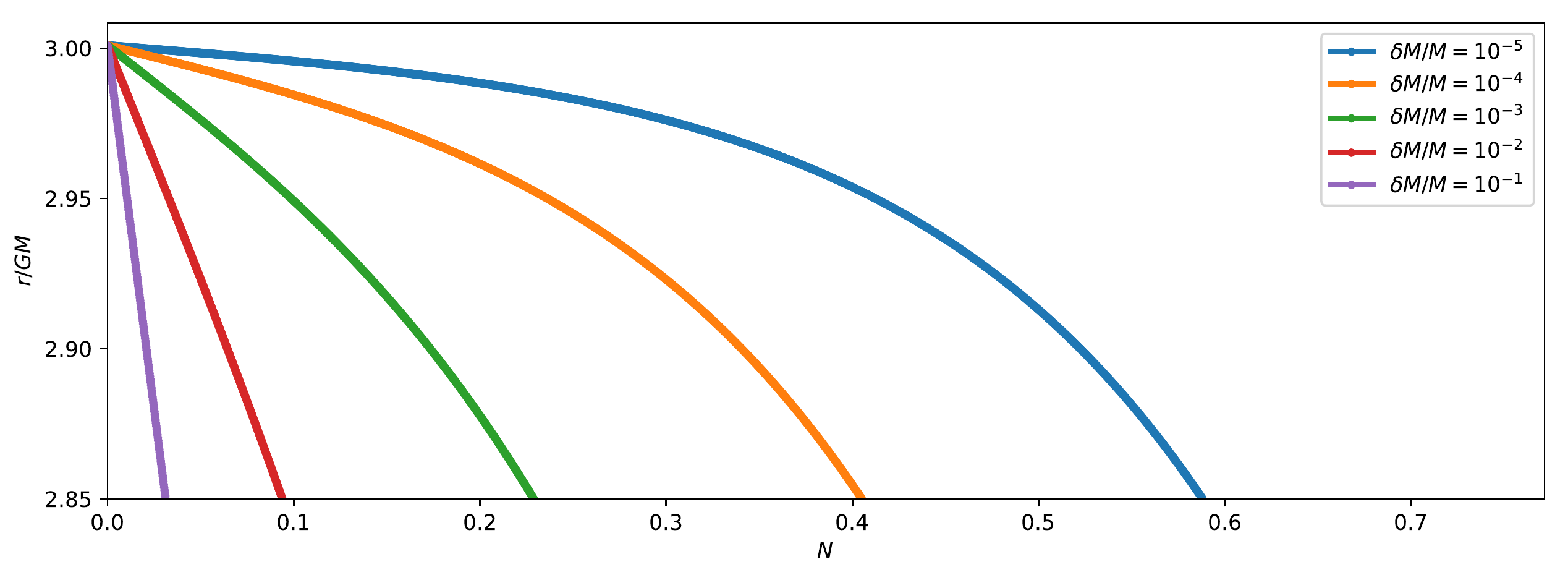}
	\\
	\includegraphics[scale=0.44]{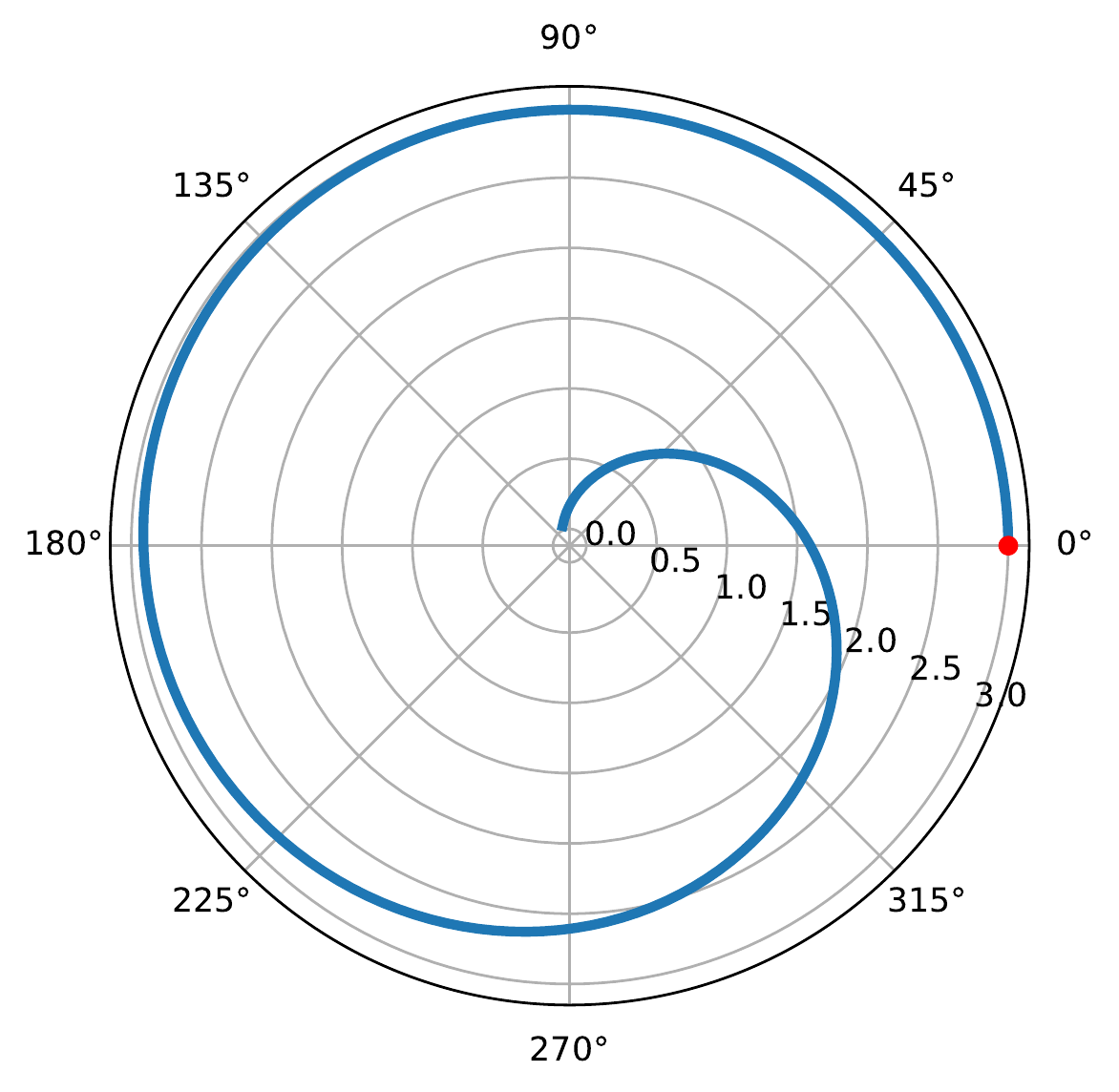}%
	\includegraphics[scale=0.44]{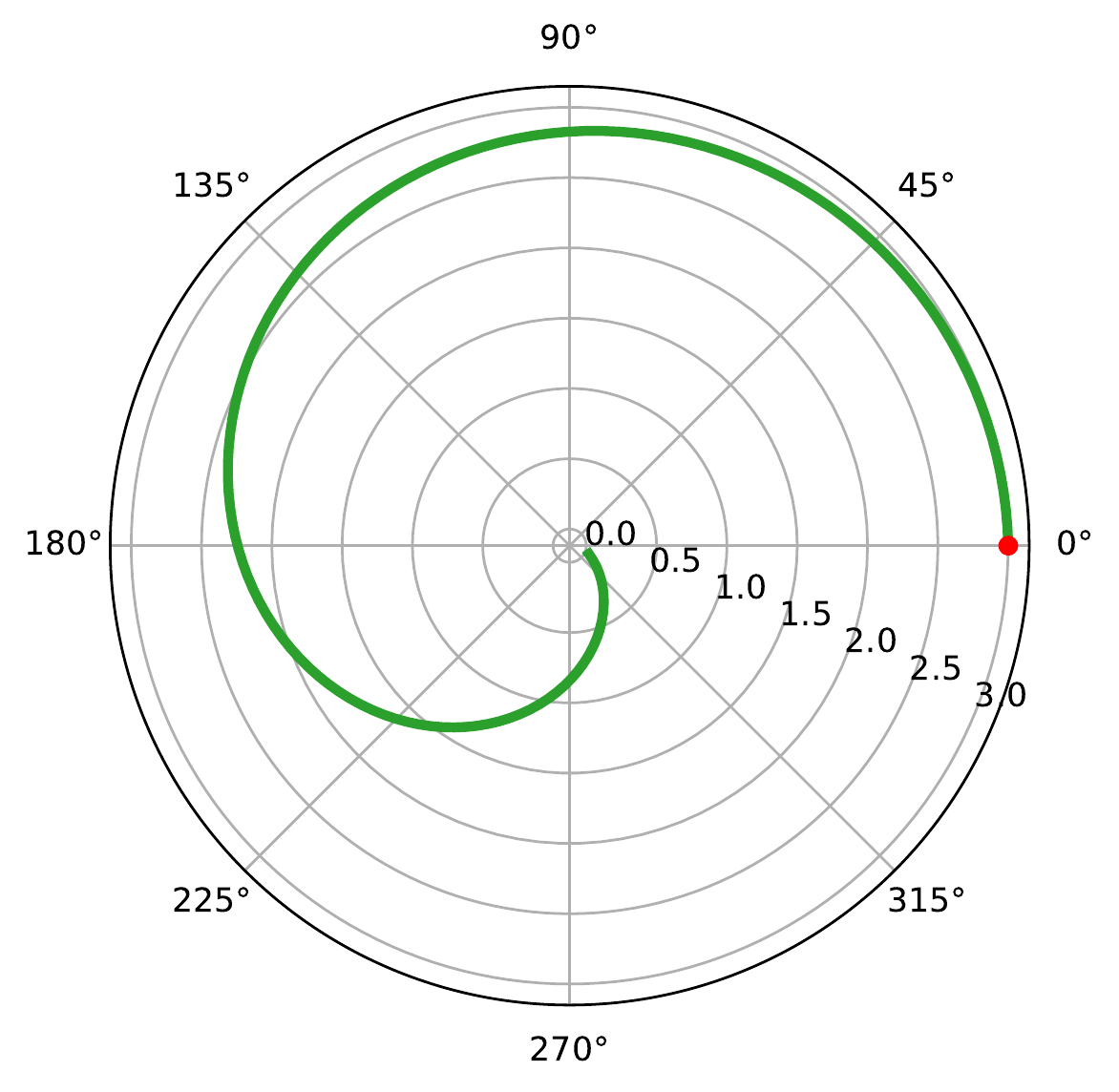}%
	\includegraphics[scale=0.44]{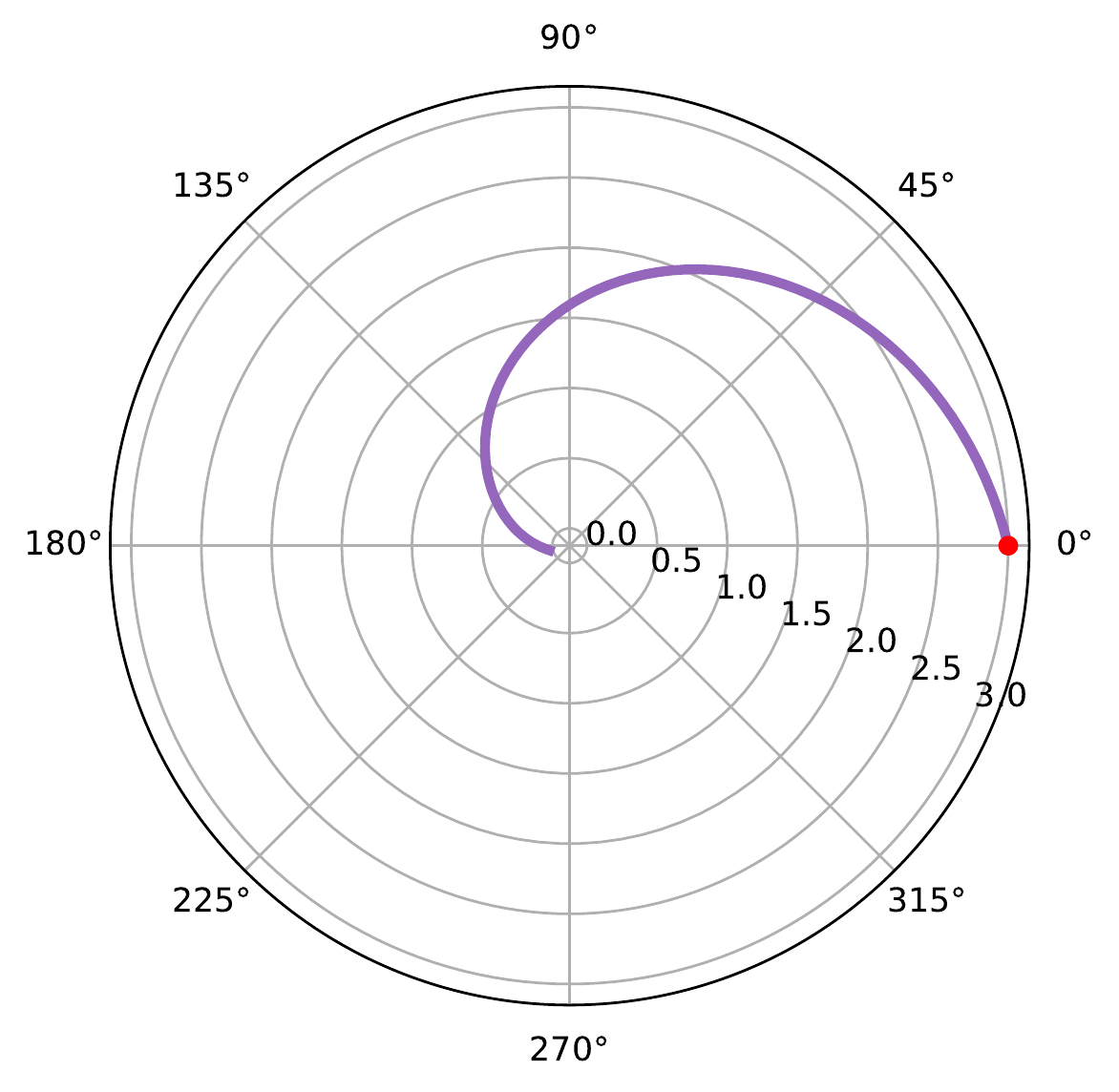}
	\caption{Radial coordinate $\tilde r = r/G\,M$ of trajectories subject to perturbations
	$\delta M $ starting from the unstable circular orbit for $\tilde L = L / G\,M = 100$ (top panel);
	trajectories subject to the perturbation $\delta M / M = 10^{-5}$ (bottom left panel),
	$\delta M / M = 10^{-3}$ (bottom center panel) and $\delta M / M = 10^{-1}$ (bottom right panel).
	\label{fig:caseN_1c} }
\end{figure}
\begin{figure}
	\includegraphics[scale=0.55]{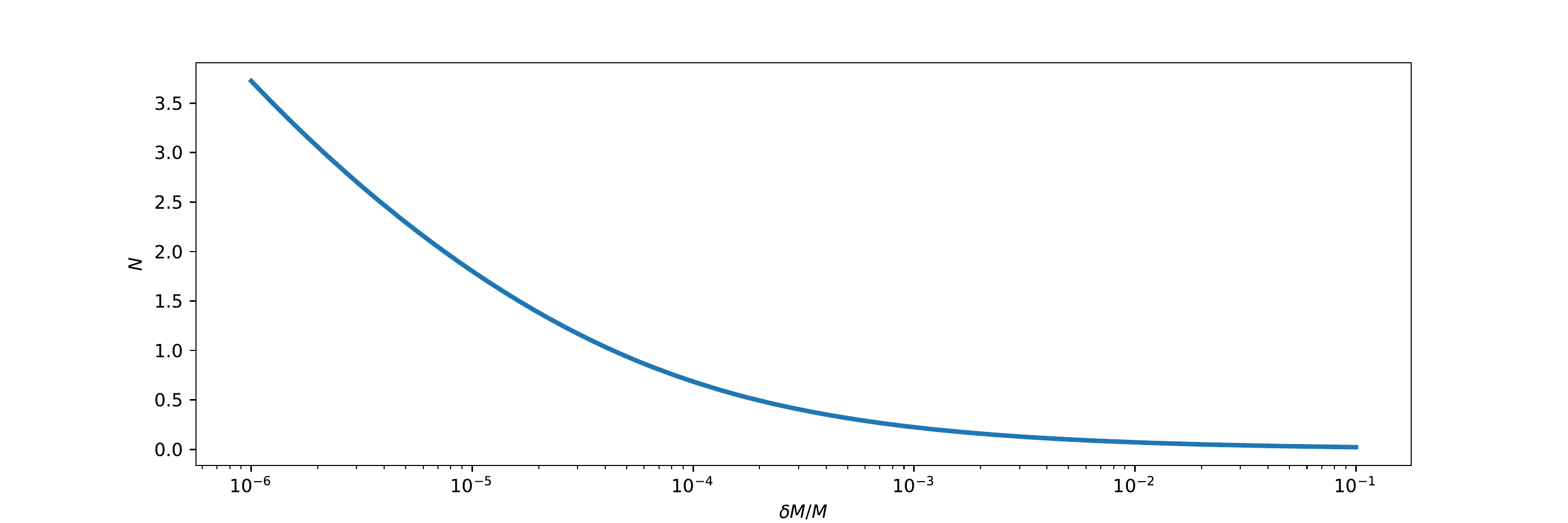}
	\caption{Number of revolutions $N$ corresponding to a change of $5\%$ in the radial coordinate
	as a function of the perturbation $\delta M /M$ ranging from $10^{-6}$ to $10^{-1}$.
	\label{fig:caseN_2} }
\end{figure}

The change of the radial coordinate of the particle as a function of the angle $\varphi$ following
an instantaneous perturbation $\delta M > 0$ are shown in Fig.~\ref{fig:caseN_1a},
Fig.~\ref{fig:caseN_1b} and Fig.~\ref{fig:caseN_1c} for three values of the angular momentum,
namely
\begin{enumerate}
\item
$\tilde L = \sqrt{12}$, which is the minimum possible value for which the circular orbit exists.
The radius of the circular orbit in this case is $\tilde r_{\rm in} = 6$ (Fig.~\ref{fig:caseN_1a});
\item
$\tilde L = 4$, for which the radius of the circular orbit is $\tilde r_{\rm in} = 4$
(Fig.~\ref{fig:caseN_1b});
\item
$\tilde L = 100$, for which the radius of the circular orbit is $\tilde r_{\rm in} \approx 3.0009$,
close to the minimum possible value $\tilde r_{\rm in} = 3$ (Fig.~\ref{fig:caseN_1c}).
\end{enumerate}
For all the three cases analysed here, the radial position of the particle as a function of the
number $N$ of revolutions ($N = \varphi / 2\,\pi$) is shown for several values of the perturbation
$\delta M / M$, ranging from $10^{-5}$ to $10^{-1}$.
The results show that the number of revolutions after which the particle decrease its radial
coordinate by 5\% of the original value is always less than 2, and decreases as the perturbation
increases or the angular momentum increases.
For each of the three cases, the polar representations of the orbits are shown for three
selected values of the perturbation, namely $\delta M / M = 10^{-5},\,10^{-3},\,10^{-1}$.
The number of revolutions $N$ after which the variation of the radial position of the particle amounts
to 5\% of the initial value has then been systematically investigated for values of the perturbation parameter
$\delta M/M$ varying between $10^{-6}$ and $10^{-1}$.
The result is shown in Fig.~\ref{fig:caseN_2}. 
\par
\begin{figure}
\centering
	\includegraphics[scale=0.53]{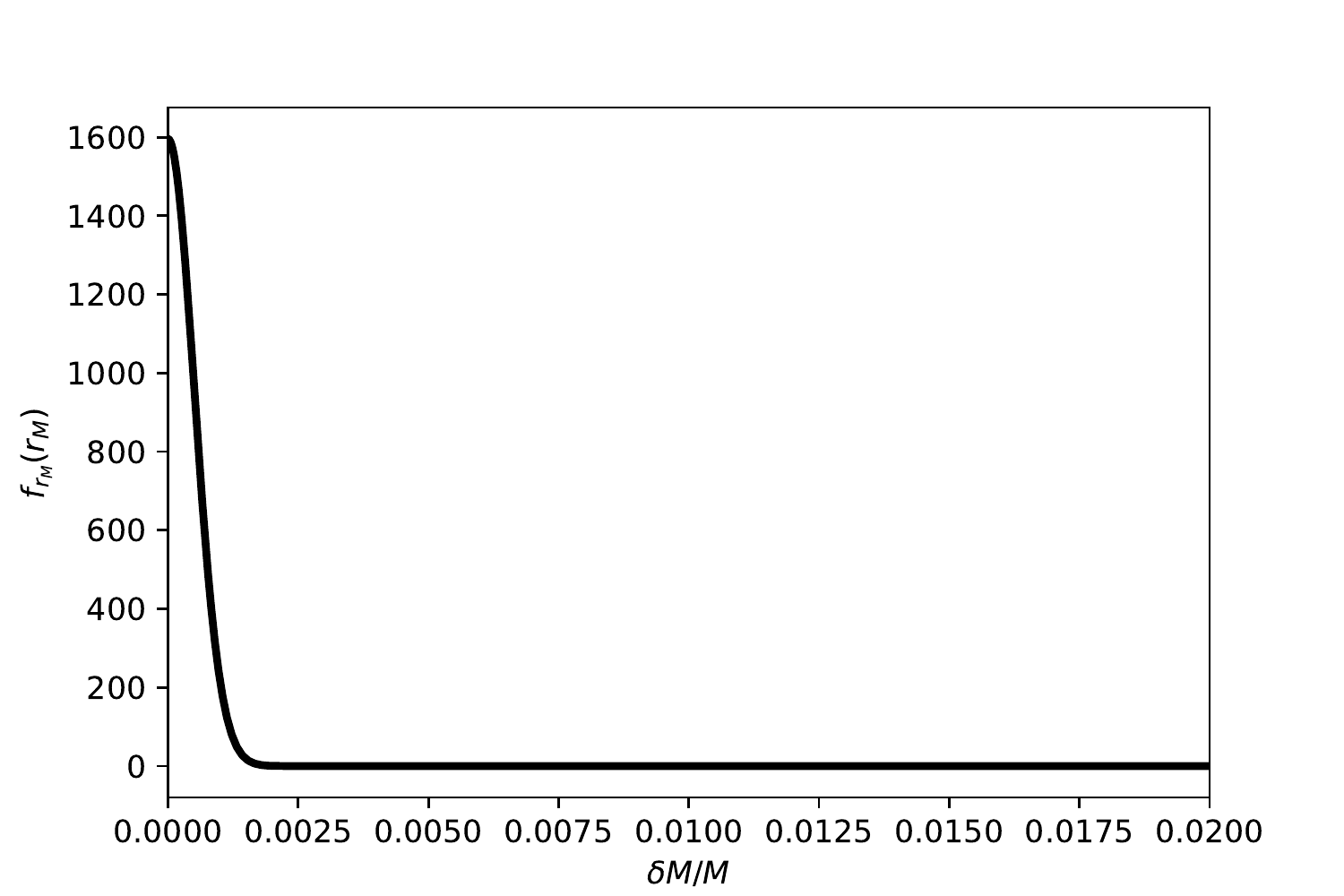}
	\includegraphics[scale=0.53]{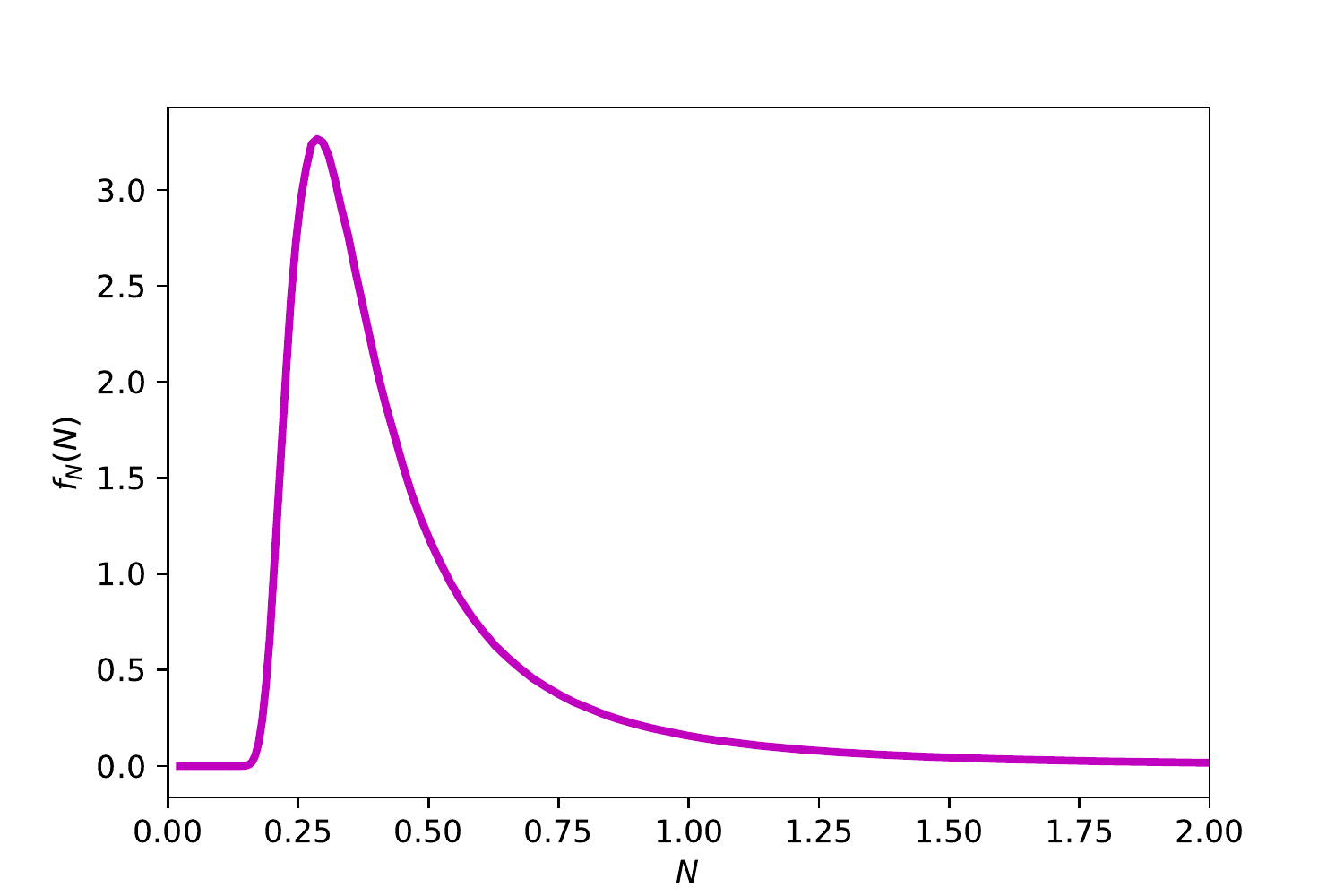}
	\caption{Probability density function of the perturbation
	$\delta M \sim \normalx{\mu_{\delta M}=0}{\sigma_{\delta M}^2 = 0.0005^2}$ (left panel)
	and corresponding probability density function of the number of revolutions $N$ after
	which the variation of the radial position of the particle amounts to $5\%$ of the initial value
	(right panel).
	\label{fig:caseN_3a} }
\end{figure}
\begin{figure}
\centering
	\includegraphics[scale=0.53]{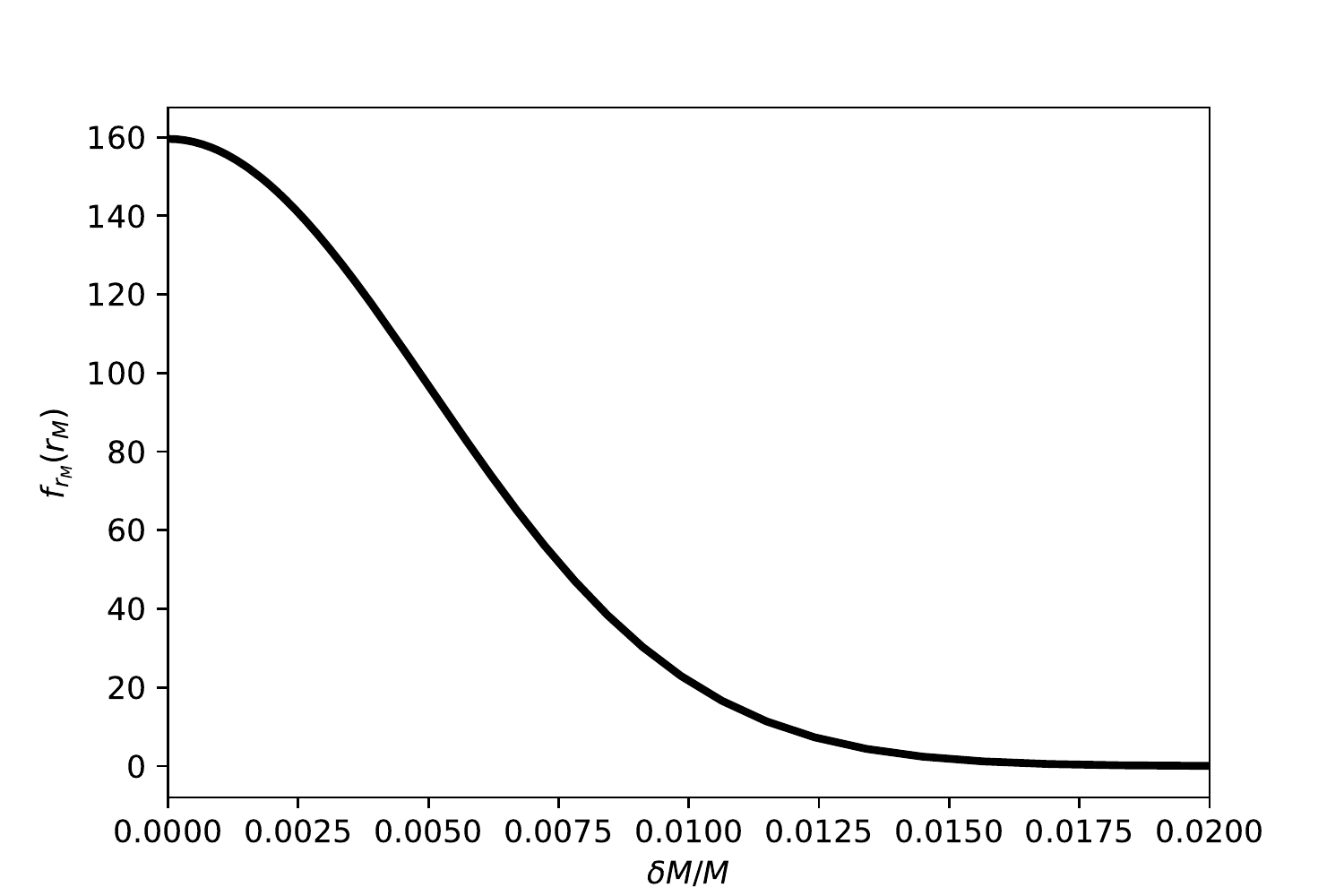}
	\includegraphics[scale=0.53]{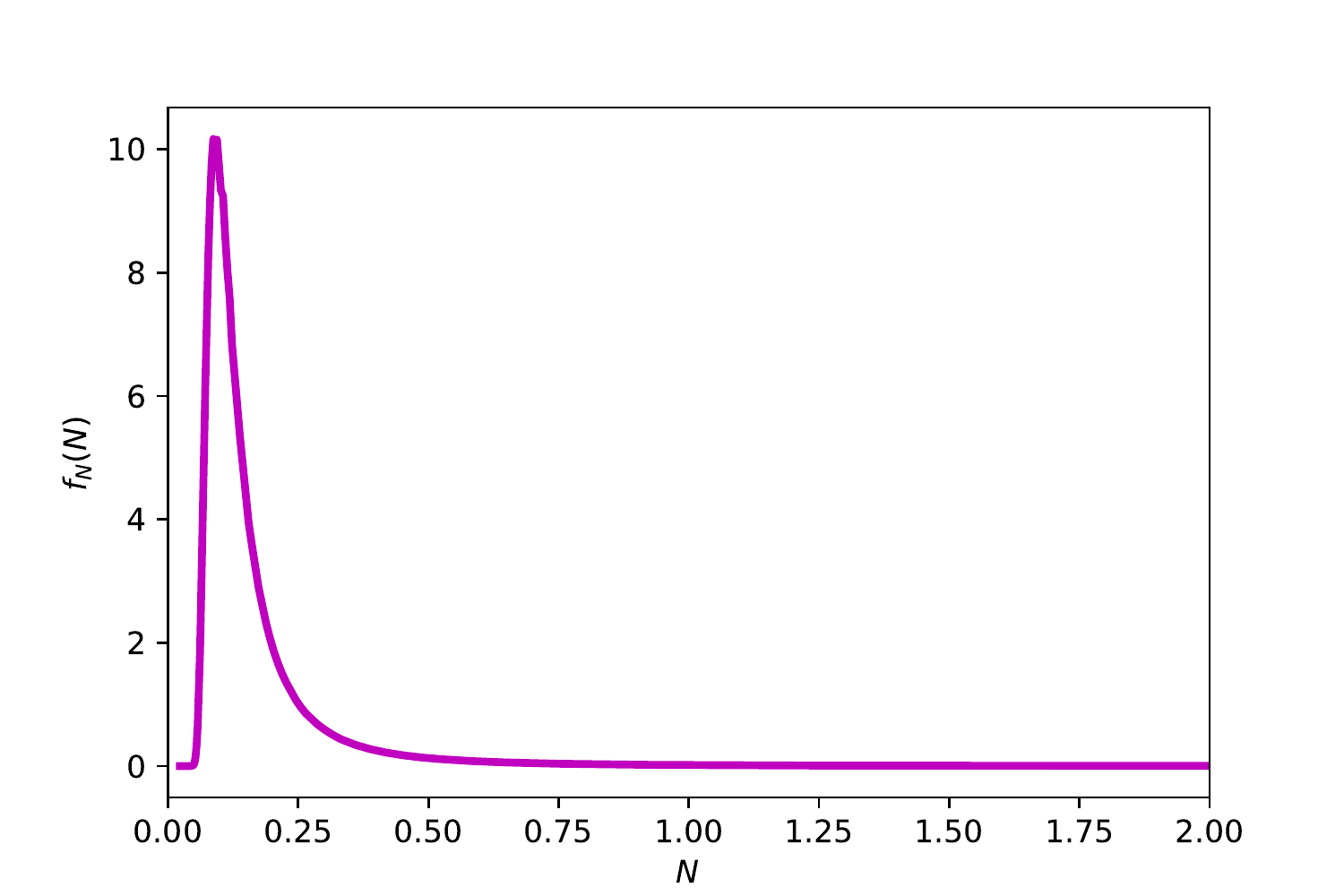}
	\caption{Probability density function of the perturbation
	$\delta M \sim \normalx{\mu_{\delta M}=0}{\sigma_{\delta M}^2 = 0.005^2}$ (left panel)
	and corresponding probability density function of the number of revolutions $N$ after
	which the variation of the radial position of the particle amounts to $5\%$ of the initial value
	(right panel).
	\label{fig:caseN_3b} }
\end{figure}
\begin{figure}
\centering
	\includegraphics[scale=0.53]{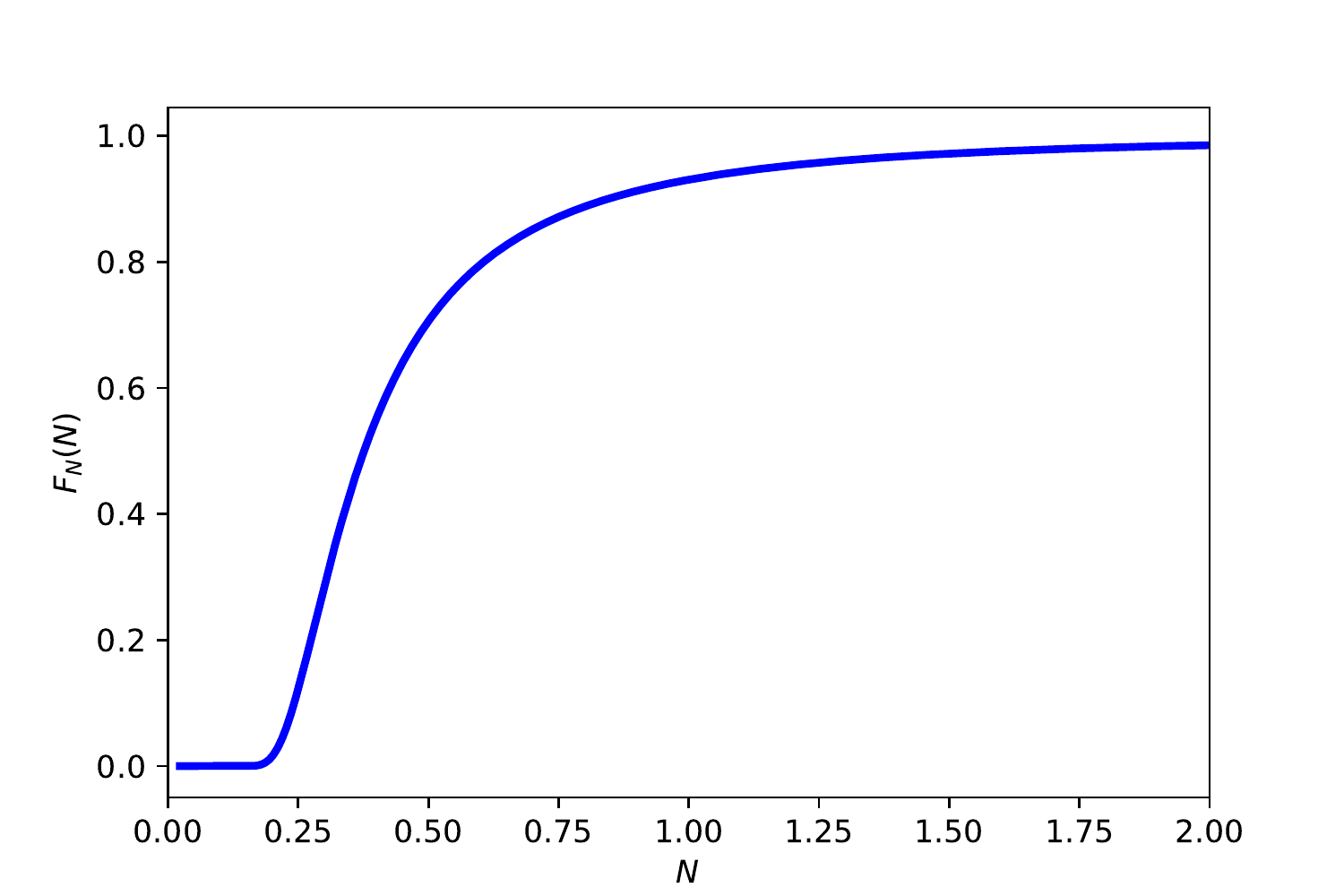}
	\includegraphics[scale=0.53]{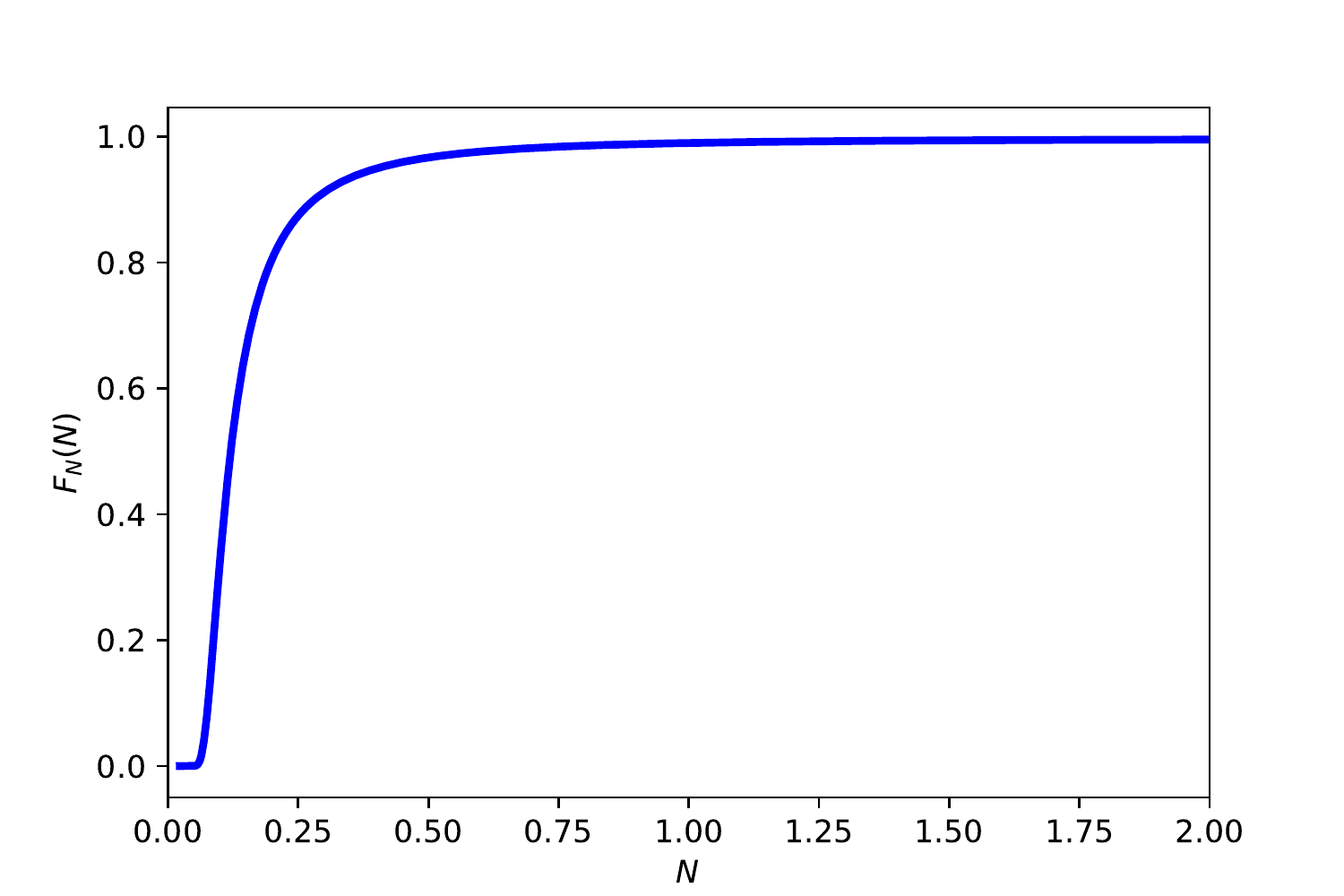}
	\caption{Cumulative distribution functions of the random variable $N$ resulting from
	a probability density function of the perturbation
	$\delta M \sim \normalx{\mu_{\delta M}=0}{\sigma_{\delta M}^2 = 0.0005^2}$ (left panel)
	and $\normalx{\mu_{\delta M}=0}{\sigma_{\delta M}^2 = 0.005^2}$ (right panel).
	\label{fig:caseN_4} }
\end{figure}

Let us next assume the perturbation $\delta M>0$ is a random variable with distribution
$\normalx{\mu_{\delta M}}{\sigma_{\delta M}}$, i.e.~a random variable with probability density function
\begin{equation}
f_{\delta M}(x)
=
\begin{cases}
0
&
x < 0
\\
\strut\displaystyle\frac{2}{\sqrt{2\pi\sigma^2}}\, \exp\left[-\frac{\left(x-\mu_{\delta M}\right)^2}{2\, \sigma_{\delta M}^2}\right]
\qquad
&
x \geq 0
\ .
\end{cases}
\end{equation}
Setting $\mu_{\delta M} = 0$ and $\sigma_{\delta M} = 0.0005$, the corresponding random variable
representing the number of revolutions $N$ after which the variation of the radial coordinate amounts
to $5\%$ of the initial value is shown in Fig.~\ref{fig:caseN_3a}.
The same analysis for a ten times larger value of $\sigma_{\delta M}= 0.005$
is also shown in Fig.~\ref{fig:caseN_3b}.
The cumulative distribution functions of the random variable $N$ for the two cases $\sigma_{\delta M} = 0.0005$
and $\sigma_{\delta M} = 0.005$ are shown in Fig.~\ref{fig:caseN_4}.
\section{Discussion and outlook}
\label{S:conc}
In this work we started to consider in detail the effects of uncertainties in the determination of the
black hole mass on geodesics, mainly motivated by studies of quantum aspects of black holes~\cite{HQM-1,HQM-2,HQM-3,HQM-4,hu}.
In particular, we analysed the consequences of a randomly distributed mass on qualitatively different
orbits and we were interested in the possible differences between such effects and those
simply stemming from the experimental uncertainties in the measurements of masses and positions.
It is in fact important to be able to tell apart the two sources of uncertainties if one eventually wishes
to look for experimental evidence of quantum gravity in black hole physics.
\par
A comparison between the two kinds of uncertainties for circular orbits was considered in Section~\ref{S:perturbedCircular},
where we showed that they appear functionally very similar, which points to the fact that it would be very difficult to identify
quantum effects from experimental data~\cite{Falcke:2018uqa} about such orbits of test bodies obtained, for instance,
by the Event Horizon Telescope~\cite{Ricarte:2014nca} and 	
BlackHoleCam~\cite{Goddi:2017pfy, Dokuchaev:2018ibr}.
In the search for more significant signatures of quantum effects, beside considering more explicitly 
models of quantum
black holes~\cite{giusti,Casadio:2013ulk,Casadio:2014vja,Casadio:2015bna,Carr:2015nqa,Mureika:2018gxl,Rovelli:2014cta,DeLorenzo:2015taa},
we also intend to study other effects occurring on black hole space-times.
In particular, it will be interesting to investigate the red-shift of signals emitted by sources either falling towards
the black hole along (perturbed and unperturbed geodesics) as well as following perturbed 
unstable circular orbits.
Given the precision with which the red-shift can be measured, this might be a more promising
route towards experimental quantum gravity.
\subsection*{Acknowledgments}
This work has been carried out in the framework of the activities of the National Group for Mathematical Physics (GNFM/INdAM).
R.C.~and A.G.~are partially supported by the INFN grant FLAG.
\appendix
\section{Equivalent analytic solutions}
\label{A:equivalent}
Introducing the change of variable $\tilde x = (3\,\rho\,x - 1)/{12}$, Eq.~\eqref{eq:dx_dphi_2} can be written
as~\cite{whittaker-watson, Pastras}
\begin{equation}
\label{eq:orbit_ord1_w}
	\left( \frac{\d\tilde x}{\d\varphi}\right)^2
	=
	4\, \tilde x^3 - g_2\, \tilde x - g_3
	\ ,
\end{equation}
with
\begin{equation}
	g_2
	=
	\frac{1}{12} - \frac{\rho^2}{4\,\alpha^2}
	\ ,
	\qquad
	g_3
	=
	\frac{1}{216} + \left(\frac{1}{24\, \alpha^2} - \frac{1}{16\, \beta^2} \right) \rho^2
	\ , 
\end{equation}
whose general solution is given by the Weierstrass elliptic function
$\wp\left(z; g_2, g_3\right)$ of parameters $\left(g_2, g_3\right)$. 
Hence, the general solution of Eq.~\eqref{eq:dx_dphi_2} can be expressed as
\begin{equation}
\label{eq:orbit_sol_w}
	x(\varphi)
	=
	\frac{1}{\rho} \left[ 4\, \wp\left( \varphi + \tilde \delta;\, g_2, g_3\right) + \frac{1}{3} \right]
	\ ,
\end{equation}
where the constant $\tilde\delta$ is again obtained from the initial condition $x(\varphi_0) = x_0$.
\par
Moreover, the Weierstrass elliptic function $\wp$ can be written in terms of Jacobi's elliptic function sn as
\begin{equation}
	\wp(z)
	=e_1
	+\frac{e_3 - e_1}{\sn{2}{z\, \sqrt{e_3 - e_1}}{\bar k}}
	\ ,
	\qquad
	{\rm with}
	\
	\bar k
	=
	\sqrt{\frac{e_2 - e_1}{e_3 - e_1}}
	\ ,
\end{equation}
where $e_1$, $e_2$, $e_3$ are the roots of the polynomial on the right-hand side of Eq.~\eqref{eq:orbit_ord1_w}, i.e.
\begin{equation}
	4\, \tilde x^3 - g_2\, \tilde x - g_3
	=
	4\left(\tilde x - e_1\right) \left(\tilde x - e_2\right)\left(\tilde x - e_3\right)
	\ ,
\end{equation}
with $e_1 + e_2 + e_3 = 0$.
Taking into account that, for $j=1,2,3$, 
\begin{equation}
	e_j
	=
	\frac{\rho}{4}\, \x_j - \frac{1}{12}
	\ , 
\end{equation}
we see that $\bar k = k$ in Eq.~\eqref{k}.
Making use of the relation~\cite{whittaker-watson}
\begin{equation}
	\wp(z)
	=
	e_1
	+\frac{e_3 - e_1}{\sn{2}{z\,\sqrt{e_3 - e_1}}{k}}
	=
	e_1
	+
	\frac{e_2 - e_1}{k^2\,\sn{2}{z\sqrt{e_3 - e_1}}{k}}
	\ ,
\end{equation}
the solution~\eqref{eq:orbit_ord1_w} can be also written as 
\begin{equation}
\label{eq:orbit_sol_w2}
	x(\varphi)
	= 
	x_1 + \frac{x_2 - x_1}{k^2\,\sn{2}{\frac{\varphi}{2}\, \sqrt{\rho\left(x_3 - x_1\right)} + \bar\delta}{k}}
\ ,
\end{equation}
where $\bar\delta = \tilde\delta \sqrt{\rho\left(\x_3 - \x_1\right)}/2$.
Finally, making use of the relation~\cite{whittaker-watson}
\begin{equation}
	\wp(z)
	=
	e_1
	+
	\left(e_2 - e_1\right) \sn{2}{\left(z-\omega_3\right) \sqrt{e_1 - e_3}}{k}
	\ ,
\end{equation}
where $\wp\left(\omega_3/2\right) = e_3$, Eq.~\eqref{eq:orbit_ord1_w} can be finally written as
\begin{equation}
\label{eq:orbit_sol_sn2}
	x(\varphi)
	=
	x_1
	+
	\left(x_2 - x_1\right) \sn{2}{\frac{\varphi}{2}\, \sqrt{\rho\left(x_3 - x_1\right)} + \delta^*}{k}
	\ ,
\end{equation}
where $\delta^* = (\tilde\delta - \omega_3)\,\sqrt{\rho\left(x_3 - x_1\right)}/2$.
Of course, the three expressions~\eqref{eq:orbit_sol_sn}, \eqref{eq:orbit_sol_w2} and \eqref{eq:orbit_sol_sn2}
are perfectly equivalent.
%
%
%
%
%
%%%
%%%
%%% Bibliography
%%%

%
%
%

%
\end{document}